\documentclass{aa}  

\usepackage{graphicx}
\usepackage{epstopdf}
\usepackage{color}
\usepackage{amsmath}
\usepackage{subcaption}
\usepackage{booktabs}

\usepackage{txfonts}
\newcommand{\rxte}{{\it RXTE}}

\newcommand{\rxtepca}{{\it RXTE}/PCA}

\newcommand{\nustar}{{\it NuSTAR}}

\newcommand{\maxi}{{\it MAXI}/GSC}

\newcommand{\chandra}{{\it Chandra}}

\begin{document}

   \title{The obscured X-ray binaries V404 Cyg, Cyg X--3, V4641 Sgr, and GRS 1915$+$105}

   \author{K.~I.~I.~Koljonen
          \inst{1,2}
          \and
          J.~A.~Tomsick\inst{3}
          }

   \institute{Finnish Centre for Astronomy with ESO (FINCA), University of Turku, V\"ais\"al\"antie 20, 21500 Piikki\"o, Finland \\
              \email{karri.koljonen@utu.fi}
         \and
             Aalto University Mets\"ahovi Radio Observatory, PO Box 13000, FI-00076 Aalto, Finland \\
         \and    
             Space Sciences Laboratory, 7 Gauss Way, University of California, Berkeley, CA 94720-7450, USA \\           
             }

   \date{Received ; accepted }

% \abstract{}{}{}{}{} 
% 5 {} token are mandatory

  \abstract
  {}  
    {V404 Cyg, Cyg X-3, V4641 Sgr, and GRS 1915$+$105 are among the brightest X-ray binaries and display complex behavior in their multiwavelength emission. Except for Cyg X-3, the other three sources have large accretion disks, and there is evidence of a high orbital inclination. Therefore, any large-scale geometrical change in the accretion disk can cause local obscuration events. On the other hand, Cyg X-3 orbits its Wolf-Rayet companion star inside the heavy stellar wind obscuring the X-ray source. We study here whether the peculiar X-ray spectra observed from all four sources can be explained by local obscuration events.}
   {We used spectra obtained with the \textit{Nuclear Spectroscopic Telescope Array} and \textit{Rossi X-ray Timing Explorer} to study the spectral evolution of the four luminous hard X-ray sources. We fit the time-averaged spectra, and also time-resolved spectra in case of V404 Cyg, with two physically motivated models describing either a scenario where all the intrinsic emission is reprocessed in the surrounding matter or where the emitter is surrounded by a thick torus with variable opening angle.}
   {We show that the X-ray spectra during specific times are very similar in all four sources, likely arising from the high-density environments where they are embedded. The fitted models suggest that a low-luminosity phase preceding an intense flaring episode in the 2015 outburst of V404 Cyg is heavily obscured, but intrinsically very bright (super-Eddington) accretion state. Similar spectral evolution to that of V404 Cyg is observed from the recent X-ray state of GRS 1915$+$105 that presented unusually low luminosity. The modeling results point to a geometry change in the (outflowing) obscuring matter in V404 Cyg and GRS 1915$+$105, which is also linked to the radio (jet) evolution. Within the framework of the models, all sources display obscured X-ray emission, but with different intrinsic luminosities ranging from lower than 1\% of the Eddington luminosity up to the Eddington limit. This indicates that different factors cause the obscuration. This work highlights the importance of taking the reprocessing of the X-ray emission in the surrounding medium into account in modeling the X-ray spectra. This may well take place in other sources as well.}
   {} 


   \maketitle
%
%________________________________________________________________

\section{Introduction} \label{introduction}

Cyg X-3, V4641 Sgr, V404 Cyg and GRS 1915$+$105 are unique sources even in the fairly non-homogenous group of X-ray binaries (XRBs). They present very different companion stars: Cyg X-3 harbors a Wolf-Rayet companion \citep{vankerkwijk96,koljonen17}, making it a high-mass XRB, while V4641 Sgr, V404 Cyg, and GRS 1915$+$105 are low-mass XRBs with a late B-type \citep{orosz01} star, a K-type subgiant \citep{casares92,king93,khargharia10}, and a K-type giant star \citep{greiner01a} as donors, respectively. However, they share some similarities that are unique among the XRB population. They are all very powerful X-ray emitters, with Cyg X-3 persistently emitting a luminosity of 10$^{38}$ erg s$^{-1}$ in the X-ray band, V4641 Sgr and V404 Cyg exhibiting luminous outbursts where the X-ray luminosity can exceed the Eddington luminosity for a 10 solar mass black hole \citep{revnivtsev02,motta17a}, and GRS 1915$+$105 being in outburst for the past 27 years \citep{castrotirado92,fender04}, with luminosities reaching and surpassing the Eddington limit \citep[e.g.,][]{done04}. 

Except for Cyg X-3, all these sources have long orbital periods and thus large accretion disks, and there is evidence of a high orbital inclination. Therefore, any large-scale geometrical change in the accretion disk such as puffing-up or warping of the accretion flow, or an equatorial outflow can cause local obscuration events. Evidence of this could be seen in the June 2015 outburst of V404 Cyg, which showed highly variable high column density material absorbing the X-ray continuum that remained hard throughout the outburst \citep{motta17a}. Especially so-called X-ray plateaus with diminished X-ray luminosity and softer spectra suggested a heavy obscuration of the intrinsic emission \citep{motta17b,sanchez17}. Similarly, the high-luminosity active accretion phases of V4641 Sgr can be very rapid with a heavily absorbed hard X-ray continuum \citep{munozdarias18,revnivtsev02}. In both systems, there is evidence that the intense likely super-Eddington X-ray emission drives a strong disk wind, thus expelling a significant amount of mass to surround the systems \citep{king15,munozdarias18}. 

On the other hand, Cyg X-3 orbits its companion star at a close distance with a short 4.8-hour orbital period \citep{parsignault72}. Because the companion is a Wolf-Rayet star exhibiting a heavy stellar wind that extends much farther than the binary orbit, Cyg X-3 is constantly embedded in a high-density environment that affects its X-ray spectra in all accretion states \citep{szostek08,zdziarski10,koljonen18}. This material is optically thick in X-rays as a result of the absorption of metals and Compton scattering, causing iron absorption edges \citep{koljonen18} and Compton downscattering \citep{zdziarski10} and/or Compton scattering out of the line of sight to the intrinsic X-ray continuum. Highly ionized iron lines are resolved with \chandra\/ and reveal a distinct component of gas at much higher ionization, in addition to a component from fluorescence by neutral or near-neutral material. Attempts to unify the iron emission with that of lower-Z elements implied a need for an additional absorption component, possibly associated with a disk wind \citep{kallman19}.  

GRS 1915$+$105 is known to have ionized accretion disk wind \citep{neilsen09}, but a high obscuration like that in the other three sources has not been observed. However, GRS 1915$+$105 recently entered a new accretion state that presents lower fluxes throughout its spectral energy distribution than ever before during its 27-year-long outburst. In this state, sporadic X-ray flares have been observed \citep[e.g.,][]{iwakiri19,neilsen19,jithesh19} in addition to the X-ray spectra, indicating heavy obscuration \citep{miller19}. Strong radio flares were also observed in the flaring period \citep{motta19,trushkin19,koljonen19}, indicating episodic jet emission that is also similar to the multiwavelength evolution of V404 Cyg and Cyg X-3. The similarity of the X-ray spectra in Cyg X-3 and GRS 1915$+$105 has previously been noted by \citet{vrtilek13} and \citet{zdziarski16}, who studied the color-color-intensity diagrams of XRBs and found that GRS 1915$+$105 and Cyg X-3 occupy an area that is different from that of other black hole or neutron star XRBs. This underlines the connection between the two and their likely `messy' surroundings.    

In this paper, we study spectra obtained with the \textit{Nuclear Spectroscopic Telescope Array} (\nustar) and \textit{Rossi X-ray Timing Explorer} (\rxte) of the four luminous hard X-ray sources Cyg X-3, V404 Cyg, V4641 Sgr, and GRS 1915$+$105, which are all likely surrounded or occluded in the line of sight by dense material that affects the intrinsic X-ray emission at specific times. The data processing of all sources is described in Section \ref{observations}. In Section \ref{results} we show that the X-ray spectra are very similar and peculiar at specific times in the evolution in all four sources, which is caused by the high-density environments in which they are embedded. The mutual spectral characteristics include a low-energy cutoff of the hard X-ray spectra, absorption edges of highly ionized iron, ionized or Doppler-shifted iron emission or absorption lines, and high absorption. We furthermore fit the time-averaged spectra, and also time-resolved spectra in case of V404 Cyg, with two physically motivated models that either describe a scenario in which all the intrinsic emission is reprocessed in the surrounding matter or in which the emitter is surrounded by a thick torus with variable opening angle. This underlines the assumption of X-ray obscuration in all sources. Using the results from both fits, we discuss in Section \ref{discussion} that the (outflowing) obscuring matter in V404 Cyg and GRS 1915$+$105 shows a change in geometry that is linked to the radio (jet) evolution observed from the sources, in addition to a change in the intrinsic X-ray emission. The sources display X-ray obscuration with varying intrinsic luminosities from lower than 1\% of the Eddington luminosity up to the Eddington limit within the framework of the models. This indicates that different factors cause the obscuration. We therefore further discuss the effect of these results for other sources. Finally, we conclude in Section \ref{conclusions}.      

\section{Observations and data reduction} \label{observations}

\begin{table*}
\centering
\caption{Observation log.}
\label{obslog}
\begin{tabular}{lccccccc}
\toprule
Source & Instrument & Pointing & Date$^{a}$ & MJD$^{b}$ & Obs. length & Exposure & Count rate$^{c}$ \\
& & & & & (s) & (s) & (cts/s) \\
\midrule
Cyg X-3 & \nustar & 10102002002 & 2015/11/13 & 57339.53742 & 20789 & 10168 & 113 (51--188) \\
V404 Cyg & \nustar & 90102007002 & 2015/06/24 & 57197.93615 & 64383 & 17721 & 588 (10--8305) \\
GRS 1915 & \nustar & 90501321002 & 2019/05/05 & 58608.30055 & 63813 & 28700 & 31 (9--68) \\
& & 30502008002 & 2019/05/19 & 58622.52845 & 52451 & 25403 & 15 (4--51) \\
& & 30502008004 & 2019/07/31 & 58695.87901 & 53420 & 23243 & 17 (1--81) \\
V4641 Sgr & \rxtepca & 70119-01-01-14 & 2002/05/23 & 52417.81503 & 736 & 736 & 73 (65--87) \\
& & 80054-08-01-01$^{d}$ & 2003/08/06 & 52857.43170 & 2992 & 2992 & 67 (45--90) \\
\bottomrule
\end{tabular}
\tablefoot{
\tablefoottext{a}{Year/month/day of data start time.}
\tablefoottext{b}{Modified Julian Date of data start time.}
\tablefoottext{c}{Mean countrate in 3--79 keV band (\nustar) and 3--40 keV band (\rxtepca) with the data range in parenthesis.}
\tablefoottext{d}{Only the second part of the lightcurve were studied here.}
}
\end{table*}

\subsection{\nustar}

In the case of V404 Cyg, we selected the \nustar\/ observation that was taken during the June outburst of 2015 nearing the end of the 12-day flaring activity that contained both a plateau spectral state with slow spectral evolution and low flux density and a flaring state with rapid spectral changes and highly variable flux density (pointing 90102007002). This observation was previously analyzed in \citet{walton17}, although they mostly concentrated on analyzing the low-absorption  flaring periods with high count rates. For GRS 1915$+$105, we selected observations that were taken when the source descended to a very anomalous low-flux state in June 2019 that was interspersed by very luminous flares (pointings 90501321002, 30502008002, and 30502008004). For Cyg X-3, we selected the only \nustar\/ observation that was taken in the hard state (pointing 10102002002). All observations used in this paper are tabulated in Table \ref{obslog}.

We reduced the \nustar\/ data from the two focal plane modules (FPMA and FPMB) using \textsc{nupipeline}. We used a circular source region with a 100 arcsec radius centered on the location of the source, and circular background regions with a 100 arcsec radius that were selected from a sourceless region in the detector image. The source region size was a compromise of including most of the point-spread function but avoiding to confuse it with the possible contribution from the scattering halo that is present in the data of V404 Cyg \citep{beardmore16,heinz16,vasilopoulos16}. The pipeline was run with the parameters \textsc{tentacle=`yes'} and \textsc{saamode=`optimized'}. The former requires a simultaneous increase in the CdZnTe detector event count rates and the observed shield single rates, and the latter allows identification and flagging of time intervals in which the CdZnTe detector event count rates show an increase when the spacecraft enters the South Atlantic Anomaly (SAA). The data reduction was performed with \textsc{heasoft 6.26.1}.

We extracted an averaged spectrum from the two detectors using whole pointings in the case of Cyg X-3 and GRS 1915$+105$, and spectra with several intervals in the case of V404 Cyg (see below). The broadband (3--79 keV) \nustar\/ count rate ranged between 51--188 cts/s for Cyg X-3, which mostly arises from the orbital modulation (a factor of 2--3). The total exposure ($\sim$21 ks) is longer than the orbit ($\sim$17.3 ks), therefore we can expect the effect of orbital modulation on the spectral components to average out. For GRS 1915$+105$, the \nustar\/ count rate was found to vary between 1 and 81 cts/s during the $\sim$30--60 ks long pointings, which is considerably lower than usually observed. GRS 1915$+105$ is famous for its plethora of X-ray variability states \citep{belloni13}; the flux and hardness vary on short timescales. For the observations considered here, the hardness ratio between 3--5 keV and 10--79 keV remained relatively constant with 0.8$\pm$0.1, 0.22$\pm$0.03, and 0.2$\pm$0.1 for pointings 90501321002, 30502008002, and 30502008004, respectively. We therefore consider the averaged spectrum as a relatively accurate representation of the spectral shape, although we cannot rule out fast changes in the source spectrum.       

For V404 Cyg, we concentrated on analyzing the times preceding and in between the intense X-ray flaring with the total count rate not exceeding $\sim$1000 cts/s in order to study spectral characteristics when the intrinsic X-ray emission was likely obscured (the low-absorption intense flaring spectra were studied in detail in \citealt{walton17}). Because the spectrum changed rapidly during the pointing, we analyzed the spectral evolution in several ways. We divided the pointing into 70 segments, where the extracted spectrum has 30000 counts to ensure sufficient spectral quality. We excluded spectra from the analysis that contained count rates exceeding 1050 cts/s in order to fully include the preflaring period (the first $\sim$21 ksec of the pointing) and periods with low count rates between the luminous flares. This resulted in 25 individual time bins ranging from 420 s to 3980 s in exposure time. We furthermore binned time-resolved spectra that were similar in shape to spectral epochs in order to increase the spectral quality for detailed modeling. 

For X-ray modeling, we binned the data to a minimum signal-to-noise ratio (S/N) of 30 in the full band 3--79 keV. The spectral fitting was performed using the Interactive Spectral Interpretation System \textsc{(isis;}  \citealt{houck02}). In the modeling, a constant factor was added to account for the flux difference of the \nustar\/ detectors. In some pointings (e.g., 30502008004 of GRS 1915$+$105), the discrepancies in the FPMA and FPMB data cannot be explained by a simple constant, especially in the 6--10 keV region. This can affect the fit quality significantly. Therefore we also fit the same models to the spectra from a single detector and present its fit quality as well. However, all the model parameters we present here are estimated from fits to data from both detectors. All the \nustar\/ fluxes are normalized to the FPMA detector.

For the X-ray timing analysis, we extracted 2$^{-6}$-s light curves from three energy bands: 3--10 keV, 10--79 keV, and 3--79 keV. The cospectra, which can be used as a proxy for white-noise-subtracted power spectral densities (PSDs), were calculated from 512-s long segments averaged over each good time interval (GTI) using Matteo's Libraries and Tools in Python for \nustar\/ timing (\textsc{maltpynt}; \citealt{bachetti15b}). The cospectrum is used to mitigate instrumental effects in the \nustar\/ light curves \citep{bachetti15}. We used the rms normalization and binned the cospectra geometrically by a factor of 1.1--1.5 before importing them to \textsc{isis} for model fitting. 

\subsection{\rxte}

We downloaded all the proportional counter array (\rxte/PCA) data from the High Energy Astrophysics Science Archive Research Center (HEASARC) during outbursts of V4641 Sgr in 1999, 2002, 2003, and 2005, and selected two representative observations for spectral modeling that resemble the data from Cyg X-3 and V404 Cyg (pointings 80054-08-01-01 and 70119-01-01-14). The former was taken during the outburst of 2003 and was analyzed in \citet{maitra06}. However, only the first $\sim$2 ksec were studied in their paper, and here we concentrate on the latter part of the light curve. The other selected pointing was taken during the outburst of 2002, and we are not aware that this spectrum has been studied in detail elsewhere. The pointing 80054-08-01-01 was taken at MJD 52857.37 with an exposure of 2.9 ksec and a mean count rate of the proportional counter unit 2 (PCU2) of 67 cts/s, while the pointing 70119-01-01-14 was taken at MJD 52417.81 with an exposure of 0.7 ksec and a mean PCU2 countrate of 73 cts/s. Because neither light curve presented significant spectral changes during the pointings, we extracted the average spectrum for spectral modeling in both cases.

\rxtepca\/ data were reduced using the methods described in the \rxte\/ cookbook with \textsc{heasoft 6.26.1}. The 128-channel energy spectra were extracted from the standard-2 data using all available PCUs and all layers. For the spectral fitting, we ignore bins below 3 keV and above 40 keV, binned the data to S/N=5.5, and added 0.5\% systematics to each channel. For the timing analysis, we extracted 0.125-s light curves from standard-1 data. We calculated the averaged PSD using 512-s light-curve segments and binned them geometrically by a factor of 1.1 before importing them to \textsc{isis} for model fitting.     

\section{Results} \label{results}

\subsection{X-ray spectra: Overview}

\begin{table*}
\centering
\caption{Source parameters.}
\label{sourceparam}
\begin{tabular}{lccccccc}
\toprule
Source & Distance & Mass & Period & Inclination & ISM abs. \\
& (kpc) & (M$_{\odot}$) & (days) & (deg) & (10$^{22}$ cm$^{-2}$) \\
\midrule
Cyg X-3 & 7.4$\pm$1.1 (1) & 2.4$^{+2.1}_{-1.1}$ (2) & 0.2 (3) & 30--50 (2,4,5) & 3.5 (6,7) \\
V404 Cyg & 2.39$\pm$0.14 (8) & 9.0$^{+0.2}_{-0.6}$ (9) & 6.5 (10) & 67$^{+3}_{-1}$ (9) & 0.83 (11) \\
GRS 1915$+$105 & 8.6$^{+2.0}_{-1.6}$ (12) & 12.4$^{+2.0}_{-1.8}$ (12) & 33.9 (13) & 60$\pm$5 (12) & 3.5 (14) \\
V4641 Sgr & 6.2$\pm$0.7 (15) & 6.4$\pm$0.6 (15) & 2.8 (16) & 72 (15) & 0.23 (17) \\
\bottomrule
\end{tabular}
\tablebib{
(1) \citet{mccollough16}; (2) \citet{zdziarski13}; (3) \citet{parsignault72}; (4) \citet{vilhu09}; (5) \citet{zdziarski12}; (6) \citet{koljonen18}; (7) \citet{kallman19}; (8) \citet{millerjones09}; (9) \citet{khargharia10}; (10) \citet{casares92}; (11) \citet{motta17b}; (12) \citet{reid14}; (13) \citet{steeghs13}; (14) \citet{chapuis04}; (15) \citet{macdonald14}; (16) \citet{orosz01}; (17) \citet{maitra06}.
}
\end{table*}

\begin{figure}
 \centering
 \includegraphics[width=\linewidth]{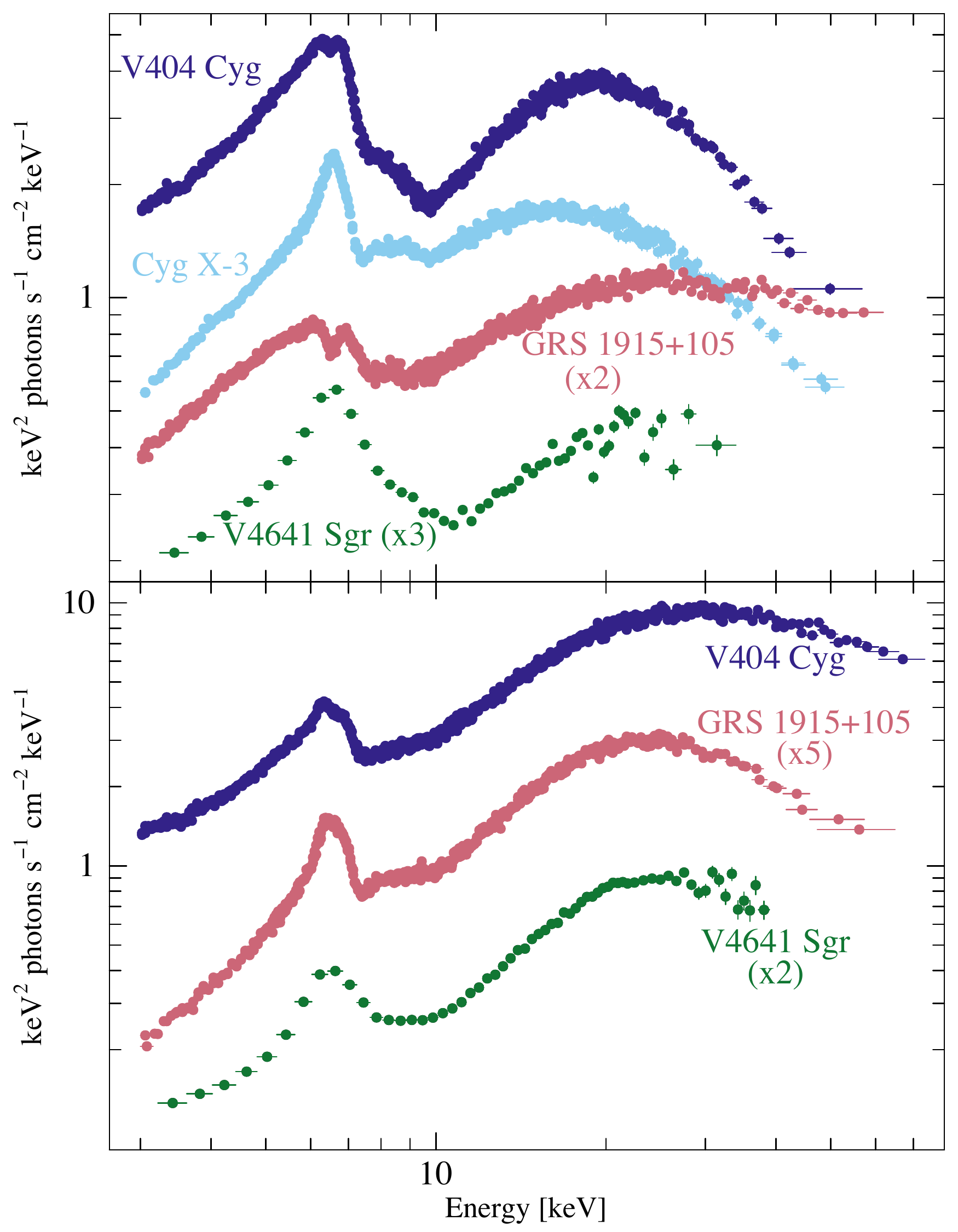}
 \caption{Collection of X-ray spectra from the four sources ordered into two panels according to their spectral shape. \textit{Top:} \nustar\/ spectra (FPMA/FPMB data both included) of V404 Cyg during the outburst period of 2015 (preceding intense X-ray flaring period), Cyg X-3 in the hard state, GRS 1915$+$105 during the anomalous low-luminosity state of 2019--2020 (preceding a period of sporadic X-ray and radio flaring), and the \rxte/PCA spectrum of V4641 Sgr during the outburst period of 2002. \textit{Bottom:} \nustar\/ spectra of V404 Cyg during the outburst period of 2015 (in between intense X-ray flaring), \rxte/PCA spectrum of V4641 Sgr during the outburst period of 2003, and \nustar\/ spectra of GRS 1915$+$105 from the anomalous low-luminosity state (after the intense X-ray and radio flaring period). Note the spectral similarity between sources. The spectra of GRS 1915$+$105 and V4641 Sgr are renormalized by the amount shown for illustrative purposes.}
 \label{spectra}
\end{figure}

Fig. \ref{spectra} shows some of the X-ray spectra from the observations tabulated in Table \ref{obslog}. In the top panel, the hard-state spectra of V404 Cyg (from the beginning of the pointing before spectral softening, see Section \ref{v404}), Cyg X-3, and a 2002 outburst peak spectrum of V4641 Sgr show a striking similarity in spectral shape with a similar absorption profile, a broad peak at the iron line region, a sharp drop at the iron edge energies, and a low-energy cutoff in the hard X-rays. This type of spectrum is not observed from any other XRB. The closest comparison can be found in Compton-thick active galactic nuclei \citep[AGN; e.g.,][]{balokovic14,bauer15}. The GRS 1915$+$105 spectrum from the anomalous low-luminosity X-ray state before the X-ray or radio flaring shows similar features, but with a prominent iron absorption line and a higher X-ray cutoff energy. The spectra in Fig. \ref{spectra} (top panel) display different flux densities: the V404 Cyg is a factor of $\sim$2, $\sim$6, and $\sim$20 brighter than Cyg X-3, GRS 1915$+$105, and V4641 Sgr, respectively. However, when the distances (see Table \ref{sourceparam}) are taken into account, the X-ray luminosity of V404 Cyg is a factor of $\sim$3 brighter than V4641 Sgr, but a factor of $\sim$2 and $\sim$4 dimmer than GRS 1915$+$105 and Cyg X-3, respectively, in this state. 

In the bottom panel of Fig. \ref{spectra}, a harder X-ray spectrum is shown for V404 Cyg (during X-ray flaring) and V4641 Sgr (2003 outburst peak spectrum), which share approximately the same shape, in addition to a spectrum of GRS 1915$+$105 after a high-intensity X-ray and radio-flaring period in the low-luminosity state. However, no similar spectrum can be found for Cyg X-3 because all the other spectral states are softer (see all the different spectra from different accretion states in \citealt{koljonen10}). This might therefore indicate that these accretion states are less absorbed or obscured, and the spectral shape might be explained by strong reflection from the accretion disk surface. However, we show here that the observed spectral evolution is also compatible with a change in the geometry of the obscuring material, or with a change in the ionization structure.

All the spectra show a strong iron line with a line width above 200 eV. This has been resolved as a combination of a neutral iron K$\alpha$ line (6.4 keV) and ionized iron lines, with the strongest component arising from the Fe XXV K$\alpha$ line (6.7 keV) and possibly from Fe XXVI Ly$\alpha$ line (7.0 keV) with \chandra\/ in the case of Cyg X-3 \citep{paerels00,kallman19} and V404 Cyg \citep{king15}. The energy band from $\sim$7 keV to $\sim$10 keV appears to be affected by the absorption of the above-mentioned species of iron in all sources. Their ionization energies are 7.1 keV, 8.8 keV, and 9.2 keV for Fe K$\alpha$, Fe XXV K$\alpha$, and Fe XXVI Ly$\alpha$, respectively. In addition, there is a strong visible iron absorption line around 6.5 keV in the preflare spectrum of GRS 1915$+$105, and a weaker line in the preflare spectrum of V404 Cyg. Moreover, there might be an indication of emission from an iron Fe XXV K$\beta$ line (7.8 keV) or Ni XXVII K$\alpha$ (7.8 keV) in the Cyg X-3 spectrum and the Fe XXVI Ly$\beta$ line (8.3 keV) in the GRS 1915$+$105 preflare spectrum. Curiously, some of the V404 Cyg spectra show iron line centroids close to 6.3 keV, similar to findings of \citet{motta17a}, indicating that the redshifted neutral iron K$\alpha$ line arises either very close to the compact object (gravitational redshift) or in a medium moving away (Doppler redshift). The strong photoionized emission and absorption lines point to a significant amount of absorbing matter in the line of sight to all sources.

All spectra exhibit a strong curvature or a cutoff in the hard X-rays around 20--30 keV. This is atypical for a hard-state XRB. This cutoff might be caused by a heavy absorption of the intrinsic (cutoff) power-law spectrum, or by a Compton downscattering in the accretion disk (i.e., reflection spectrum) or the surrounding medium. In addition, the spectra shown in Fig. \ref{spectra} (top panel) present a rather sharp upturn at  $\sim$ 10 keV, possibly indicating a location where two model components meet that can have very different spectral slopes. In the following, we consider these scenarios by fitting the data with appropriate spectral models.

\subsection{X-ray spectra: Initial modeling} \label{inimod}

\begin{figure*}
 \centering
 \includegraphics[width=\linewidth]{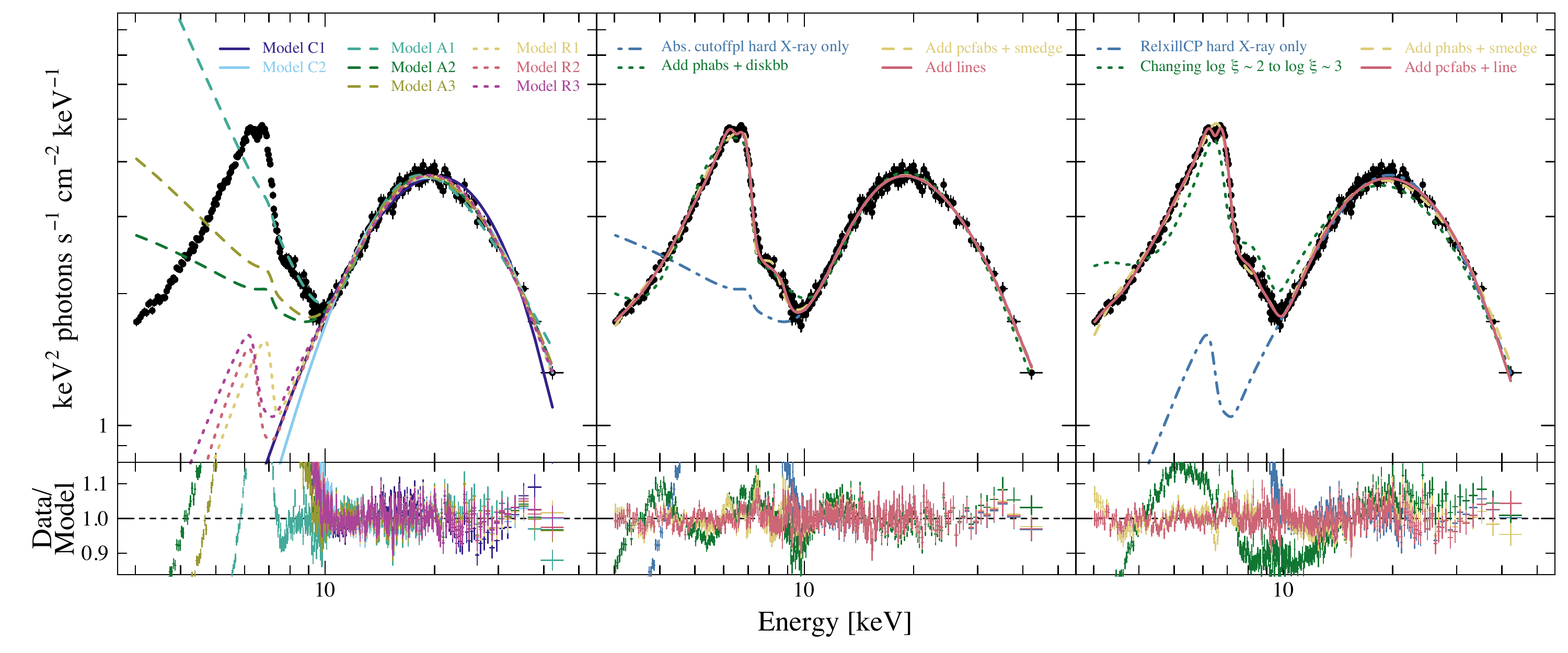}
 \caption{Initial modeling of the V404 Cyg preflare spectrum. \textit{Left:} Best-fit models fit to the hard X-ray data (10--79 keV), but plotted in the full data range. Different models are labeled, and the parameters are tabulated in Table \ref{phenom}. Solid lines refer to pure continuum models, dashed lines to absorbed continuum models, and dotted lines to reprocessed continuum models. \textit{Middle:} Fitting an absorbed cutoff power-law model to the full data range. The dot-dashed blue line corresponds to model A2 fit to the hard X-ray data (see left panel), the dotted green line corresponds to absorption and disk blackbody components added to the model, the dashed yellow line corresponds to partial absorption and smeared edge components added to the model, and the solid red line shows emission and absorption lines added to the model (parameters of the final model are tabulated in Table \ref{phenom2}). See the text for more details. \textit{Right:} Fitting a reprocessed thermal Comptonization model to the full data range. The dot-dashed blue line corresponds to model R3 fit to the hard X-ray data (see the left panel), the dotted green line shows an increased value of the ionization parameter, the dashed yellow line corresponds to absorption and smeared edge components added to the model, and the solid red line shows partial absorption and an absorption line added to the model (parameters of the final model are tabulated in Table \ref{phenom2}). See the text for more details.}
 \label{initial}
\end{figure*}

In the hard X-ray spectral state, the hard X-ray spectrum of XRBs can typically be fit with a cutoff power law or a Comptonized continuum with a power-law photon index in the range of $\Gamma=$1.5--2.0 and cutoff energy of $\sim$100 keV. The spectra presented in Fig. \ref{spectra}, especially in the top panel, are quite unlike typical XRB hard-state spectra. Thus, we started by finding and fitting curved models to the hard X-ray data (above 10 keV) of V404 Cyg (preflare or plateau spectrum, as shown in Fig. \ref{spectra}, top panel). The resulting model parameters are tabulated in Table \ref{phenom} and the models are shown in Fig. \ref{initial} (left panel).

To model the strong spectral curvature, we first tried a cutoff power-law model (C1 in Fig. \ref{initial} and Table \ref{phenom}). The best-fit model is highly inverted, with a low-energy cutoff (power-law index, $\Gamma\sim-1.6$, cutoff energy, $E_{\mathrm{cut}}\sim6$ keV), and the fit is poor ($\chi^{2}_{\mathrm{red}}=2.6$). Obviously, the power-law index is too low for any reasonable physical scenario. Next, we tried a thermal Comptonization model in a spherical plasma cloud (\textsc{compTT}; \citealt{titarchuk94}). The resulting model (C2 in Fig. \ref{initial} and Table \ref{phenom}) has an electron temperature of $kT_{e}\sim24$ keV, a seed photon temperature of $kT_{s}\sim5$ keV, and an optical depth of $\tau\sim0.7$. Model C2 provides a better fit to the data ($\chi^{2}_{\mathrm{red}}=1.8$) than model C1, but it is still poor. A tendency for a high seed-photon temperature in fitting thermal Comptonization models to the hard X-ray data of V404 Cyg has been noted earlier by \citet{jenke16}, \citet{roques15}, and \citet{natalucci15}, who discussed that the high-temperature seed photons might arise from a synchrotron emission in the jet base. However, the radio luminosity during the time of the V404 Cyg preflare observation was very low, $\sim$10 mJy \citep{munozdarias16,gandhi17}, with a steep spectrum that indicates optically thin emission from the jet ejecta \citep{millerjones19}. The core jet therefore was likely quenched and the synchrotron scenario is not plausible.   

The considerations above show why the simple models are not sufficient for modeling the highly unusual hard X-ray spectrum. \citet{natalucci15} also fit the hard X-ray data from the \textit{INTErnational Gamma-Ray Astrophysics Laboratory} with a pure reflection model and a partially covering absorber model. The first model was found to fit most of the data only by assuming very high values for the reflection factor, and the latter by assuming very high column densities for the absorber. The authors regarded these models as unphysical. However, based on the analyses of X-ray and optical data as detailed in Section 1, the spectral similarity to the Cyg X-3 hard X-ray state spectrum and the strong X-ray emission lines indicate that the absorption and reprocessing and reflection scenarios are plausible at least for the low-luminosity phases. We therefore continued to fit the V404 Cyg preflare hard X-ray data with absorption and reflection models. We first tried an absorbed power-law model, but found that a fully absorbed model (\textsc{phabs} $\times$ \textsc{powerlaw}) is completely inadequate to fit the data ($\chi^{2}_{\mathrm{red}}=8.2$). Instead, a partially absorbed power law (\textsc{pcfabs} $\times$ \textsc{powerlaw}; model A1 in Fig. \ref{initial} and Table \ref{phenom}) fits the hard X-ray data much better ($\chi^{2}_{\mathrm{red}}=1.5$), although the intrinsic spectrum is very soft ($\Gamma\sim3.9$) and partially absorbed ($f_{\mathrm{cov}}\sim0.93$) by a dense medium ($N_{\mathrm{H}} \sim 3.6 \times 10^{24}$ cm$^{-2}$). An even better fit ($\chi^{2}_{\mathrm{red}}=1.1$) can be found by changing the intrinsic spectrum to a cutoff power-law model (A2 in Fig. \ref{initial}/Table \ref{phenom}) with a value for the power-law photon index more in line with a typical XRB in a hard X-ray state ($\Gamma\sim2.2$) and cutoff energy of $E_{\mathrm{cut}}\sim18$ keV. The parameters of the absorption component are slightly lower than those of model A1, but they are still comparable. A similar fit quality is achieved by changing the intrinsic emission component to a thermal Comptonization component (model A3 in Fig. \ref{initial} and Table \ref{phenom}). The best-fit model has an electron temperature of $kT_{e}\sim$10 keV, the seed-photon temperature is fixed to 0.1 keV, and the optical depth is $\tau\sim3.6$. The absorption component has similar parameter values as models A1 and A2.

For the reflection scenario, we began by fitting the data with \textsc{pexrav} \citep{magdziarz95}; a cutoff power-law continuum reprocessed in a neutral medium (model R1 in Fig. \ref{initial} and Table \ref{phenom}). We fixed the inclination to the value presented in Table \ref{sourceparam} and the abundances to solar. The resulting model has a power-law photon index $\Gamma\sim1.8$, a cutoff energy $E_{\mathrm{cut}}\sim20$ keV, and a reflection factor $R_{f}\gtrsim400$. The fit quality is moderate ($\chi^{2}_{\mathrm{red}}=1.3$). We also fit the V404 Cyg hard X-ray spectrum with a relativistic reflection model \citep[\textsc{relxill}][]{garcia14,dauser14}. The resulting model (R2 in Fig. \ref{initial} and Table \ref{phenom}) also has a very high reflection factor ($R_{f}\sim90$) in a moderately ionized matter (log $\xi\sim2.3$). The intrinsic spectrum is a very hard cutoff power law with $\Gamma\sim1.1$ and cutoff energy of $E_{\mathrm{cut}}\sim$20 keV. The spin of the black hole is low: $a\sim0.2$. Fixing the inclination to the value presented in Table \ref{sourceparam} did not result in a good fit, therefore we left it free to vary, which returned rather low values of $\theta_{\mathrm{inc}}\lesssim10^\circ$. All the other parameters were fixed in the default values. The fit quality is slightly better than in model R1 ($\chi^{2}_{\mathrm{red}}=1.1$). In the reflection models, the amount of the radiation that ionizes the reflecting material is typically defined as the ratio of the photon intensity that illuminates it to the direct photon intensity that reaches the observer. When instead of the accretion disk, the reflecting medium is surrounding the X-ray source, the maximum reflection factor can be much higher than in the unobscured case, where usually $R_{f} \lesssim 1$. We note, however, that high values for the reflection factor, $R_{f}\lesssim 10$, can be accommodated in the unobscured case as well when strong light bending deep in the gravitational potential of the compact object is assumed (\citealt{dauser16}). Clearly, the reflection factors in the fits are much higher, and we can assume that the majority of the incident emission is reprocessed in a surrounding medium. From the \textsc{relxill} model family, we also tried the lamppost geometry (\textsc{relxilllp}), which resulted in similar parameters as for model R2 (coronal geometry), but with an even higher reflection factor. Finally, we selected a model with a Comptonized incident spectrum: \textsc{relxillCp} (model R3 in Fig. \ref{initial}/Table \ref{phenom}). Because the reflection factor found above is very high, we fit only the reflection component to the data. The parameters and fit quality are very similar to the cutoff power-law \textsc{relxill} model, except for the power-law index, which is $\Gamma\sim2$. This initial model fitting shows that the absorption or reprocessing models describe the hard X-ray data better than normal continuum models. In addition, when we discard models where the incident continuum is not too soft or too hard for a hard X-ray state, we are left with models A2, A3, R1, and R3.

\begin{table*}
\centering
\caption{Model parameters from the initial fits to the V404 Cyg preflare hard X-ray spectrum (10--79 keV).}
\label{phenom}
\begin{tabular}{lccccccccc}
\toprule
& & \multicolumn{2}{c}{Pure continuum model} & \multicolumn{3}{c}{Absorbed continuum model} & \multicolumn{3}{c}{Reprocessed continuum model} \\
\cmidrule(lr){3-4} \cmidrule(lr){5-7} \cmidrule(lr){8-10}
& & \textsc{cutoffpl} & \textsc{compTT} & \textsc{powerlaw} & \textsc{cutoffpl}& \textsc{compTT} & \textsc{pexrav} & \textsc{relxill} & \textsc{relxillCp} \\
Param. & Unit & (C1) & (C2) & (A1) & (A2) & (A3) & (R1) & (R2) & (R3) \\
\midrule
N$_{\mathrm{H}}$ & 10$^{22}$ cm$^{-2}$ & & & 363$\pm$7 & 290$\pm$15 & 299$\pm$13 &  \\
f$_{\mathrm{cov}}$ & & & & 0.931$\pm$0.002 & 0.86$\pm$0.01 & 0.89$\pm$0.01 & \\
$\Gamma$ & & -1.64$\pm$0.05 & & 3.85$\pm$0.03 & 2.2$\pm$0.3 & & 1.84$\pm$0.04 & 1.11$^{+0.07}_{-0.05}$ & 1.96$^{+0.05}_{-0.04}$ \\
E$_{\mathrm{cut}}$ & keV & 5.6$\pm$0.1 & & & 18$\pm$3 & & 20$\pm$1 & 23.3$\pm$0.6 \\
kT$_{\mathrm{seed}}$ & keV & & 4.6$^{+0.06}_{-0.08}$ & & & 0.1 & & \\
kT$_{e}$ & keV & & 24$^{+42}_{-10}$ & & & 10$\pm$1 & & & 13.7$^{+0.8}_{-0.5}$ \\
$\tau$ & & & 0.7$^{+1.0}_{-0.6}$ & & & 3.6$\pm$0.5 & & \\
R$_{f}$ & & & & & & & $>$411 & $>$480 & -2 \\
$\theta_{\mathrm{inc}}$ & deg & & & & & & 67 & $<$11 & $<8$ \\
$a$ & & & & & & & & 0.2$\pm$0.2 & 0.6$^{+0.1}_{-0.2}$ \\ 
log $\xi$ & & & & & & & & 2.30$^{+0.02}_{-0.26}$ & 2.00$^{+0.01}_{-0.06}$\\
\midrule
$\chi^{2}$/d.o.f & & 713/279 & 491/278 & 429/278 & 293/277 & 299/277 & 356/278 & 314/275 & 340/276 \\
$\chi_{\mathrm{red}}^{2}$ & & 2.56 & 1.77 & 1.54 & 1.06 & 1.08 & 1.28 & 1.14 & 1.23 \\
\bottomrule
\end{tabular}
\end{table*}

\begin{table*}
\centering
\caption{Model parameters from the initial fits to the V404 Cyg preflare X-ray spectrum (3--79 keV).}
\label{phenom2}
\begin{tabular}{cccccccccc}
\toprule
\multicolumn{9}{c}{Model: \textsc{phabs} $\times$ \textsc{smedge} $\times$ (\textsc{pcfabs1} $\times$ \textsc{cutoffpl} + \textsc{pcfabs2} $\times$ \textsc{diskbb} + \textsc{gauss1} + \textsc{gauss2})} \\
\midrule
\textsc{phabs} & \multicolumn{3}{c}{\textsc{smedge}} & \multicolumn{2}{c}{\textsc{pcfabs1}} & \multicolumn{3}{c}{\textsc{cutoffpl}} \\
\cmidrule(lr){1-1} \cmidrule(lr){2-4} \cmidrule(lr){5-6} \cmidrule(lr){7-9}
N$_{\mathrm{H}}$ & E & $\tau$ & $\sigma$ & N$_{\mathrm{H}}$ & f$_{\mathrm{cov}}$ & norm & $\Gamma$ & E$_{\mathrm{cut}}$ \\
(10$^{22}$ cm$^{-2}$) & (keV) & & (keV) & (10$^{24}$ cm$^{-2}$) & & & & (keV) \\
2.8$\pm$0.7 & 8.7$\pm$0.1 & 0.30$^{+0.07}_{-0.05}$ & 0.5$\pm$0.2 & 2.4$\pm$0.2 & 0.92$\pm$0.01 & 33$^{+26}_{-15}$ & 2.2$\pm$0.2 & 19$\pm$3\\
\addlinespace
\multicolumn{2}{c}{\textsc{pcfabs2}} & \multicolumn{2}{c}{\textsc{diskbb}} & \multicolumn{3}{c}{\textsc{gauss1}} & \multicolumn{3}{c}{\textsc{gauss2}} \\ 
\cmidrule(lr){1-2} \cmidrule(lr){3-4} \cmidrule(lr){5-7} \cmidrule(lr){8-10}
N$_{\mathrm{H}}$ & f$_{\mathrm{cov}}$ & norm & kT & E$_{1}$ & $\sigma_{1}$ & norm & E$_{2}$ & $\sigma_{2}$ & norm \\
(10$^{22}$ cm$^{-2}$) & & & (keV) & (keV) & (keV) & ($\times$10$^{-3}$) & (keV) & (keV) & ($\times$10$^{-3}$) \\
46$\pm$3 & 0.92$\pm$0.01 & 518$^{+105}_{-95}$ & 1.34$\pm$0.03 & 6.4 & 0.46$\pm$0.02 & 36$\pm$3 & 6.5 & 0.002 & -3.6$\pm$0.7 \\
\midrule
\multicolumn{9}{c}{Both \nustar\/ detectors: $\chi^{2}$/d.o.f = 871/617 \hspace{0.2cm} $\chi_{\mathrm{red}}^{2}$ = 1.41 \hspace{0.2cm} FPMA-only: $\chi^{2}$/d.o.f = 347/305 \hspace{0.2cm} $\chi_{\mathrm{red}}^{2}$ = 1.14} \\
\bottomrule
\addlinespace
\multicolumn{9}{c}{Model: \textsc{phabs} $\times$ \textsc{smedge} $\times$ \textsc{pcfabs} $\times$ (\textsc{gauss} + \textsc{relxillcp})} \\
\midrule
\textsc{phabs} & \multicolumn{3}{c}{\textsc{smedge}} & \multicolumn{2}{c}{\textsc{pcfabs}} & \multicolumn{3}{c}{\textsc{gauss}} \\
\cmidrule(lr){1-1} \cmidrule(lr){2-4} \cmidrule(lr){5-6} \cmidrule(lr){7-9}
N$_{\mathrm{H}}$ & E & $\tau$ & $\sigma$ & N$_{\mathrm{H}}$ & f$_{\mathrm{cov}}$ & E & $\sigma$ & norm \\
(10$^{22}$ cm$^{-2}$) & (keV) & & (keV) & (10$^{22}$ cm$^{-2}$) & & (keV) & (keV) & ($\times$10$^{-3}$) \\
2.7$^{+0.6}_{-0.7}$ & 7.4$\pm$0.07 & 0.32$^{+0.05}_{-0.04}$ & 1 & 39$\pm$2 & 0.82$\pm$0.01 & 6.5 & 0.002 & -11$\pm$2 \\
\addlinespace
\multicolumn{9}{c}{\textsc{relxillcp}}  \\ 
\cmidrule(lr){1-9}
norm & $\Gamma$ & kT$_{e}$ & $\theta_{\mathrm{inc}}$ & R$_{f}$ & R$_{\mathrm{in}}$ & $a$ & log $\xi$ & A$_{\mathrm{Fe}}$\\
($\times$10$^{-3}$) & & (keV) & (deg) & & & & & (solar) \\
27$\pm$2 & 1.85$\pm$0.01 & 3.9$\pm$0.2 & 20$^{+2}_{-4}$ & -2 & 2.6$^{+13.8}_{-0.4}$ & -0.998--0.998 & 3.43$\pm$0.03 & 1.8$\pm$0.2 \\
\\
\midrule
\multicolumn{9}{c}{Both \nustar\/ detectors: $\chi^{2}$/d.o.f = 873/617 \hspace{0.2cm} $\chi_{\mathrm{red}}^{2}$ = 1.41 \hspace{0.2cm} FPMA-only: $\chi^{2}$/d.o.f = 335/306 \hspace{0.2cm} $\chi_{\mathrm{red}}^{2}$ = 1.09} \\
\bottomrule
\end{tabular}
\end{table*} 

Next, we included the soft X-ray (3--10 keV) data in the model fitting. Fig. \ref{initial} (left panel) shows the above models for the whole data range. It is clear that models with partially absorbed but soft intrinsic spectra would need further absorption components to bring the spectrum down to match the data in the soft X-rays, while models with a hard intrinsic spectrum need an additional soft component to account for the data. We selected two models to continue fitting the full data range: an absorbed cutoff power-law continuum (A2), and a fully reprocessed thermal Comptonization continuum (R3). The reasoning behind this selection is that models A2 and A3 are likely very similar, therefore we selected the slightly better fit of model A2. Model R3 was selected because an ionization parameter is included in the model, because of relativistic effects to the spectral shape, and because the fit quality is slightly better.

For the absorbed thermal Comptonization continuum (i.e., model A2), we first added a soft component that we modeled with an absorbed disk blackbody component (\textsc{phabs}$\times$\textsc{diskbb}). However, any model producing a Planckian-type spectrum, for instance, thermal Comptonization or bremsstrahlung, produced equally good fits. The absorbed disk with $kT\sim1.1$ keV and $N_{\mathrm{H}} \sim 6.2 \times 10^{23}$ cm$^{-2}$ can adequately model the soft X-rays, but the soft X-ray slope below 6 keV is not well fit, and large residuals can be found in the 9--10 keV region as well (Fig. \ref{initial}, middle panel). We therefore added another partial covering absorber and a smeared edge to improve the model. The resulting fit is much better in the soft X-rays, but residuals are still found in the 5--10 keV energy range, likely due to missing line components. Therefore we added two Gaussian lines to the model: an emission line fixed to 6.4 keV (neutral iron), and an absorption line fixed to 6.5 keV (ionized iron). This resulted in a fit quality of $\chi^{2}_{\mathrm{red}}=1.4$. All the parameters of this model can be found in Table \ref{phenom2}. A very similar but physically more accurate model was fit to all datasets and is discussed in more detail in the following section.

For the fully reprocessed thermal Comptonization continuum (i.e., model R3), we first fit the same model again for the whole range. The soft X-rays are better taken into account by increasing the value of the ionization parameter from log $\xi\sim2$ to log $\xi\sim3$ (Fig. \ref{initial}, right panel). Thus, there is no need to add a soft component. Clear residuals are left for the soft X-ray regions below 4 keV and between 7--10 keV. We continued to add an absorption (\textsc{phabs}) and a smeared edge (\textsc{smedge}) component to bring the model down in these regions. The resulting fit was already much better with $\chi^{2}_{\mathrm{red}}=2.4$, but some small residuals remained below 4 keV and around $\sim$6.5 keV. Therefore we further added a partial covering absorption component (\textsc{pcfabs}) and an absorption line (\textsc{egauss}) to the model, resulting in a fit quality of $\chi^{2}_{\mathrm{red}}=1.4$. A very similar model was fit to all datasets and is discussed in more detail in the following section.

\subsection{X-ray spectra: Physical modeling} \label{modeling}

Because of the observational evidence of a high-density environment described above, we considered the possibility that all sources are embedded in a dense medium ($N_{\mathrm{H}} \gtrsim 10^{23-25}$ cm$^{-2}$) that surrounds the X-ray source and causes significant absorption and scattering that affects the X-ray spectrum up to $\sim$30--40 keV. To facilitate this scenario for spectral fitting, we considered two models: model (A) consisting of a partially absorbed thermal Comptonization component reprocessed in a highly ionized plasma (\textsc{xillverCp,} or \textsc{relxillCp} when relativistic effects are important) and model (B) consisting of an intrinsic cutoff power-law component (mimicking the thermal Comptonization process) reprocessed in a surrounding neutral uniform-density sphere with polar cutouts of various sizes resembling a torus of different opening angles (\textsc{borus02}; \citealt{balokovic18}). Model (B) is similar to the Compton-thick AGN scenario where the X-ray source is surrounded by a thick torus with the emission received from a highly absorbed line-of-sight component and a reflected component from the surface of the torus (to follow the discussion of the resulting torus geometries for each source, we refer to Fig. \ref{drawing}). In model (A), the reprocessing occurs in a shell or shells of ionized matter surrounding the X-ray source.

Model A corresponds to a scenario in which the spectra are dominated by a reflection or scattered component, and the contribution of the incident continuum is severely diminished. This model consists of one or two partially absorbed reflection models (\textsc{xillverCp} and/or \textsc{relxillCp}), where the incident photons arise from thermal Comptonization (\textsc{nthComp}). Following the indication from the initial modeling in Section \ref{inimod} that the majority of incident photons are reprocessed in a medium encompassing the incident photon source, we fixed the reflection factor to a negative value in the model (all radiation was reprocessed). The parameters of the two reflection components (if needed) were kept at the same values, except for the ionization parameter and normalization, which were left free to vary separately for both components. This is to account for changes in ionization parameter in the scattering component, which is evident in the high-resolution X-ray spectra observed from V404 Cyg, Cyg X-3, and GRS 1915$+$105, which show neutral as well as ionized iron lines. For V404 Cyg and GRS 1915$+$105, we allowed the redshift to vary freely for the scattering components in order to fit the iron line centroids of $\sim$6.3 keV, indicating either gravitational or Doppler redshift of the neutral iron line (for Cyg X-3, we fixed this to 1000 km/s, which is approximately the wind speed of the Wolf-Rayet companion, and to zero for V4641 Sgr because the spectral resolution is far lower). In addition, a narrow iron absorption line and a smeared iron edge with variable absorption energy are needed to successfully fit the first two epochs of V404 Cyg, and epoch 1 of GRS 1915$+$105. This might indicate an additional absorbing medium during these epochs. Thus, the total model can be described as \textsc{constant} $\times$ \textsc{phabs} $\times$ \textsc{smedge} $\times$ \textsc{pcfabs} $\times$ (\textsc{xillverCp$_{1}$}/\textsc{relxillCp} + \textsc{xillverCp$_{2}$} + \textsc{gauss}). Here, constant is the instrument cross-normalization. We did not fix the inclination of the scattering components because the reflecting surface might be inclined away from the disk inclination angle, for example, for an equatorial outflow disk wind with an opening angle of several tens of degrees, or in the case of spherical obscuration, the reflection angle would correspond to some mean angle from all scattering processes. To reduce the parameter space, we fixed the black hole spin to 0. If it is let free, the spectral fits do not restrict the parameter well, which has also been found by \citet{walton17} in the case of V404 Cyg, where the spin was estimated as $a>-0.1$. It can be also expected that the reflecting medium can lie farther away than the innermost stable circular orbit (ISCO) when reprocessing in the surrounding media is assumed. We therefore fixed the outer radius to 1000 gravitational radii in all sources.

For model B, we used the \textsc{borus02} model component \citep{balokovic18}, which instead of a disk reprocessing allows a variety of geometries from a uniform sphere to torus-like shapes through polar cutouts. The reprocessing torus in \textsc{borus02} is considered to be cold, neutral, and static. To take a moving reprocessor into account, we therefore allowed the redshift of the scattering component to vary. Like in model A, we also introduced a highly ionized iron edge component and an ionized iron absorption or emission line component to find acceptable fits to the data. We can expect that the reprocessed photons are either redshifted (behind the source, i.e., moving away) or blueshifted (in front of the source, i.e., moving toward) because they arise in the fast outflow or wind or in the accretion flow. The total model can be described as \textsc{constant1 $\times$ phabs1 $\times$ smedge1 $\times$ (constant2 $\times$ borus02(red) + constant3$ \times$ borus02(blue)  + phabs2 $\times$ cabs1 $\times$ cutoffpl + constant4 $\times$ cutoffpl)}. \textsc{constant1} is the instrument cross-normalization, \textsc{constant2} is the relative normalization of the redshifted scattered component, \textsc{constant3} is the relative normalization of the blueshifted scattered component, \textsc{constant4} is the relative normalization of the leaked (unabsorbed) intrinsic spectrum, \textsc{phabs2 $\times$ cabs1} is the line-of-sight absorption including beam-scattering with the column densities linked between the components, and \textsc{cutoffpl} represents the intrinsic continuum of the accretion flow (mimicking thermal Comptonization spectrum). For the scattered component, we fixed the inclination angle according to Table \ref{sourceparam} because the scattering angle is now taken into account in the model. We also fixed the iron abundance to solar. All the parameters are linked between \textsc{borus02(red)} and \textsc{borus02(blue),} except for the redshift for \textsc{borus02(blue),} which is determined as being the negative of the value in \textsc{borus02(red)}. In addition, for the first two epochs of V404 Cyg, an additional soft component is needed, which we modeled with a partially absorbed blackbody component. However, any component resembling a Planckian spectrum might be inserted instead, such as a low-temperature thermal Comptonization component (cf. model A), a disk blackbody component, or a Wien spectrum. The physical interpretation of this component is discussed in Sects. \ref{v404} and \ref{v404_soft}. In addition, as mentioned above, we added an emission line component to the preflare spectrum of GRS 1915$+$105 and to the Cyg X-3 spectrum. In the following, we concentrate on the fitting results of these two models for individual sources. 

\subsubsection{Cyg X-3} \label{cygx3}

The peculiarity of the hard-state spectrum in Cyg X-3 has been known for more than a decade \citep{hjalmarsdotter04}. The observed low-energy cutoff has previously been attributed to either strong absorption in the stellar wind, pure Compton reflection in a medium that covers the emitting source, or nonthermal Comptonization of a steep electron population by \citet{hjalmarsdotter08}. They preferred the latter scenario, although the former two produce better fits and require either an unusual accretion state or a very massive black hole as a primary star. Later, \citet{zdziarski10} showed that a low-energy cutoff in the X-ray spectrum can be obtained when Compton downscattering is considered in an optically thick plasma cloud, likely arising from the interaction of the accretion disk and the strong stellar wind of the Wolf-Rayet companion. In addition, there is evidence that the primary star in Cyg X-3 is a black hole with a relatively low mass \citep{zdziarski13,koljonen17}. Because of the different types of companion stars in V404 Cyg, V4641 Sgr, GRS 1915$+$105, and Cyg X-3, it seems unlikely that the similar X-ray spectra would be a result of different accretion mechanisms; wind versus Roche-lobe accretion. In addition, the hard X-ray state of Cyg X-3 is a relatively stable state that can last up to several years. Therefore it seems unlikely that it would present a peculiar accretion state for such a long time. Rather, the spectral similarity to V404 Cyg, V4641 Sgr, and GRS 1915$+$105 likely arises from some type of radiation reprocessing.   

The interstellar absorption for Cyg X-3 is relatively high as a result of its location in the plane of the Galaxy, and likely because it is located behind two spiral arms and the Cygnus X star-forming region \citep{mccollough16}. We fixed the lower limit of the hydrogen column to 3.5$\times$10$^{22}$ atoms cm$^{-2}$, which is approximately the value found in studies where instruments with softer X-ray response were used \citep{koljonen18,kallman19}. Because the reprocessing matter in the model is neutral, we added a smeared edge in model B to account for the absorption of highly ionized iron. The energy of the smeared-edge component is about 9 keV, indicating that the Fe XXVI Ly$\alpha$ line (with an ionization energy of 9.2 keV) is the most dominant absorber. We found that including a narrow line at the energy of 7.8 keV (either from Fe XXV K$\beta$ and/or Ni XXVII K$\alpha$) improves the fit as well. We also fixed the redshift in both models to 1000 km/s ($z = 0.003$), which is approximately the wind speed of the Wolf-Rayet companion \citep{koljonen17}, although this has only a slight effect on the fit.

The resulting parameters for model A fits can be found in Table \ref{modela2}. The Cyg X-3 hard-state spectrum can be adequately described by a rather soft, thermally Comptonized ($\Gamma\sim2.4$, kT$_{\mathrm{e}}\sim33$ keV) spectrum reprocessed in a highly ionized medium (log $\xi\sim3.8$) and further in a lower ionization medium (log $\xi\sim2.9$). This probably is a scattering cloud of decreasing ionization parameter. Similar modeling with two ionized absorbers was successfully used in \citet{kallman19} to fit the X-ray emission lines from \chandra\/ data with a medium-ionization component having log $\xi\sim2.9$ and a high-ionization component fixed to log $\xi\sim4.2-5.0$. The emission is further partially absorbed in a relatively dense environment (N$_{\mathrm{H}}\sim4\times10^{23}$ atoms cm$^{-2}$) with a covering fraction of f$_{\mathrm{cov}}\sim0.5$. This is consistent with the estimates of the wind column of $\sim10^{23}$ atoms cm$^{-2}$ acquired from fitting the X-ray emission lines \citep{kallman19}. Furthemore, the covering fraction can be understood as half of the emission back-scattering and passing through the cloud. 

Model B (parameters can be found in Table \ref{modelb2}) delivers similar results, with the surrounding dense torus essentially a sphere (cos($\theta_{\mathrm{tor}})>0.96$, N$_{\mathrm{H,tor}}\sim10^{24}$ atoms cm$^{-2}$; see also Fig. \ref{drawing}). The incident spectrum is very similar to that of model A ($\Gamma\sim2.4$), with a cutoff energy at 28 keV (although if the mechanism is thermal Comptonization, this corresponds to an electron temperature of kT$_{e}\sim9-14$ keV). The line-of-sight component is absorbed by a column of N$_{\mathrm{H}}\sim3\times10^{23}$ atoms cm$^{-2}$ and comprises 65\% of the flux received, while the 17\% and 18\% of the flux come from the scattered and leaked (or direct) component, respectively. The unabsorbed 3--79 keV luminosity is similar for both models and corresponds to $\sim$10\% of the Eddington luminosity\footnote{Because most of the accreted matter from the Wolf-Rayet companion is helium, the Eddington limit is twice the value of the pure hydrogen Eddington limit.} for a 2.5 solar mass black hole, which is fairly high and corresponds to values found for V404 Cyg. However, because Cyg X-3 is orbiting the Wolf-Rayet star inside its photosphere, there is always ample matter to accrete and sustain the high luminosity. On the other hand, assuming a 10 solar mass black hole (an upper limit for the allowed mass of the compact object; \citealt{koljonen17}), the luminosity would be $\sim$2--3\% of the Eddington luminosity, which is similar to what has been observed from Cyg X-1 \citep{basak17}.    

\subsubsection{V404 Cyg} \label{v404}

\begin{figure*}
 \centering
 \includegraphics[width=\linewidth]{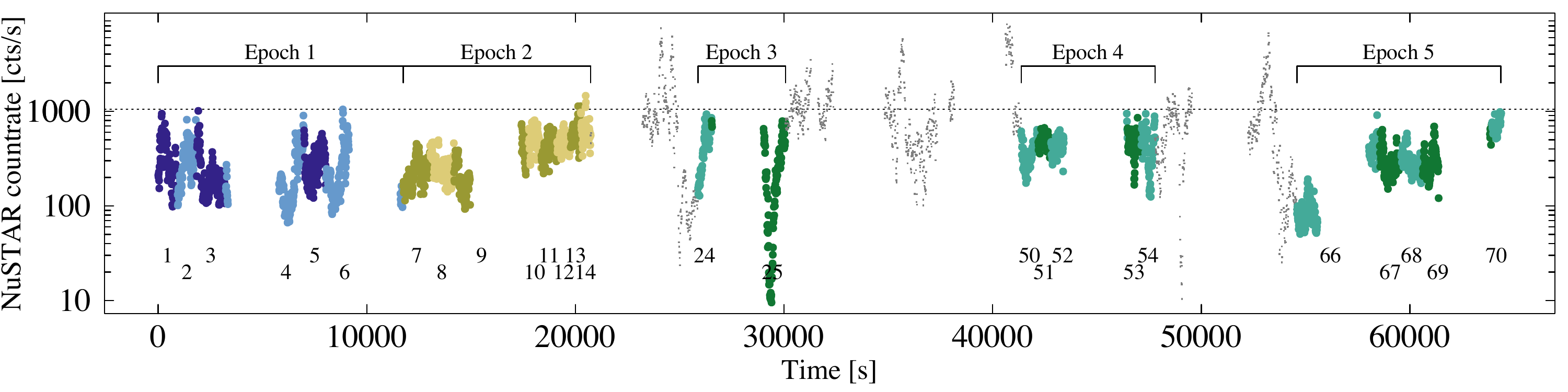}
 \caption{\nustar\/ 3--79 keV light curve of V404 Cyg divided into GTIs that total 30000 counts (numbered). The numbers for GTIs that include count rates exceeding 1050 cts/s are not shown (excluding GTIs 13 and 14), and the data points are shown in gray with a reduced point size. The coloring scheme corresponds to the spectral shape shown in Fig. \ref{groups}, with every other GTI shown in a different hue for clarity.}
 \label{v404_gti}
\end{figure*}

\begin{figure}
 \centering
 \includegraphics[width=\linewidth]{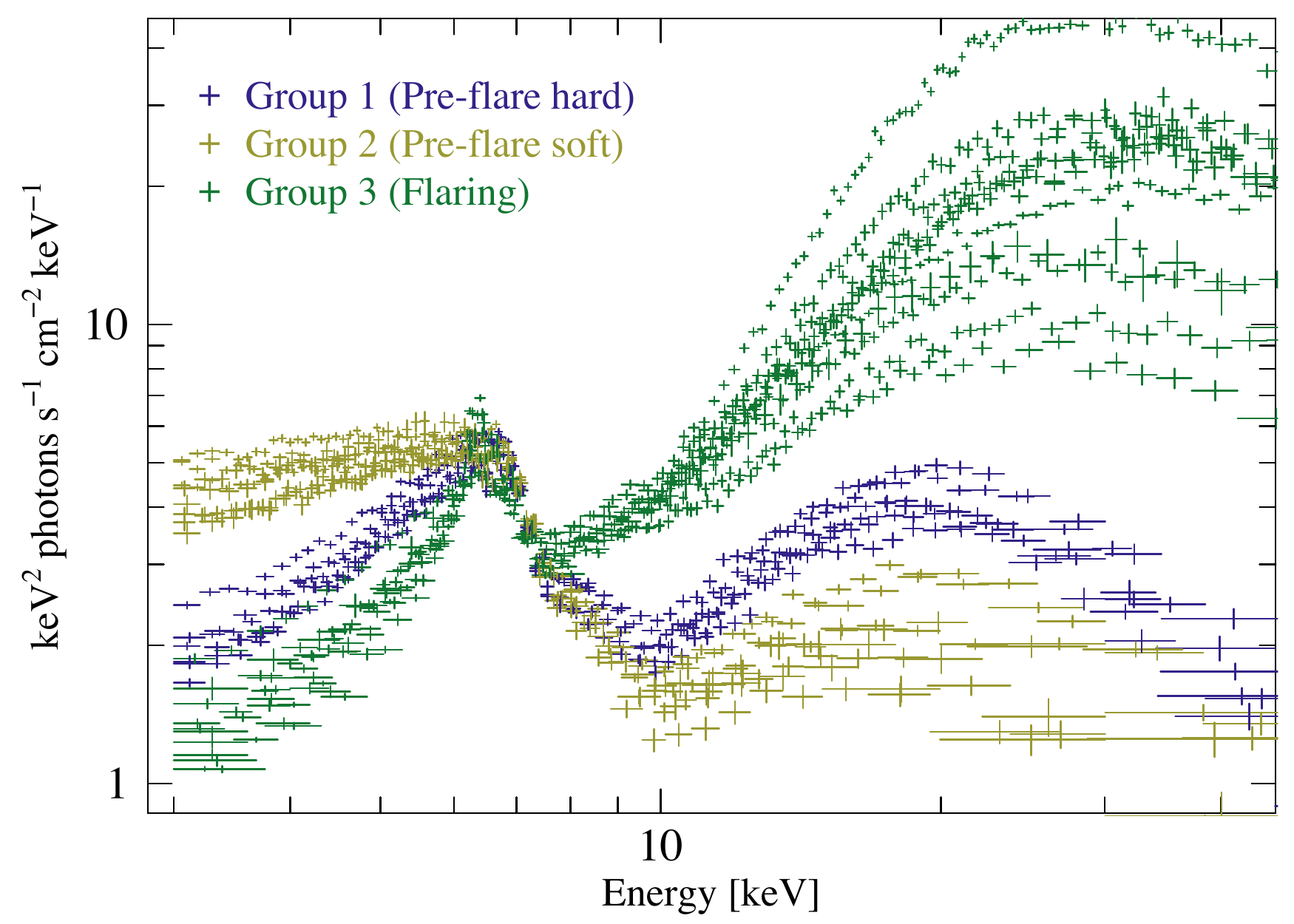}
 \caption{Changes in the spectral shape of V404 Cyg during the GTIs shown in Fig. \ref{v404_gti}. All the spectra have been normalized to the 7 keV flux in the first spectrum to show the difference in the spectral shape.}
 \label{groups}
\end{figure}

\begin{figure}
 \centering
 \includegraphics[width=\linewidth]{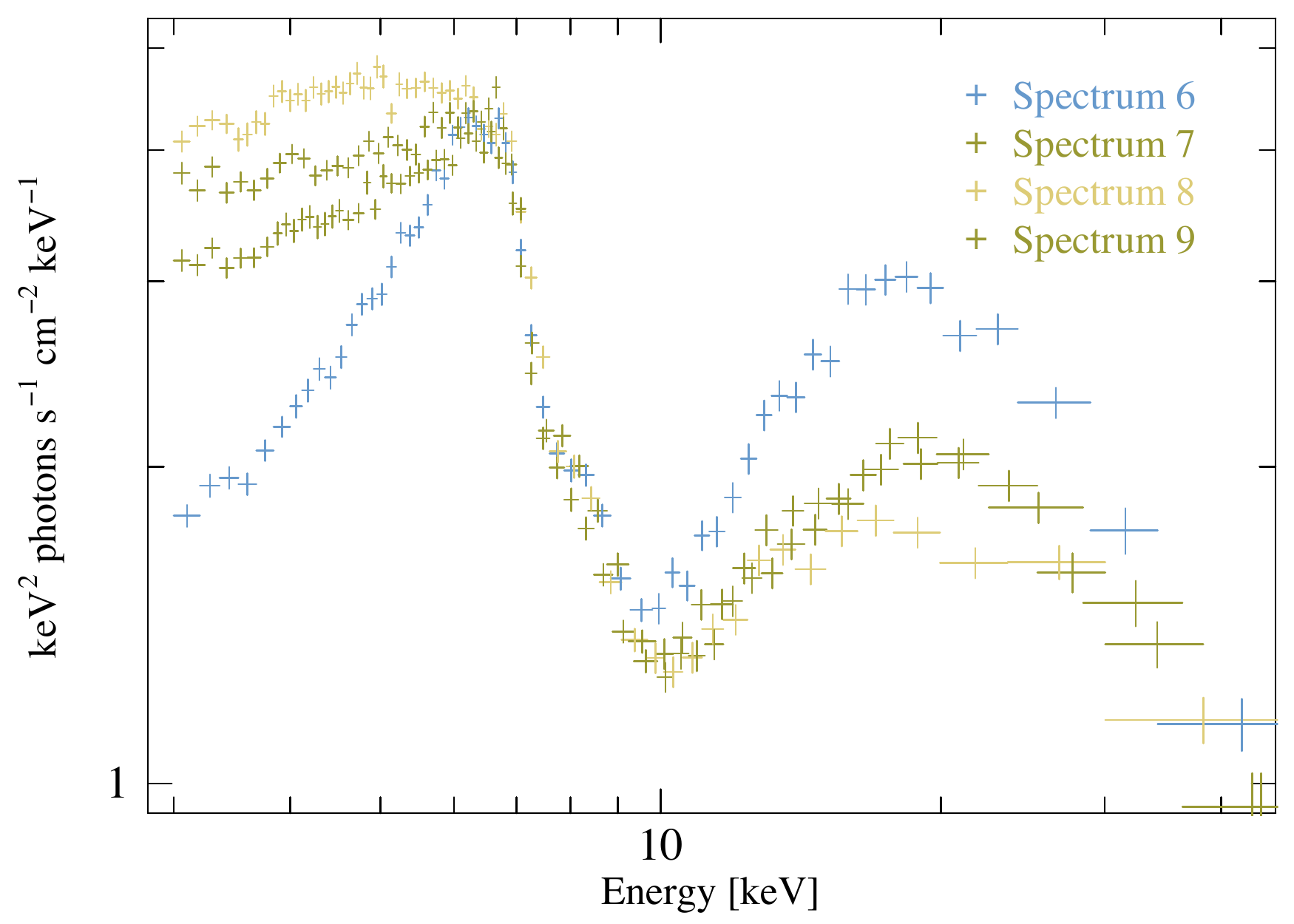}
 \caption{Spectra of V404 Cyg from GTIs 6--9 in Fig. \ref{v404_gti}, showing the change from a harder (6) to a softer spectrum (7--9) preceding the flaring period. The spectra are not normalized, thus showing the variable regimes below 6 keV and above 10 keV.}
 \label{transit}
\end{figure}

V404 Cyg (GS 2023$+$338) is one of the closest XRB (2.39$\pm$0.14 kpc; \citealt{millerjones09}) with a relatively long orbit (6.473$\pm$0.001 d; \citealt{casares92}), implying a large accretion disk and a long outburst recurrence time. Overall, two outbursts have been detected from V404 Cyg with X-ray instruments in 1989 \citep{kitamoto89,oosterbroek97,zycki99} and 2015 (\citealt{rodriguez15,motta17a,motta17b,sanchez17,walton17,kajava18}; considering the June and December 2015 flaring events to be parts of the same outburst), although in retrospect, additional optical outbursts have been detected in 1938 and 1956 \citep{richter89,wagner91}. The two X-ray outbursts in 1989 and 2015 were both hard-state outbursts with no excursion into a soft X-ray state, although the luminosities reached or exceeded the Eddington luminosity. During the 2015 outbursts, a plethora of X-ray spectral behavior was observed from V404 Cyg, with some applicable to absorption events with intermediate flux densities and X-ray spectra not consistent with a Comptonization model, and some to intrinsic variations with very high or low flux densities consistent with a Comptonization model \citep{motta17a,motta17b,sanchez17,kajava18,walton17,hynes19}. In both cases, the X-ray spectra exhibit fast changes from one state to another in a matter of seconds to minutes \citep{motta17a,kajava18,sanchez17,walton17}. 

We identify three different spectral states in the set of 30000 cts spectra during the \nustar\/ pointing (Figs. \ref{v404_gti}, \ref{groups}). At the beginning of the pointing, the source presents Cyg X-3 hard-state-like spectra, that is, spectra that likely exhibit strong absorption or reprocessing, as evidenced by the curved hard X-ray spectra (epoch 1 in Fig. \ref{v404_gti}, blue spectra in Fig. \ref{groups}). This spectrum was used in the initial model fitting in Section \ref{inimod}. The spectrum further evolves to a softer state with an increase in the fluxes below 6 keV and a decrease in the fluxes above 10 keV (epoch 2 in Fig. \ref{v404_gti}, yellow spectra in Fig. \ref{groups}). Both epochs are approximately 10 ksec long, and the change takes place in a gap between spectra 6 and 7 (Fig. \ref{transit}). During the transition, the energy region between 7--10 keV remains remarkably constant. This might be interpreted as the soft and hard part of the spectrum arising from two different components that are anticorrelated (spectral pivoting is not enough to explain the whole spectral change). After epochs 1 and 2, V404 Cyg entered into a high count rate flaring period and presented significantly harder spectra and a variable cutoff energy higher than that of the preflare spectra (epochs 3--5 in Fig. \ref{v404_gti}, green spectra in Fig. \ref{groups}).

\begin{figure}
 \centering
 \includegraphics[width=1.0\linewidth]{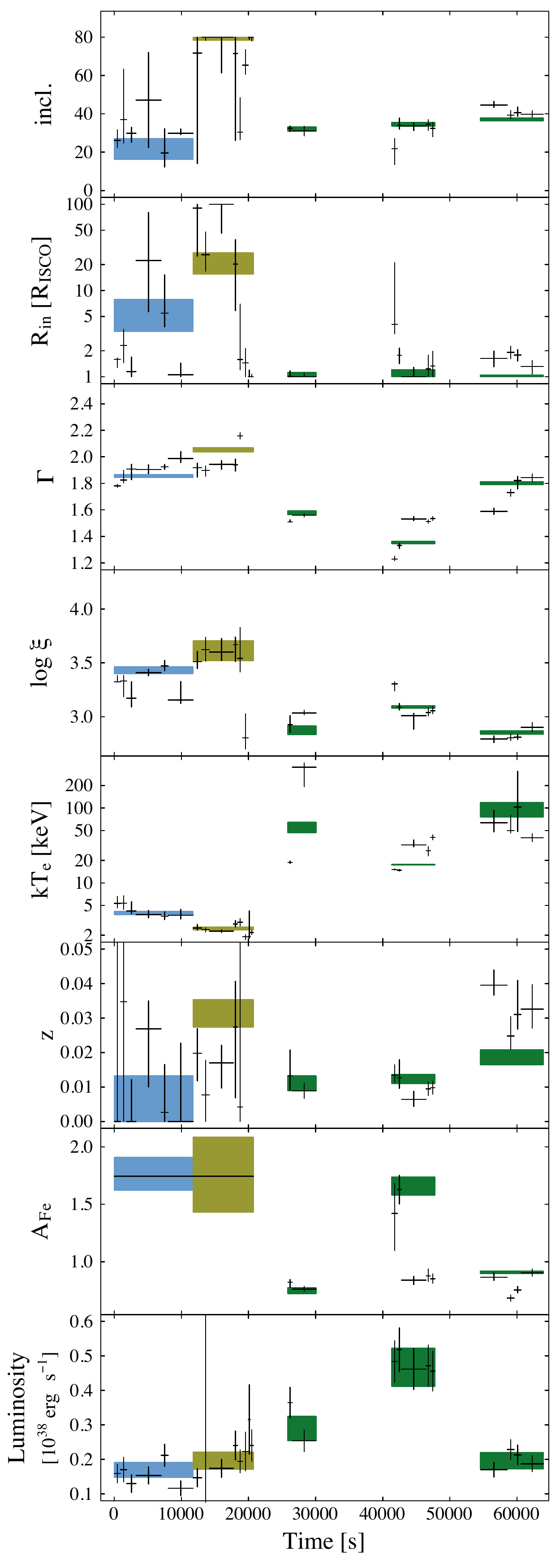}
 \caption{Model A parameter values for V404 Cyg spectra (individual GTIs in black, epochs in color).}
 \label{v404_params_A}
\end{figure}

\begin{figure}
 \centering
 \includegraphics[width=1.0\linewidth]{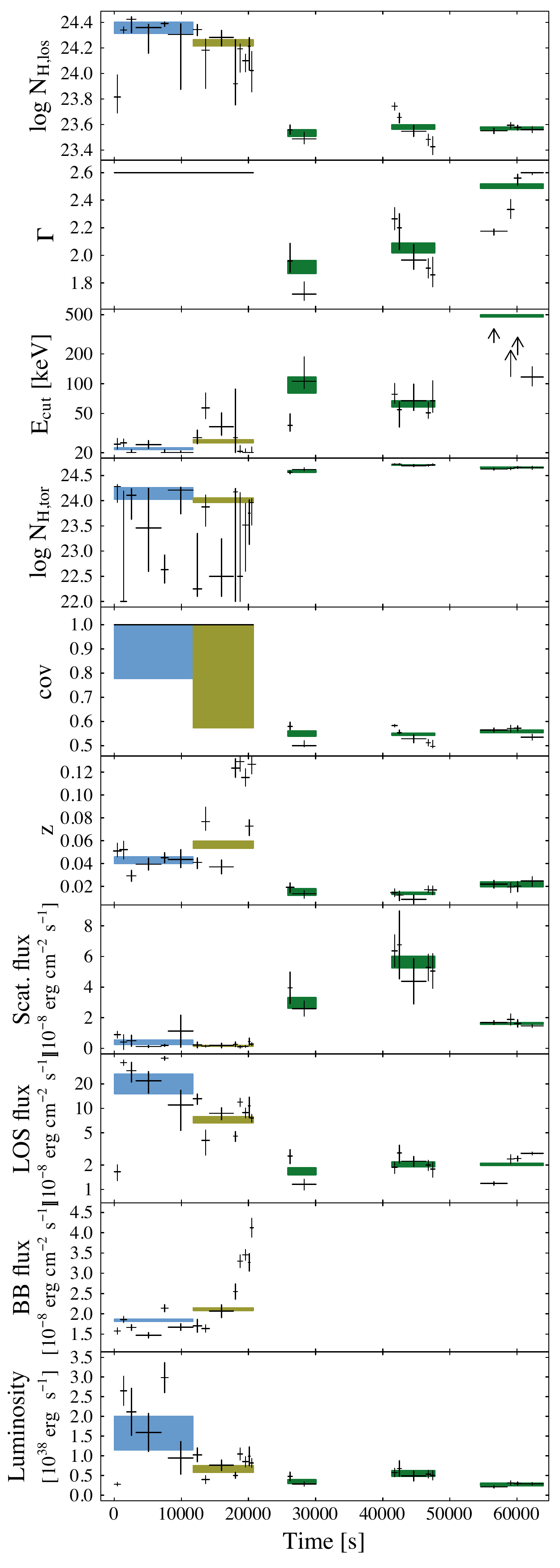}
 \caption{Model B parameter values for V404 Cyg spectra (individual GTIs in black, epochs in color).}
 \label{v404_params_B}
\end{figure}

\begin{figure*}
 \centering
 \includegraphics[width=\linewidth]{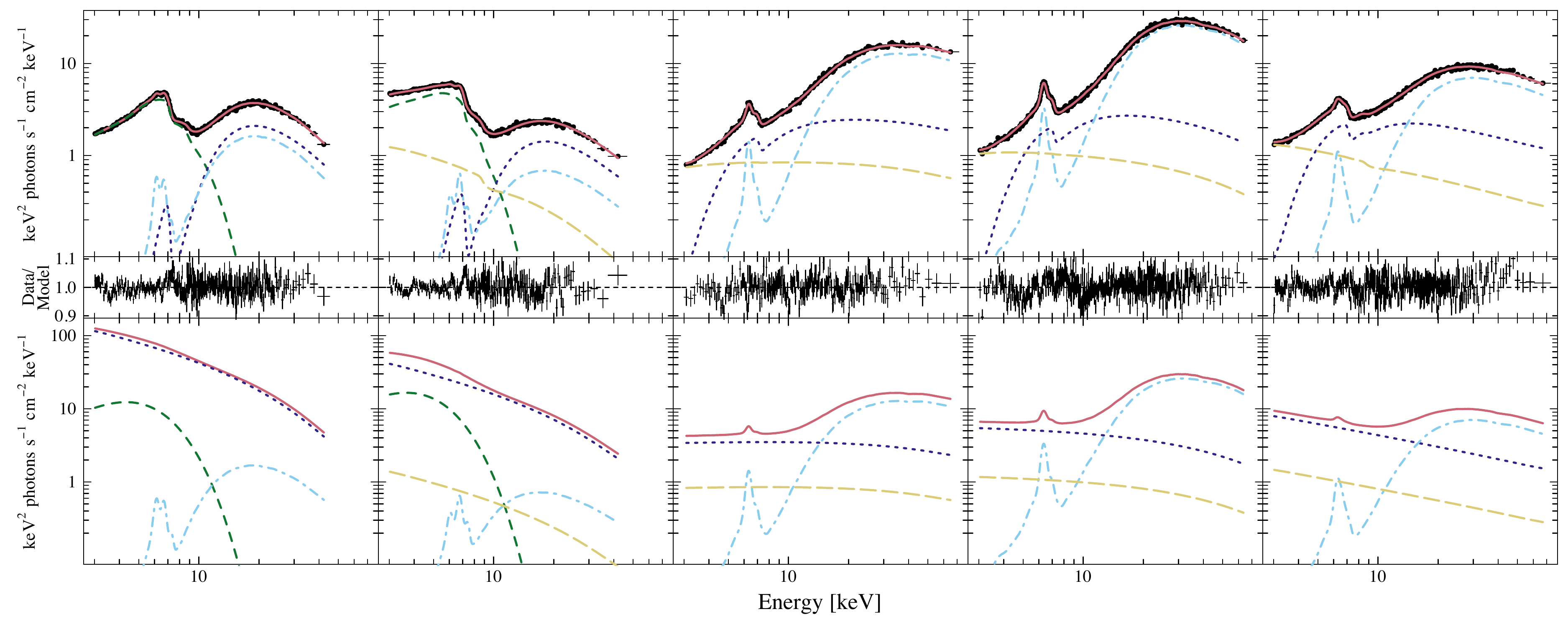}
 \caption{Averaged \nustar\/ FPMA data from V404 Cyg in the preflare (epochs 1--2) and outburst (epochs 3--5) stages as shown in Fig. \ref{groups}, together with absorbed (top row) and intrinsic (bottom row) model B components. The total model (solid red line) consists of a sum of a blackbody component (dashed green line), a cutoff power-law component absorbed in and scattered off from the material in the line of sight (dotted blue line), scattered into the line of sight by the surrounding medium (dot-dashed light blue line), and unabsorbed or direct component (long-dashed yellow line). The middle panels show the residuals of the models to the data.}
 \label{models}
\end{figure*}

We fit the averaged spectra from epochs 1--5 and the individual GTIs with models A and B. The model parameters, corresponding fluxes, and the fit quality for the averaged spectra can be found in Table \ref{modela1} for model A and in Table \ref{modelb1} for model B, while a selection of the model parameters and fluxes is shown for the individual GTIs in Figs. \ref{v404_params_A} and \ref{v404_params_B} (with a fit quality ranging between $\chi^{2}_{\mathrm{red}}$ = 0.9--1.6 with a mean of $\chi^{2}_{\mathrm{red}}$ = 1.2). In addition, the average spectra and the corresponding model B fits divided into different model components (both absorbed and unabsorbed) are shown in Fig. \ref{models}.

Considering model A fits, the spectra of all epochs can be fit with the fully reprocessed thermal Comptonization emission; a model that is essentially very similar to the basic model used in \citet[][their Table 3]{walton17}, except that here we used the version of \textsc{relxill} with a spherically symmetric corona and an intrinsic emission arising from thermal Comptonization (\textsc{relxillCp}) instead of a lamp-post geometry and an intrinsic emission modeled as a cutoff power-law spectrum (\textsc{relxilllp}). In Section \ref{inimod} the lamppost model fit to the epoch 1 hard X-ray spectrum (essentially very similar to the fit with coronal geometry) would require a very steep incident power-law spectrum. In addition, we used a simple absorption component instead of an \textsc{xstar} reprocessor. We also let the ionization parameter and the redshift of the \textsc{xillvercp} component vary freely instead of fixing them to 1 and 0, respectively, to fit the redshifted iron line with energies $\sim$6.3 keV. The results of the model A fits are very similar to those in \citet{walton17} for epochs 3--5 (the flaring state), as expected, with low intrinsic absorption, an inclination close to 30$^{\circ}$, a photon index of $\sim$1.6 on average, and an ionization parameter of $\sim$1000. Some differences do arise, however, with the iron abundance close to solar in our fits except for a few GTIs in epoch 4, compared to twice the solar value in \citet{walton17}. On another note, our fits for the epoch 1--2 averaged spectra also show elevated abundances (for individual GTIs, the abundances were fixed to the averaged value of the epoch because they were not well constrained in the fits). Because the geometry and intrinsic emission were modeled differently, the remaining parameters are more difficult to compare, but \citet{walton17} reported very high values for the reflection factor ($R_{f}\sim$1--3). This might also indicate reprocessing in the surrounding medium, as speculated in this paper, instead of a strong gravitational bending and a scenario with a high black hole spin.

On the other hand, epochs 1--2 display much softer spectra, as shown in Fig. \ref{groups}, which is reflected in the fits with increased values for the photon index ($\Gamma\sim$ 2) and ionization parameter (log $\xi\sim$ 3.5), and very low values for the electron temperature ($kT_{e}\sim3-4$ keV, corresponding to an optical depth of $\tau\sim10$). In addition, a partially covered absorption component with a high column density ($4\times10^{23}$ cm$^{-2}$, f$_{\mathrm{cov}}\sim$ 0.8), a narrow iron absorption line at 6.5 keV, and a smeared iron edge at 7.4 keV likely arising from an additional variable absorption component are needed to fit the spectra successfully with \textsc{relxillCp} (an additional \textsc{xillverCp} component is not needed in these epochs). The physical explanation for the low electron temperature is a challenge in this model. Because the cutoff energy in the epoch 1--2 spectra is low ($<$20 keV), all models with thermal Comptonization are expected to give electron temperatures that are approximately lower than 6--10 keV (assuming E$_{\mathrm{cut}}\sim$ 2--3 kT$_{e}$). We recall that the electron temperature in the model is given in the frame of the observer. A very efficient cooling mechanism, such as radiative cooling by soft photons from the strong outburst or reprocessing in an optically thick medium, can therefore thermalize and Compton downscatter the intrinsic emission to lower energies (similar to what has been proposed for Cyg X-3 in \citealt{zdziarski10}, see also Section \ref{cygx3}). The increase in soft X-ray emission between epochs 1 and 2, as mentioned above, is mirrored in the model A parameter evolution as a change in the absorption parameters (decrease in the column density and covering fraction), as an increase in the power-law photon index and normalization, and a decrease in the electron temperature. In addition, there is an increase in R$_{\mathrm{in}}$, log $\xi$, $z$, and inclination. In Section \ref{v404_soft} we discuss the hypothesis that this parameter evolution can arise from a geometry change gearing toward jet ejection.  

In epochs 1--2, model B fits are consistent with the torus completely covering the source (cos($\theta_{\mathrm{tor}}$) pegged to 1; see also Fig. \ref{drawing}), with an average density of the torus of $N_{\mathrm{H,tor}}\sim10^{24}$ atoms cm$^{-2}$ and a similar if slightly higher (a factor of two) line-of-sight column density through the torus. The flux from the line-of-sight component dominates the spectrum during these epochs, comprising of 75--90\% of the total flux, while scattered and leaked flux contribute only up to 2\% (the remaining flux comes from the blackbody component discussed below). In epochs 3--5, the geometry changes from spherical to more disk-like with cos($\theta_{\mathrm{tor}})\sim0.55$, and an increase of a factor of four in $N_{\mathrm{H,tor}}$, while the line-of-sight column density decreases by a factor of about six (see also Fig. \ref{drawing}). The flux from the line-of-sight component decreases to 25--50\% of the total flux, while the scattered flux increases to 40--70\% of the total flux and causes the majority of the spectral variability in this state. Moreover, the flux from the leaked component increases significantly, up to 10\% of the total flux, indicating that the geometry has changed so that the direct intrinsic emission is partly visible. Similar to the parameter evolution in model A, the intrinsic cutoff power-law spectrum is soft in epochs 1--2, with the power-law photon index $\Gamma$ pegged to 2.6 and cutoff energy values of about 21--26 keV. The spectrum hardens in epochs 3--5 with $\Gamma\sim2$ and much higher cutoff energies, although overall the intrinsic spectrum is softer than in model A. In contrast to model A, no absorption line is required in the model for epochs 1--2 because the double-horned line profile can be explained by reprocessing in an outflowing medium, which is seen as both redshifted and blueshifted ($z\sim0.05c\sim$15000 km/s). Previously, \citet{motta17a} have shown that the iron line exhibit redshift and blueshift with velocities $v<0.1c,$ in line with the \nustar\/ values. For epochs 3--5, the resulting line speeds ($\sim$4000 km/s) are comparable to the values measured from the P Cygni profiles of He I $\lambda$5876 (a well-known accretion wind tracer), while for epochs 1--2, they are still well within of what can be achieved with super-Eddington accretion (usually on the order of $0.1c-0.2c$; e.g., \citealt{pinto19}). In epochs 3--5, the back-illumination dominates the scattering (i.e., c2 $>$ c3 in Table \ref{modelb1}), so that the scattered emission is seen through material that is more translucent than that of the line of sight. The scattered line therefore exhibits mainly redshift and reduces the observed iron K$\alpha$ line energy to 6.3 keV without producing the blueshifted line.

The strong absorption of the intrinsic emission in model B requires an additional component  to account for the elevated soft X-ray emission in epochs 1--2. We modeled this as a partially absorbed blackbody component, as discussed in Section \ref{modeling}. The temperature of the blackbody component decreases from 1.1 keV to 0.95 keV, while its normalization (and flux) increases from epoch 1 to epoch 2, indicating an increase in the size of the emitting medium. While the temperature of this component might arise from the hot disk, it does not seem plausible that the emitting area would increase and temperature decrease when the source is gearing toward the flaring state when emission from an accretion event is considered.  \citet{zycki99} briefly discussed that the blackbody component needed to provide the soft excess in the 1989 outburst might arise from incident disk photons that are thermalized in multiple scattering processes in the surrounding medium. Thus, a dense medium, whether a disk with a  large scale-height or a stellar or accretion disk wind, might then explain both the large column and the soft component, similar to what was discussed for model A above. 

For both models, the unabsorbed luminosities in the 3--79 keV \nustar\/ band are sub-Eddington for all epochs (2--4\% for model A, and 2--14\% for model B), although because the intrinsic spectrum is steep in epochs 1--2, the Eddington limit would be reached by extrapolating the spectrum down to $\sim$0.2 keV (for model B at least). The luminosity of epochs 1--2 is much higher for model B because the highly absorbed intrinsic emission is included, while model A presents only the flux for the scattered component. When the scattered flux is only taken into account in model B, the resulting luminosities agree with those for model A.

\subsubsection{V4641 Sgr}

\begin{figure}
 \centering
 \includegraphics[width=\linewidth]{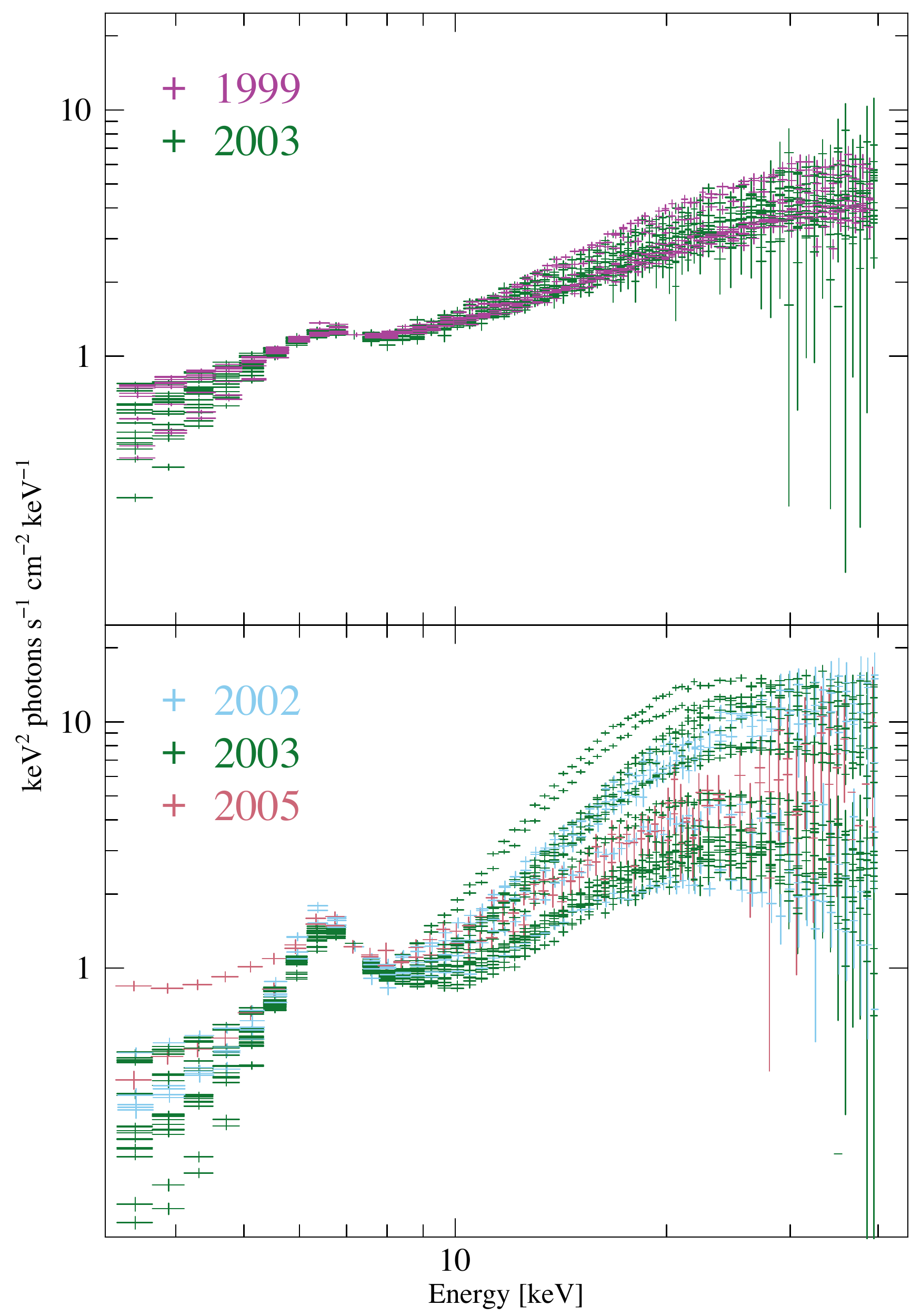}
 \caption{Changes in the \rxtepca\/ spectral shape of V4641 Sgr during the outburst periods of 1999, 2002, 2003, and 2005. All the spectra have been normalized to the 7 keV flux in the first spectrum to show the difference in spectral shape. The top panel shows outburst spectra reminiscent of the spectra with high count rate and low absorption of V404 Cyg \citep{walton17}, while the bottom panel shows  spectra reminiscent of the flaring spectra with lower count rate and higher absorption of V404 Cyg in Fig. \ref{groups}.}
 \label{v4641_groups}
\end{figure}

V4641 Sgr is a very peculiar source, presenting a dynamically confirmed black hole with a high-mass companion \citep{orosz01}, but showing transient outbursts similar to low-mass XRBs. The outbursts of V4641 Sgr can be very short ($\text{about}$ a week) and intense, reaching super-Eddington levels, as in 1999 outburst \citep{revnivtsev02}, or longer but much weaker \citep[e.g.,][]{uemura02,maitra06,munozdarias18}. There is some evidence that the inner accretion disk is misaligned to the orbital plane of the binary ($i \sim 70^{\circ}$; \citealt{orosz01,macdonald14,pahari15}), and V4641 Sgr might instead be a low-inclination source in X-rays and radio ($i \sim 10^{\circ}$; \citealt{hjellming00,orosz01,gallo14}). It has previously been suggested that a \chandra\/ spectrum observed during an outburst decline is very similar to those observed from Seyfert-2 AGN \citep{morningstar14}. 

The \rxtepca\/ spectra of V4641 Sgr gathered from the 1999, 2002, 2003, and 2005 outbursts are plotted in Fig. \ref{v4641_groups}. In the top panel, the spectra from 1999 and the latter part of 2003 are very reminiscent of the spectra with low absorption and high count rate of V404 Cyg that were studied in \citet{walton17}, while in the bottom panel, the spectra from the 2002, 2003, and 2005 outburst are more reminiscent of the flaring spectra with lower count rate of V404 Cyg (Fig. \ref{groups}). The only outburst containing both types of spectra is the 2003 outburst, which began with those presented in the bottom panel (epochs 1 and 2 in \citealt{maitra06}, including the spectra shown in Fig. \ref{spectra}) and continued with those presented in the top panel (epochs 3 and 4 in \citealt{maitra06}). 

The two models (A and B) were fit to the spectra from two pointings observed during the 2002 (hereafter epoch 1) and 2003 (hereafter epoch 2) outbursts (Fig. \ref{spectra}; the top corresponds to epoch 1, and the bottom corresponds to epoch 2), and the resulting parameters are shown in Tables \ref{modela2} and \ref{modelb2}. Both observations were taken during the middle of the outburst peak with similar optical magnitudes ($\sim$11.5 mag in V and R band; \citealt{uemura02,maitra06}) and X-ray count rates. Because the statistics of \rxtepca\/ are much lower than those of \nustar\/, we fixed the inclination to $i = 70^{\circ}$ in both models to reduce the number of free parameters. In the case of model A, a successful fit could be achieved with a single \textsc{relxillCp} model, while for model B, the spectra are dominated by the (back-)scattered component (75--80\% of the total flux) for both epochs. The main difference in the parameters of both models for the two epochs can be found in the intrinsic spectrum with the power-law photon index being higher for epoch 1 ($\Gamma_{A} = 2.2$, $\Gamma_{B} = 2.6$) with no high-energy cutoff, while for epoch 2, the power-law photon index is much lower ($\Gamma_{A} = 1.5$, $\Gamma_{B} = 1.6$) and the spectrum has a low-energy cutoff ($kT_{e} \sim$ 9 keV, E$_{\mathrm{cut}} \sim$ 20 keV).    

Unlike in other sources, the covering fraction and the line-of-sight absorption column in model B are low for both epochs. The scattered component dominates the flux in model B, while pure scattering in a plasma with a single-ionization parameter is consistent according to model A. In contrast to V404 Cyg epochs 1--2 and GRS 1915+105 epoch 1 (discussed in the next section), there is evidence of a strong jet in the two V4641 Sgr epochs. The 8.5 GHz radio observations of the Very Large Array show flux densities of 80--170 mJy  \citep{rupen02} and 550--570 mJy \citep{rupen03} coinciding with epochs 1 and 2, respectively. This means that during both observations, any surrounding matter was likely evacuated by the jet. Because the inclination of the system is likely high, we might observe the X-ray source through a disk wind or geometrically thick accretion flow (the column density of the torus is the highest of all the four sources for model B fits; log N$_{\mathrm{H,tor}}$ $\sim$ 10$^{25}$ cm$^{-2}$), and most of the emission received is from backscattering (i.e., c2 = 0 in Table \ref{modelb2}; see also Fig. \ref{drawing}). 

The luminosities of the two V4641 Sgr epochs are much lower ($\sim$5$\times$10$^{36}$ erg/s, corresponding to $\sim$0.5\% of the Eddington luminosity) than the luminosities for the other sources, although the observations are from the peak of the outbursts. The 2002 and 2003 outbursts were weaker than the much more luminous outburst in 1999, where the peak luminosity was at or greater than the Eddington luminosity \citep{hjellming00,revnivtsev02}. These weak outbursts do not have any predictable periodicity, but they seem to occur roughly at intervals of 500--600 days \citep{negoro18}. On the other hand, an optical counterpart as bright as in the 2002 and 2003 outbursts has not been detected for any other weak outbursts since the 1999 outburst, which marks them as different and probably means that they included enhanced reprocessing of the X-ray emission in the accretion disk.    

\subsubsection{GRS 1915$+$105}

GRS 1915$+$105 is one of the brightest XRBs in our Galaxy in its outburst because it has the longest orbital period known among low-mass XRBs \citep[33.9 days;][]{steeghs13} and thus the largest accretion disk size because of the largest tidal truncation radius. The huge mass reservoir has lasted already three decades, powering the outburst until the drop in the X-ray flux in 2018. We fit all three \nustar\/ observations (labeled epochs 1--3) from this anomalous state with models A and B. The resulting model parameters are shown in Tables \ref{modela2} and \ref{modelb2}. With model A, epoch 1 is fit best with a single \textsc{relxillCp} model, while epochs 2 and 3 can be fit with two \textsc{xillverCp} models, one corresponding to a plasma with a higher (log $\xi \sim$ 3.4) and the other to a plasma with a lower ionization parameter (log $\xi \sim$ 2).  Epoch 1 includes a prominent narrow (up to detector resolution) absorption line at 6.56 keV that might be the Fe XXV K$\alpha$ line, although it is redshifted by 6000--7000 km/s, or the Fe K$\alpha$ line blueshifted by the same amount. In addition, an edge at 7.4 keV is needed to fit the spectra adequately. In the case of blueshifted neutral iron absorption, this implies a velocity of the absorbing material of $\sim$13000 km/s. However, the redshift needed to model the spectrum is even higher, $z\sim$0.12, which agrees with what is observed from ultraluminous X-ray sources (ULXs). For epochs 2--3, the redshift required is much lower ($\sim$ 3000 km/s). In contrast, the iron abundance increases from solar in epoch 1 to around twice the solar value in later epochs. The incident cutoff power-law spectrum changes from epoch 1 to epochs 2--3, displaying a decrease in the power-law photon index from 1.9 to 1.6, with the latter two also showing a cutoff at 14 keV, while in epoch 1, the cutoff is unconstrained ($\gtrsim$160 keV, when left free to vary). The inclination is only constrained for epoch 1, corresponding to 60--80 degrees, while for epochs 2 and 3, we fixed it to 70 degrees. The models are absorbed with N$_{\mathrm{H}}$ $\sim$ 5$\times$10$^{22}$ atoms cm$^{-2}$, which is comparable with the interstellar value of N$_{\mathrm{H}}$ $\sim$ 3.5$\times$10$^{22}$ atoms cm$^{-2}$. Additional, partial absorption is needed for epochs 2 and 3, with N$_{\mathrm{H}}$ $\sim$ 5$\times$10$^{23}$ atoms cm$^{-2}$ and a covering fraction of $\sim$0.5.

\begin{figure}
 \centering
 \includegraphics[width=\linewidth]{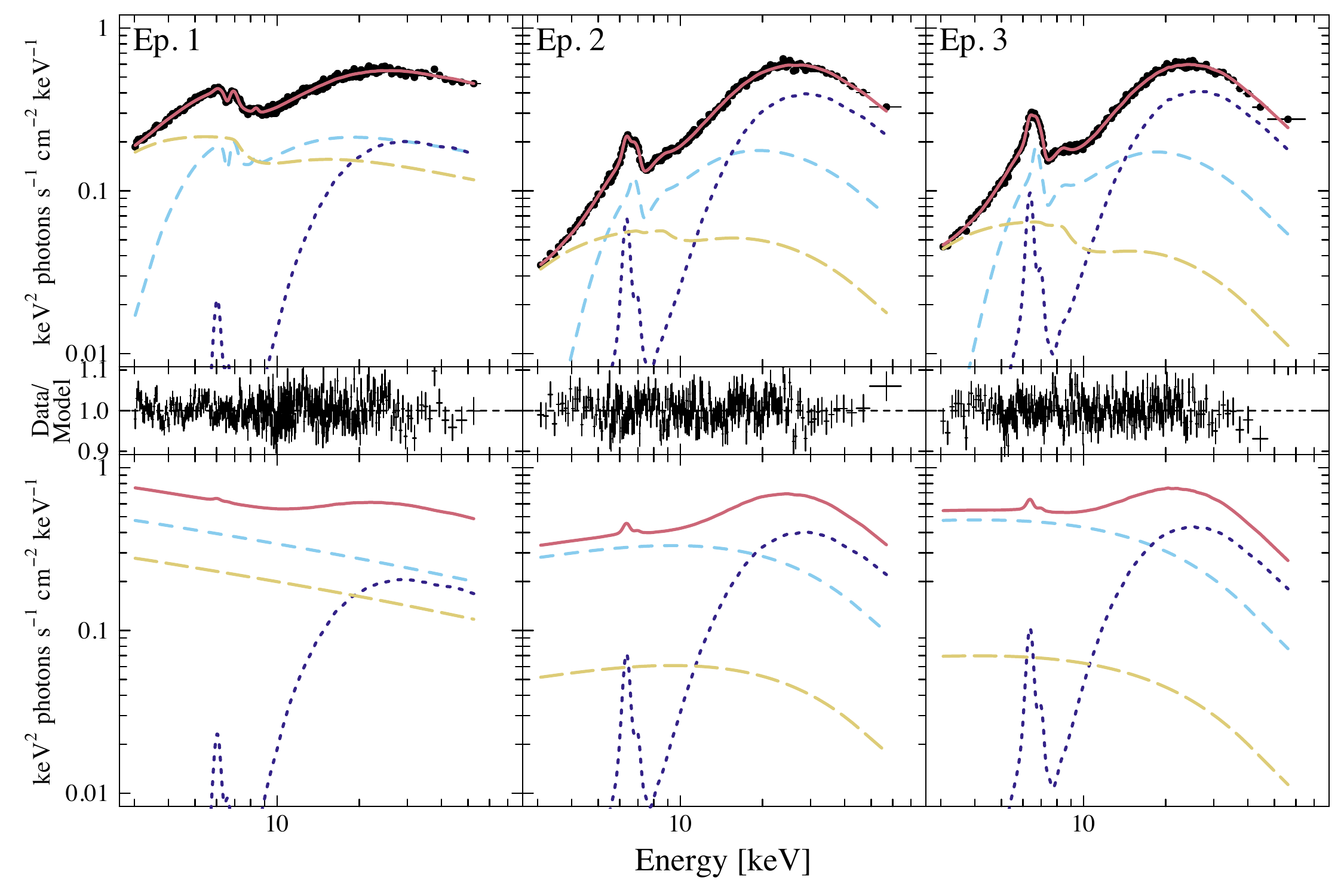}
 \caption{Averaged \nustar\/ FPMA data from GRS 1915$+$105 from the three epochs together with absorbed (top row) and intrinsic (bottom row) model B components. The total model (solid red line) consists of the sum of a cutoff power-law component absorbed in and scattered off from the material in the line of sight (dotted blue line), scattered into the line of sight by the surrounding medium (dot-dashed light blue line), and unabsorbed or direct component (long-dashed yellow line). The middle panels show the residuals of the models to the data.}
 \label{1915_params_B}
\end{figure}

\begin{figure}
 \centering
 \includegraphics[width=\linewidth]{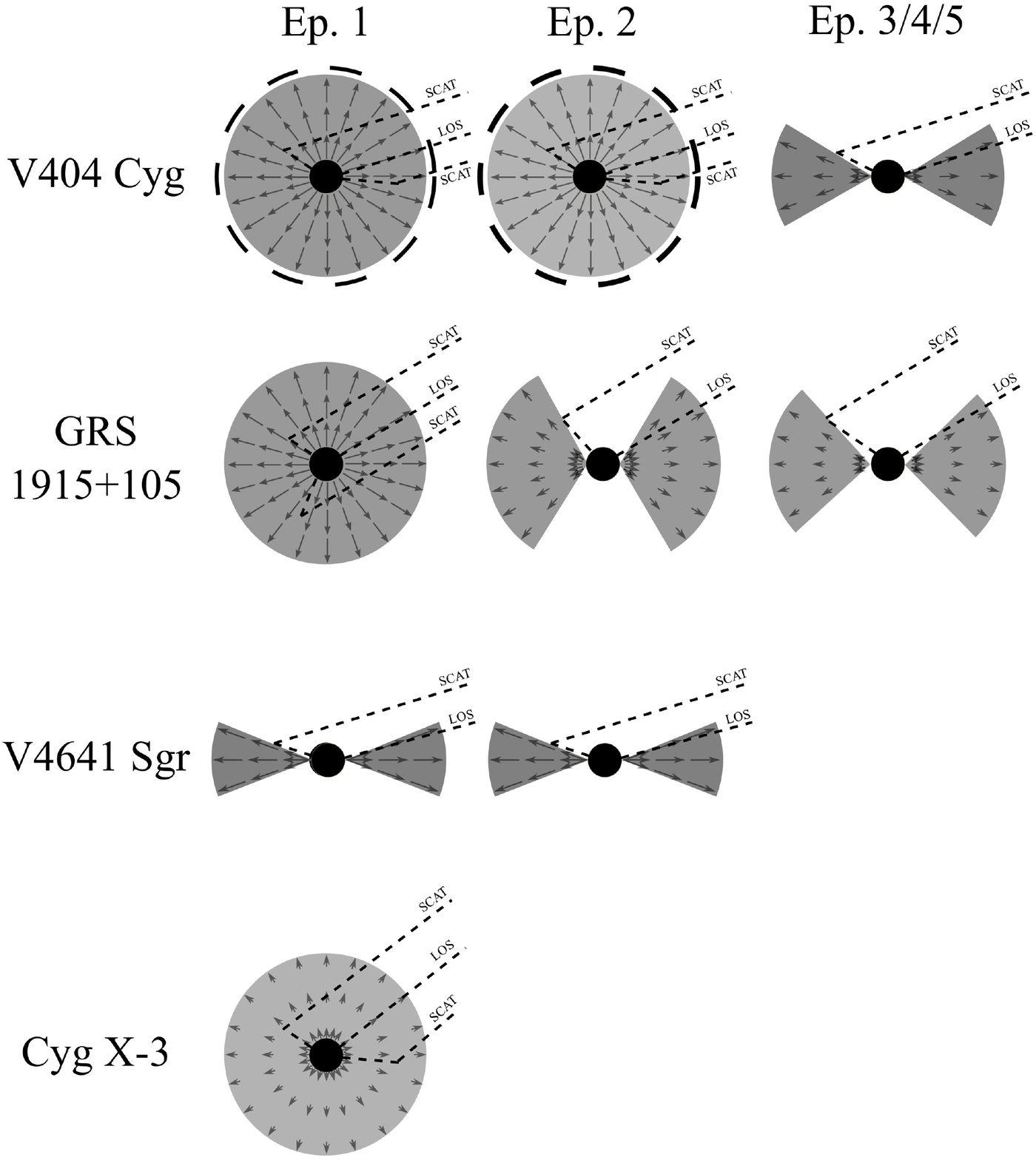}
 \caption{Geometry of the torus-shaped obscuring matter around the intrinsic X-ray source according to the parameter $\theta_{\mathrm{tor}}$ in model B fits for all sources, in addition to a graphical representation of the column densities (gray, darker for a higher column density), the wind speeds (arrows), and the lines of sight (dashed lines). The thick dashed line around V404 Cyg epoch 1 and 2 corresponds to the partially absorbed blackbody component. The model parameters for each source can be found in Tables \ref{modelb1} and \ref{modelb2}.}
 \label{drawing}
\end{figure}

With model B, epoch 1 is consistent with the torus covering the source completely, while cos($\theta_{\mathrm{tor}}$) decreases to 0.8 and 0.7 in epochs 2 and 3, respectively (Fig. \ref{drawing}). The density of the torus remains equal in all epochs ($N_{\mathrm{H,tor}}\sim2.5\times10^{24}$ atoms cm$^{-2}$). The line-of-sight component contributes about equally in all epochs (52--59\%) with a similar column density ($N_{\mathrm{H,los}}\sim2.5-5\times10^{23}$ atoms cm$^{-2}$). The scattered flux increases from 17\% to 32--35\%, while the leaked flux decreases from 31\% to 9--10\% when transiting from epoch 1 to epochs 2 and 3. The scattering component is dominated by the redshifted scattering, with epoch 1 showing high wind speeds of 0.057$c$ ($\sim$17000 km/s, which is more in line with the velocity of the absorption line if it is due to a blueshifted neutral iron line) that reduce to much lower values (600--2700 km/s) in epochs 2 and 3. The incident spectrum in epoch 1 is a steep power law ($\Gamma\sim2.3$) similar to V404 Cyg epochs 1, 2, and 5, and Cyg X-3 spectra. In epochs 2 and 3, the power-law index of the incident spectrum decreases to 1.6--1.8 and the spectrum exhibits a cutoff at 21--24 keV. Interestingly, the unabsorbed flux remains similar in all epochs and corresponds to 1--2\% of the Eddington flux. The average spectra and the corresponding model B fits divided into different model components (both absorbed and unabsorbed) are shown in Fig. \ref{1915_params_B}. The inclination is fairly well constrained in epochs 2 and 3 and corresponds to 40--60 degrees. For epoch 1, it is not well-constrained in the fit, and we froze it to 53 degrees (cos($\theta_{\mathrm{inc}}$)=0.6). Based on measurements of the jet inclination of the system, \citet{reid14} estimated the disk inclination angle as 60$^{\circ}\pm$5$^{\circ}$, which is consistent with the values derived here.

While having slightly different parameters, the two models show a similar evolution of the model parameters, consistent with a scenario of obscured emission through fast (spherical) outflowing wind in epoch 1, which flattens and decelerates in later epochs (Fig. \ref{drawing}). Epoch 1, which is observed immediately after the sudden decrease of the hard X-ray emission, shows higher wind velocities but neutral absorbing material, higher incident power-law photon indices, and no spectral cutoffs. This is consistent with a scenario where the inner accretion flow is obscured by matter that is first seen as scatterer and absorber. Later, it is mostly seen in reflection, and the regions with higher ionization are exposed. 

\subsection{X-ray timing} \label{timing}

\begin{figure*}
 \centering
 \includegraphics[width=\linewidth]{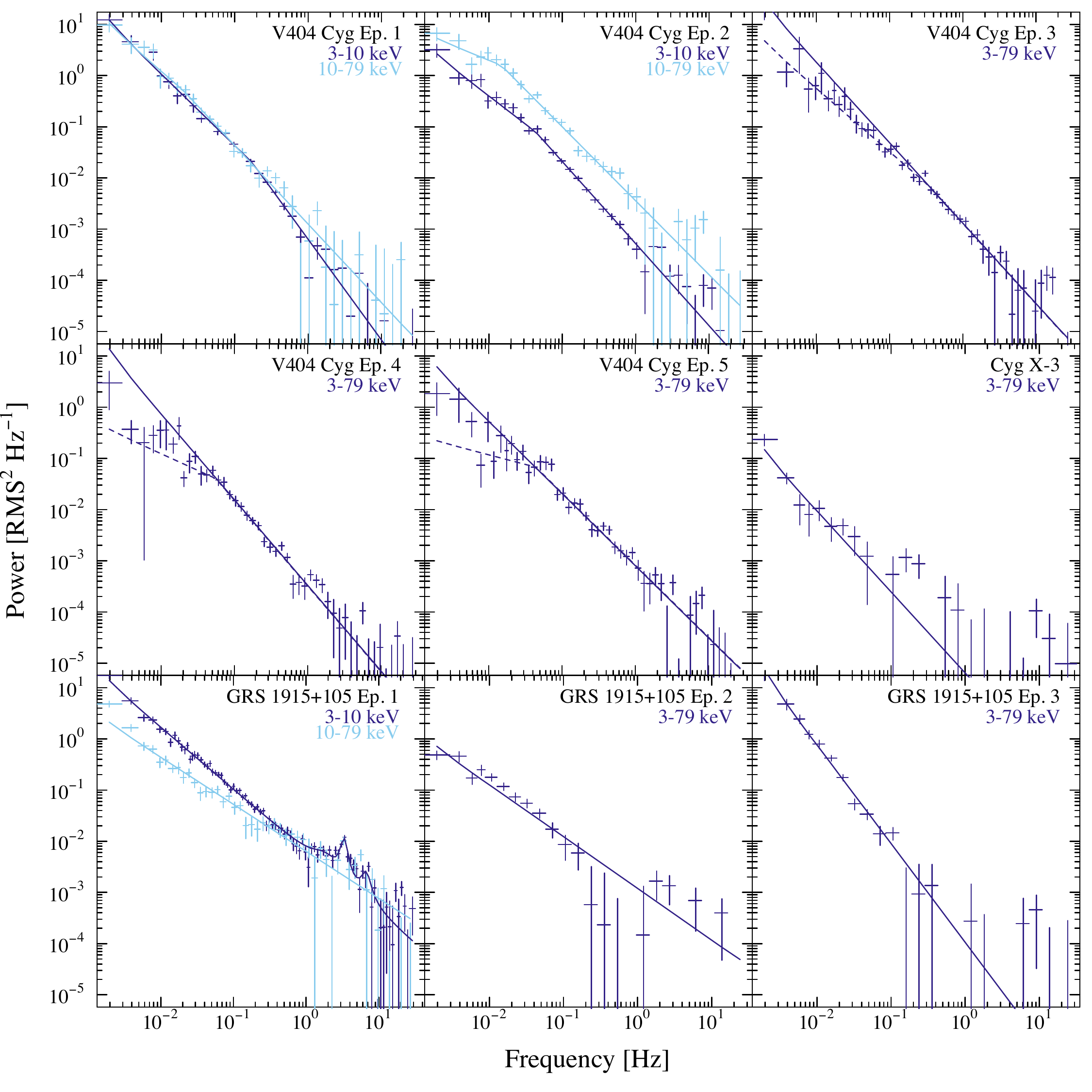}
 \caption{Cospectra of V404 Cyg, Cyg X-3, and GRS 1915$+$105 for \nustar\/ epochs. The soft- and hard-band cospectra are shown separately when they display any differences. The best-fit single or broken power-law model is also plotted for all data. Additional Lorenzian functions are needed for the QPO and harmonics in the GRS 1915$+$105 epoch 1 data.}
 \label{timing}
\end{figure*}

While we did not concentrate on the X-ray timing properties of the sources in detail, in the following we provide a quick analysis of the \nustar\/ X-ray cospectra (a proxy for the PSD). The cospectra for the \nustar\/ data for all sources are shown in Fig. \ref{timing}. As described in Section \ref{observations}, we extracted two cospectra for the soft X-ray band (3--10 keV) and the hard X-ray band (10--79 keV), and when they did not present any differences from each other, we extracted the cospectra from the full \nustar\/ range 3--79 keV. 

We can assume that the reprocessing or scattering in the obscuring matter smears out most of the high-frequency timing information of the intrinsic emission. For Cyg X-3, it has been shown that the PSD is close to a power law with an index of --2.0, independent of the accretion state of the source \citep{axelsson09,koljonen11}, and this has been speculated to be due to a suppression of the high-frequency variations by scattering in the stellar wind surrounding the X-ray source, mimicking a red noise process \citep{koljonen18}. Likewise, for the \nustar\/ pointing considered here, we obtain a power-law cospectrum with an index of --1.9$\pm$0.1.

For V404 Cyg, a timing analysis has been performed on a subset of the \nustar\/ data used in this paper in \citet{gandhi17}. The authors found that the X-ray cospectrum is consistent with a power-law spectrum with an index of --1.6; this is not as steep as in Cyg X-3 or V4641 Sgr, but steeper than the flicker noise that is typically observed from hard states of XRBs, indicating the suppression of high-frequency variability. We studied here the epoch-by-epoch variations in the cospectrum and found that a single or broken power law fits all the data sufficiently well. In all epochs, the slope of the high-frequency spectra agrees roughly with an index of $\sim$--1.6 (epoch 1: --2.0$^{+0.4}_{-0.2}$/--1.41$\pm$0.06, epoch 2: --1.58$\pm$0.08/--1.41$^{+0.08}_{-0.03}$, epoch 3: --1.6$^{+0.2}_{-0.1}$, epoch 4: --1.7$\pm$0.1, epoch 5: --1.4$\pm$0.1; when two numbers are given, they correspond to the 3--10 and 10--79 keV band), while there is some evidence of a break to flatter indices at lower frequencies. Interestingly, the epoch 2 cospectrum shows a diminished rms in the soft X-ray band compared to the hard X-ray band, which might incidate that the soft spectral component dilutes the rms. In addition, the spectral break shifts to lower frequencies from 0.2 Hz to 0.06 Hz, and if this is related to the size of the varying soft component, it indicates an increasing emitting area that is consistent with the blackbody parameter evolution of model B fits. 

GRS 1915$+$105 is famous for its complex X-ray variability. The PSDs typically consist of a band-limited noise component with one or more peaks, indicating quasi-periodic oscillations \citep[QPOs; e.g.][]{morgan97}. The cospectra from epochs 1--3 are consistent with a power-law noise (epoch 1: 1.21$\pm$0.02/0.91$\pm$0.05, epoch 2: 1.02$\pm$0.06, and epoch 3: 1.96$\pm$0.08; when two numbers are given, these correspond to the 3--10 and 10--79 keV band), with the epoch 1 cospectrum also showing a low-frequency QPO with harmonics at 3.3 Hz (at least for the soft X-ray band). Clearly, the epoch 1 cospectrum differs from epochs 2--3, which do not present power over 0.1 Hz, and likely we have a more direct view of the accretion flow. In addition, the soft X-ray band where the incident (leaked) spectral component dominates (Fig. \ref{1915_params_B}) is more variable than the hard X-ray band. For the model B fit, the leaked component presents the highest luminosity fraction in epoch 1, comprising 31\% of the total flux, and in this pointing, we might be observing a patchy outflow. For epochs 2--3, the amount of flux observed from the leaked component is diminished and the total flux is dominated by the line-of-sight and scattered components, which accounts for the loss of high-frequency power. The low-frequency power varies from flicker noise in epochs 1--2 to red noise in epoch 3.        

For V4641 Sgr, \citet{maitra06} studied the X-ray timing properties during the 2003 outburst and found red-noise-dominated PSD below 1 Hz and Poisson noise above it. Slightly more structure was found in the much brighter 1999 outburst, with the PSD showing a broken power-law shape with indices of $\sim$--1 and $\sim$--2 below and above 5 Hz, respectively \citep{wijnands00}. However, no other structure, such as QPOs, were found in the PSD reaching 100 Hz, and again the higher frequency spectrum is consistent with the red-noise process. For the \rxte\/ pointings of V4641 Sgr considered here, the PSDs are consistent with pure Poisson noise in the case of the 2002 data, and a red noise below 0.03 Hz and Poisson noise above for 2003 data. 

\section{Discussion} \label{discussion}

We have shown that the X-ray spectra of V404 Cyg, Cyg X-3, GRS 1915$+$105, and V4641 Sgr share similarities at certain evolutionary times that correspond to low X-ray flux periods during outburst events (considering that the persistent source Cyg X-3 is always `on', that is, in outburst, and that V4641 Sgr exhibits low-luminosity outbursts). We modeled these spectra successfully with two models consisting either of a fully reprocessed spectral component and/or a heavily absorbed spectral component, with the intrinsic spectra arising from a thermal Comptonization process. It is well known that Cyg X-3 orbits its Wolf-Rayet companion star inside a high-density stellar wind and that it can strongly affect the X-ray spectra. Based on the spectral similarity, we can therefore presume that similar surroundings affect the X-ray spectra of V404 Cyg and V4641 Sgr during the outbursts, and the recent anomalous accretion state of GRS 1915$+$105. Because the companion stars of V404 Cyg, V4641 Sgr, and GRS 1915$+$105 are not expected to present high-density stellar wind, we attribute this medium to either a large scale-height accretion flow (possibly due to a super-Eddington accretion rate) or an optically thick equatorial outflow or envelope either from the radiation pressure of intrinsic (super-Eddington) emission of the accretion flow or from the base of the jet. 

Both models (A and B) indicate a change in geometry in the system in the evolution of V404 Cyg and GRS 1915$+$105. Epochs 1--2 for V404 Cyg and epoch 1 for GRS 1915$+$105 are consistent with the X-ray source being obscured in an outflowing (spherical) plasma cloud that transforms into a more disk-like geometry in subsequent epochs (Fig. \ref{drawing}). This change also coincides with the start of the activity in the radio and X-ray flaring. The two V4641 Sgr epochs are consistent with a disk-like geometry, and in both cases, strong radio emission was detected at the same time. This means that the geometry change of the obscuring component seems to be linked with the radio evolution. Interestingly, a very similar evolution took place in the 1999 outburst of V4641 Sgr \citep{revnivtsev02}, with the source luminosity dropping by an order of magnitude followed by optical and subsequently radio emission and a change in the X-ray spectrum from softer (epoch 1-type) to harder (epoch 2-type).  

In V404 Cyg and GRS 1915$+$105, the intrinsic spectra first resemble that of an intermediate state of XRBs with power-law photon indices $\Gamma\sim 2.0-2.6$, and changes to lower values, $\Gamma\sim 1.4-2.0$, more typical of an XRB hard-state spectrum. However, their accretion history is very different. V404 Cyg epochs 1--2 are the softest state that the source goes through during the outburst, while for GRS 1915$+$105, epochs 1--3 are the hardest so far observed from the outburst (epoch 3 being much harder than epochs 1--2). Therefore it does not seem likely that a single accretion state would explain the similar-looking spectra. Rather, it has to do with the reprocessing of the intrinsic emission. In the following, we discuss this issue further. 

\subsection{Was V404 Cyg in a ``soft-state'' during epochs 1--2?} \label{v404_soft}

The outburst spectra of V404 Cyg do not show a soft blackbody emission component, except for the 1989 outburst, when a 0.3 keV disk component was seen; see \citet{zycki99}. We discussed in Section \ref{v404} that the parameter evolution of the blackbody component is not consistent with arising directly from the accretion disk. Instead, the soft component could arise from incident thermal (disk) photons scattered multiple times in the surrounding changing medium. On the other hand, the fits with the reflection model in Section \ref{v404} showed that the soft component could also be modeled with an increasing ionization parameter that would be consistent with the implied very high intrinsic luminosity of the source in this state. Thus, it is unlikely that the soft component needed to model epochs 1 and 2 in V404 Cyg comes from an accretion disk. In addition, V404 Cyg epoch 1 corresponds to the hardest X-ray state in Cyg X-3 \citep[e.g.,][]{koljonen10}. Cyg X-3 is a persistent wind-accreting source and thus always accreting at (fairly) similar mass accretion rates. The source likely stays in an intermediate state all the time, with an average power-law spectral index $\Gamma\geq2.0$. Thus, the spectral variation is at least partly due to changes in the surrounding medium. It has been suggested that the jet pressure can play a role in reducing the wind density in the line of sight, producing drastic variability in the spectral evolution of Cyg X-3 \citep{koljonen18}. However, in Cyg X-3, there is further evidence of a very soft state (hypersoft state), where the X-ray spectrum is dominated by a partially absorbed blackbody component with a very weak and flat hard X-ray tail \citep{koljonen18}. This might be (partly) similar to what is observed in V404 Cyg epoch 2, with the increased soft X-ray emission. The soft state in Cyg X-3 always precedes jet ejection episodes \citep{koljonen10}, which indicates that this might be the case for V404 Cyg as well. \citet{koljonen18} argued that in the hypersoft state, the jet turns off, allowing the stellar wind from the Wolf-Rayet companion to fill out the cavity created by the jet. This increases the density of the wind close to the X-ray source, providing a medium where multiple scattering can take place. When the jet is turned on later, it encounters a dense medium where an efficient energy dissipation can take place to boost the jet emission to Jy levels. 

\citet{gandhi17} showed the multiwavelength evolution of V404 Cyg during the \nustar\/ observation (their Fig. 1 and supplementary Fig. 4; their epoch 1 corresponds to our GTIs 10-14). They showed that during the change from the preflare state to the flaring state, there is a brightening in the 15 GHz flux at least by a factor of five and a change in the radio spectral index from negative values to above zero, indicating the start of the jet ejection. In model A, during epoch 2, there is a rise in the parameter evolution of $R_{\mathrm{in}}$, $\Gamma$, log $\xi$, $z,$ and inclination (Fig. \ref{v404_params_A}). This indicates that the geometry of the reprocessor changes and moves farther away (the increase in inclination and $R_{\mathrm{in}}$) with higher speeds (the increase in $z$), most likely caused by increased radiation pressure (the increase in log $\xi$) that also cools the electrons in the corona (the increase in $\Gamma$, and low values of kT$_{e}$). In model B, a similar evolution can be seen in $z$ and in the blackbody flux (Fig. \ref{v404_params_B}). Because the black body flux is proportional to the emitting area squared and temperature to the fourth (i.e., decreasing from epoch 1 to epoch 2; Table \ref{modela1}), this also implies an increase in the size of the blackbody emitter in addition to the increase in the speed of the scattering medium (even more than for model A). This means that what we may be seeing here is a jet ejection event following an accretion event that starts by pushing the reprocessing medium farther out and at the same changing its geometry from sphere-like to disk-like (Fig. \ref{drawing}) and exposing the inner accretion disk seen in the following high-luminosity flaring period \citep{walton17,gandhi17}. A similar spectral sequence can be seen as leading to a high-luminosity flare as well (see \citealt{walton17}; their Fig. 14), beginning from the $\Gamma\sim2$ low-cutoff spectrum with high absorption and evolving to a harder spectrum with a higher cutoff and low absorption (see \citealt{walton17}; their Fig. 15), but with a much faster evolutionary time (tens of seconds compared to $\sim$10000 seconds, as shown in Fig. 2).

The intrinsic luminosity in epochs 1--2 is very high in model B, with an Eddington fraction of 0.2--0.3 in the 3--79 keV band, and when the model is extrapolated to lower energies, the model luminosity reaches the Eddington limit roughly in the 0.2--100 keV band. In the case of model A, we see only the scattered emission, while the intrinsic emission would be completely absorbed or scattered, and thus there is no reliable way to estimate the intrinsic emission. Therefore we assume that model B gives a better indication of the intrinsic luminosity. Based on the evolution of the model parameters outlined above, the super-Eddington luminosities, and the radio evolution, the likeliest scenario for V404 Cyg epochs 1--2 therefore is a super-Eddington accretion rate event that resulted in a large scale-height accretion flow and a powerful optically thick accretion disk wind that launched with mildly relativistic speed and led to a jet ejection event.   

\subsection{Is GRS 1915$+$105 in the hard state?}

\begin{figure}
 \centering
 \includegraphics[width=\linewidth]{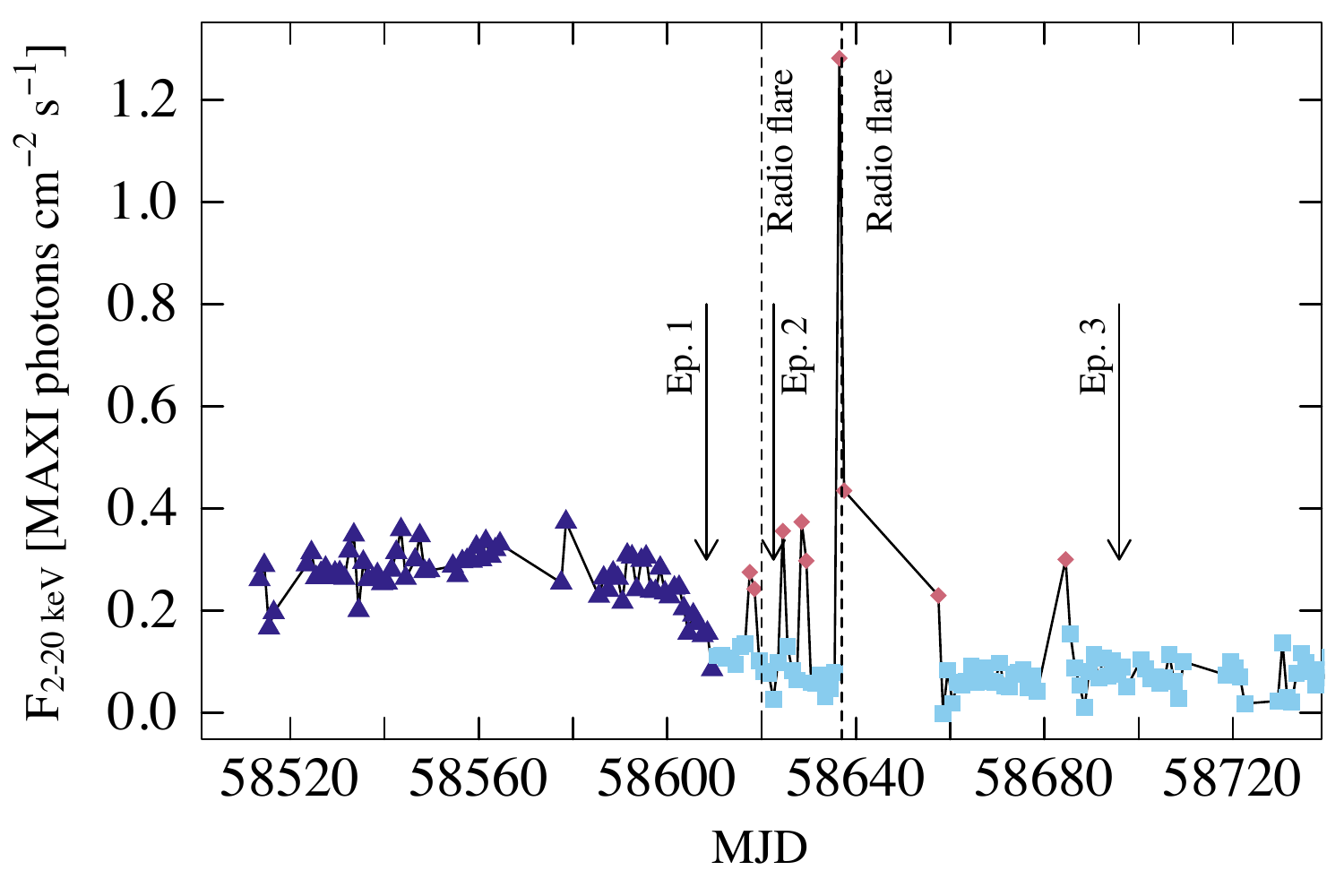}
 \caption{ 2--20 keV daily light curve of GRS 1915$+$105 from \textit{The Monitor of All-sky X-ray Image}/Gas Slit Camera (\maxi) since February 2019, with \nustar\/ observations marked as arrows and radio flare detections \citep{motta19,trushkin19,koljonen19} as vertical dotted lines. The data are colored and marked according to the spectral hardness shown in Fig. \ref{1915_hid}. }
 \label{1915_lc}
\end{figure}

\begin{figure}
 \centering
 \includegraphics[width=\linewidth]{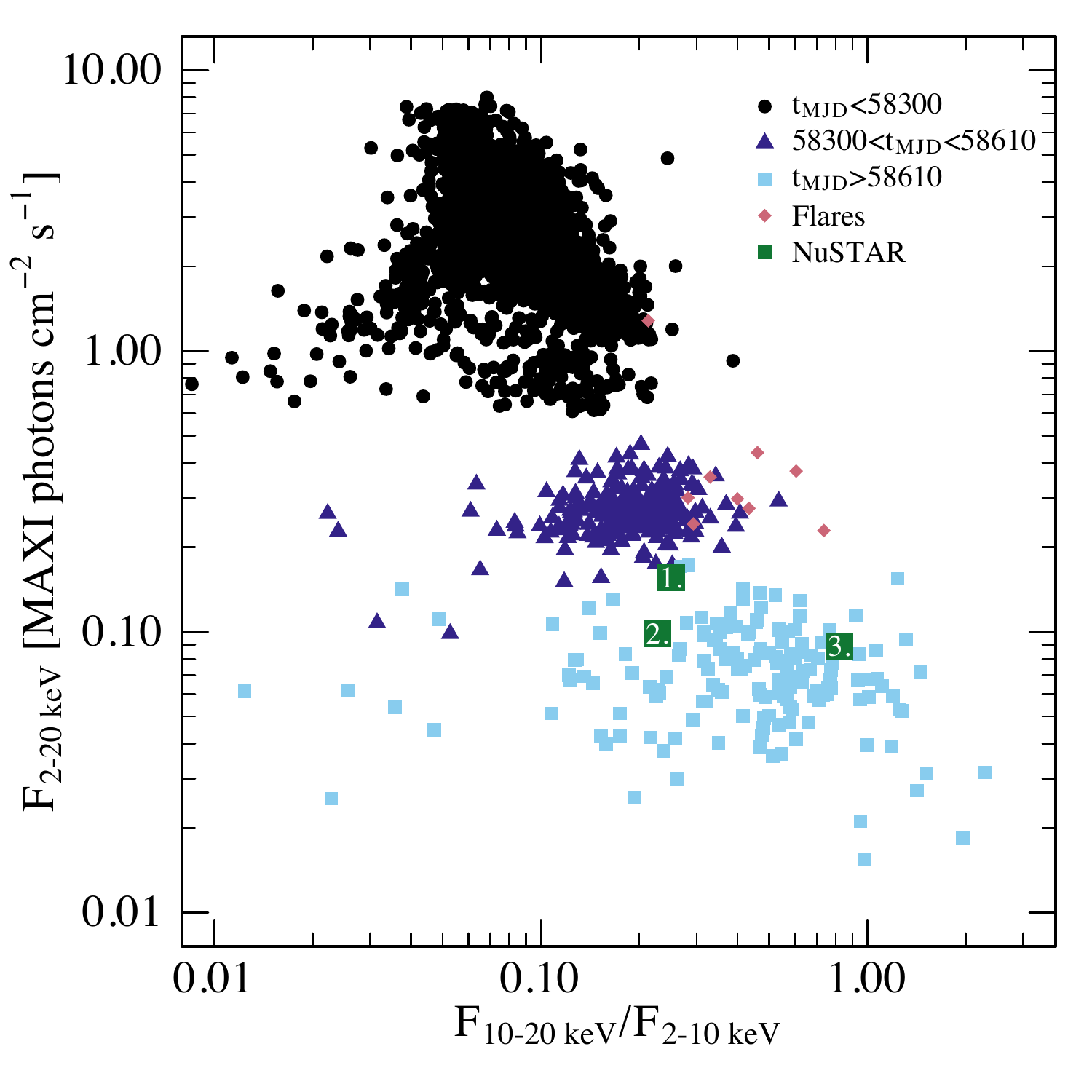}
 \caption{\maxi\/ hardness-intensity diagram of GRS 1915$+$105 from daily monitoring observations since August 2009. The blue data points (dark triangles and light blue squares) indicate the recent low-luminosity state with increased spectral hardness. The light blue squares correspond to the anomalous state with occasional strong X-ray flares (red diamonds) and highly variable radio emission. The numbered green boxes correspond to the \nustar\/ epochs.}
 \label{1915_hid}
\end{figure}

In July 2018, GRS 1915$+$105 entered an extended unusually low-flux X-ray phase followed by a change to a state with even lower average X-ray fluxes that were not seen before during the 27-year-long outburst (Fig. \ref{1915_lc}; light blue squares) but presented renewed flaring activity in radio as well as X-rays (Fig. \ref{1915_lc}; red diamonds and dashed lines). After the change to this peculiar state, radio monitoring data showed significant radio flaring \citep{motta19} that started approximately at the time of the renewed X-ray activity. The radio flaring has since continued and is still ongoing at the time of writing \citep{trushkin20}. While this radio behavior is consistent with what has frequently been observed in the past \citep[e.g.,][]{fender99,punsly13}, this is the first time that significant radio activity does not seem to be associated with a strong X-ray counterpart.

It is therefore not entirely clear whether the outburst of GRS 1915$+$105 is nearing its end or if the source is just highly obscured. The prolonged low-luminosity state since July 2018 does indicate that continuous super-Eddington accretion and Compton-thick wind production is unlikely. There seems to be no X-ray flare during the time of the flux drop (only later), which means that super-Eddington accretion and subsequent mass expulsion did probably not take place and did not cover the source. In addition, the X-ray spectrum is harder than ever observed from the source (blue points in Fig. \ref{1915_hid}), indicating that the source might have reached a regular hard state on its way toward quiescence. 

On the other hand, the radio emission from the source has become more variable and presents flux densities that are among the strongest flares ever observed from GRS 1915$+$105 \citep{trushkin20}. In addition, sporadic X-ray flares display softer spectra and reach similar X-ray hardnesses as before the anomalous low-luminosity state (red points in Fig. \ref{1915_lc} and \ref{1915_hid}), which at least in principle is consistent with varying absorption. While more detailed studies should be made in terms of radio to X-ray correlation, it seems that the X-ray emission remains at a rather constant level (Fig. \ref{1915_lc}), while large amplitude variations are evident in the radio monitoring data \citep{trushkin19,trushkin20}. The enhanced radio luminosity and variability can arise from additional energy dissipation either in merging shocks in the jet (as has been suggested as an explanation of the radio behavior of GRS 1915$+$105 in \citealt{vadawale03}) or the jet interacting with expelled matter from the accretion flow.

The unabsorbed luminosities of model A fits for V404 Cyg and GRS 1915$+$105 in the pre-flaring state are similar: $\sim$2$\times$10$^{37}$ erg/s, indicating that a similar fraction of reflected emission is received from both sources. We have speculated that epochs 1--2 of V404 Cyg might be due to super-Eddington accretion, which might be then the case for GRS 1915$+$105 as well if the intrinsic emission is completely obscured to us. However, model B fits, where the absorbed line-of-sight component is taken into account, give a similar unabsorbed luminosity for GRS 1915$+$105 as model A fits, while for V404 Cyg, it is an order of magnitude higher. The resulting Eddington luminosity of 1--2\% for both models of GRS 1915$+$105 is consistent with a regular XRB hard-state luminosity. In addition, the cospectrum for GRS 1915$+$105 in epoch 1 is markedly different from the latter epochs (or the cospectra of V404 Cyg and Cyg X-3; Fig. \ref{timing}), displaying a low-frequency QPO and spectral power at least up to 10 Hz, while in others, there appears to be no power above 0.1 Hz, indicating that we have at least a partial view to the accretion flow in epoch 1. This is also consistent with the amount of the leaked or intrinsic emission received in the model B fits. It therefore seems more likely that GRS 1915$+$105 has reached a genuine hard X-ray state. However, it is clear from the X-ray spectra and the model fits above that the source is (partly) obscured in all \nustar\/ epochs and likely continuously after MJD 58610. This may indicate that the accretion flow has changed from geometrically thin to thick, and that due to high inclination, it blocks the view to the central parts of the flow.            

\subsection{Implications for other sources}

We have argued that the intrinsic X-ray emission from V404 Cyg, Cyg X-3, V4641 Sgr, and GRS 1915$+$105 is significantly affected by the surrounding medium in scattering processes. While it is clear that these four sources are unique among the XRB population with large accretion disks and high inclination angles, there is evidence that similar scattering takes place in other sources as well: \textbf{i)} Strong and variable absorption has been found in the X-ray spectrum of Swift J1858.6$-$0814 during X-ray flaring that bears similarity to the flaring spectra from V404 Cyg and V4641 Sgr \citep{hare20}. In addition, P Cygni profiles were observed in the optical spectra, indicating a high-velocity wind \citep{munozdarias19}. \textbf{ii)} The well-known accretion disk wind source GRO 1655--40 displays similar X-ray spectra as Cyg X-3 \citep{uttley15}. \textbf{iii)} The hard X-ray emission from SS 433 has been speculated to be heavily reprocessed intrinsic X-ray emission from a supercritical accretion flow viewed through the optically thick accretion wind cone \citep{middleton18}. \textbf{iv)} Other super-Eddington accretors, such as extragalactic ULXs, typically have soft X-ray spectra that might be similar to the epoch 2 spectrum of V404 Cyg. \textbf{v)} Other high-mass XRBs, such as Cyg X-1, might exhibit reprocessing in the companion wind, and the dense environment around XRB jets can affect the properties of the jet emission through shocking, thus increasing the radiative efficiency and enhancing the radio luminosity.  

Recently, puzzling observations of a low-luminosity soft state ($<$0.01 L/L$_{\mathrm{Edd}}$) in XRBs have been reported, including V4641 Sgr \citep{pahari15}, Swift J1753.5--0127 \citep{shaw16}, and 4U 1630--47 \citep{tomsick14}. \citet{pahari15} showed that during a low-luminosity outburst or a renewed X-ray activity period in January-February 2014 \citep{tachibana14,uemura14}, the X-ray spectrum presented a soft state with reflection features including the ionized iron line and iron edges similar to the spectra presented in this paper. Swift J1753.5--0127 and 4U 1630--47 exhibited a low-luminosity soft state at the end of their outbursts in March--May 2015 and July 2010, respectively. With the low luminosity, it is difficult to attribute this component to a regular soft-state disk. The system parameters for these sources are not all well constrained, but there is some evidence for a high orbital inclination from optical variability of Swift J1753.5--0127 \citep{neustroev14} and dipping phenomena in 4U 1630--47 \citep{kuulkers98,tomsick98}, while the orbital inclination of V4641 Sgr is fairly well constrained at 72$^{\circ}\pm$4$^{\circ}$ \citep{macdonald14}. When we assume a high inclination, a significant change in the scale height of the accretion flow, for example, by disk warping or precession, might intercept our line of sight, resulting in strong absorption and reprocessing of the intrinsic emission that could thermalize the intrinsically hard spectrum. A misalignment between the black hole and accretion disk spins has been suspected for V4641 Sgr \citep{maccarone02,gallo14}. This might cause the disk precession and perhaps explain the semi-regular interval of the weak outbursts \citep{negoro18}.   

\section{Conclusions} \label{conclusions}

We have studied the \nustar\/ and \rxtepca\/ spectra of four unique XRBs, V404 Cyg, Cyg X-3, V4641 Sgr, and GRS 1915$+$105, which are known to present complex spectral evolution distinct from other XRBs. We showed that all sources have similar X-ray spectra at certain times that can be modeled by assuming that a Compton-thick medium surrounds the central X-ray source. This assumption is further enhanced by the fact that Cyg X-3 orbits its Wolf-Rayet companion star inside a high-density stellar wind that strongly affects the X-ray spectra. While the companion stars of V404 Cyg, V4641 Sgr, and GRS 1915$+$105 are not expected to present a high-density stellar wind, we attribute the obscuring medium to either a large scale-height accretion flow or to an optically thick equatorial outflow or envelope. 

The results from fitting two physically motivated scattering models suggest that a low-luminosity phase preceding a flaring episode in the 2015 outburst of V404 Cyg is a heavily obscured, but intrinsically very bright (super-Eddington) accretion state. In this state, a dense medium fully covers the X-ray source, and the majority of the received flux comes from heavily absorbed intrinsic emission. After the fully obscured phase, the geometry changes to resemble a disk-like wind, and the majority of the received emission comes from the scattered or reflected component. During this time, large-amplitude flares are observed in X-rays and radio, indicating that some of the obscuring material has been removed either by a change in the accretion flow geometry, in the accretion wind geometry, or by the jet pressure. A part of the elevated emission may be due to the interaction of the jet with the obscuring matter. The shift of the iron line energy below 6.4 keV suggests that the scattering medium is in motion. 

Similar spectral evolution to that of V404 Cyg is observed from the unusual low-luminosity state of GRS 1915$+$105, with the difference that the unabsorbed luminosity remains at a few percent of the Eddington luminosity. It is therefore more likely that the source has declined in flux and reached a regular hard X-ray state. Along with the state transition,  the accretion flow must have thickened because there is evidence that the source is absorbed. Similarly, the weaker 2002 and 2003 outbursts of V4641 Sgr present similar spectra although the unabsorbed luminosities are very low, lower than 1\% of the Eddington luminosity. In these cases, we might be seeing the sources through a disk wind or a geometrically thick accretion flow. Thus, this work highlights the importance of taking the reprocessing of the X-ray emission in the surrounding medium in the modeling of the X-ray spectra into account, which may well take place in multiple sources. 

\section*{Acknowledgements}

We thank Sara Motta for enlightening discussions and the anonymous referee for useful comments. KIIK was supported by the Academy of Finland project 320085. JAT acknowledges partial support from NASA under grant 80NSSC18K0574. This research has made use of data and software provided by HEASARC, which is a service of the Astrophysics Science Division at NASA/GSFC. The \maxi\/ data has been provided by RIKEN, JAXA and the \textit{MAXI} team.

\bibliographystyle{aa}

\bibliography{references}

\begin{thebibliography}{105}
\expandafter\ifx\csname natexlab\endcsname\relax\def\natexlab#1{#1}\fi

\bibitem[{{Axelsson} {et~al.}(2009){Axelsson}, {Larsson}, \&
  {Hjalmarsdotter}}]{axelsson09}
{Axelsson}, M., {Larsson}, S., \& {Hjalmarsdotter}, L. 2009, \mnras, 394, 1544

\bibitem[{{Bachetti}(2015)}]{bachetti15b}
{Bachetti}, M. 2015, {MaLTPyNT: Quick look timing analysis for NuSTAR data}

\bibitem[{{Bachetti} {et~al.}(2015){Bachetti}, {Harrison}, {Cook}, {Tomsick},
  {Schmid}, {Grefenstette}, {Barret}, {Boggs}, {Christensen}, {Craig},
  {Fabian}, {F{\"u}rst}, {Gandhi}, {Hailey}, {Kara}, {Maccarone}, {Miller},
  {Pottschmidt}, {Stern}, {Uttley}, {Walton}, {Wilms}, \& {Zhang}}]{bachetti15}
{Bachetti}, M., {Harrison}, F.~A., {Cook}, R., {et~al.} 2015, \apj, 800, 109

\bibitem[{{Balokovi{\'c}} {et~al.}(2018){Balokovi{\'c}}, {Brightman},
  {Harrison}, {Comastri}, {Ricci}, {Buchner}, {Gandhi}, {Farrah}, \&
  {Stern}}]{balokovic18}
{Balokovi{\'c}}, M., {Brightman}, M., {Harrison}, F.~A., {et~al.} 2018, \apj,
  854, 42

\bibitem[{{Balokovi{\'c}} {et~al.}(2014){Balokovi{\'c}}, {Comastri},
  {Harrison}, {Alexand er}, {Ballantyne}, {Bauer}, {Boggs}, {Brandt},
  {Brightman}, {Christensen}, {Craig}, {Del Moro}, {Gand hi}, {Hailey}, {Koss},
  {Lansbury}, {Luo}, {Madejski}, {Marinucci}, {Matt}, {Markwardt}, {Puccetti},
  {Reynolds}, {Risaliti}, {Rivers}, {Stern}, {Walton}, \&
  {Zhang}}]{balokovic14}
{Balokovi{\'c}}, M., {Comastri}, A., {Harrison}, F.~A., {et~al.} 2014, \apj,
  794, 111

\bibitem[{{Basak} {et~al.}(2017){Basak}, {Zdziarski}, {Parker}, \&
  {Islam}}]{basak17}
{Basak}, R., {Zdziarski}, A.~A., {Parker}, M., \& {Islam}, N. 2017, \mnras,
  472, 4220

\bibitem[{{Bauer} {et~al.}(2015){Bauer}, {Ar{\'e}valo}, {Walton}, {Koss},
  {Puccetti}, {Gandhi}, {Stern}, {Alexander}, {Balokovi{\'c}}, {Boggs},
  {Brandt}, {Brightman}, {Christensen}, {Comastri}, {Craig}, {Del Moro},
  {Hailey}, {Harrison}, {Hickox}, {Luo}, {Markwardt}, {Marinucci}, {Matt},
  {Rigby}, {Rivers}, {Saez}, {Treister}, {Urry}, \& {Zhang}}]{bauer15}
{Bauer}, F.~E., {Ar{\'e}valo}, P., {Walton}, D.~J., {et~al.} 2015, \apj, 812,
  116

\bibitem[{{Beardmore} {et~al.}(2016){Beardmore}, {Willingale}, {Kuulkers},
  {Altamirano}, {Motta}, {Osborne}, {Page}, \& {Sivakoff}}]{beardmore16}
{Beardmore}, A.~P., {Willingale}, R., {Kuulkers}, E., {et~al.} 2016, \mnras,
  462, 1847

\bibitem[{{Belloni} \& {Altamirano}(2013)}]{belloni13}
{Belloni}, T.~M. \& {Altamirano}, D. 2013, \mnras, 432, 10

\bibitem[{{Casares} {et~al.}(1992){Casares}, {Charles}, \&
  {Naylor}}]{casares92}
{Casares}, J., {Charles}, P.~A., \& {Naylor}, T. 1992, \nat, 355, 614

\bibitem[{{Castro-Tirado} {et~al.}(1994){Castro-Tirado}, {Brandt}, {Lund},
  {Lapshov}, {Sunyaev}, {Shlyapnikov}, {Guziy}, \& {Pavlenko}}]{castrotirado92}
{Castro-Tirado}, A.~J., {Brandt}, S., {Lund}, N., {et~al.} 1994, \apjs, 92, 469

\bibitem[{{Chapuis} \& {Corbel}(2004)}]{chapuis04}
{Chapuis}, C. \& {Corbel}, S. 2004, \aap, 414, 659

\bibitem[{{Dauser} {et~al.}(2014){Dauser}, {Garcia}, {Parker}, {Fabian}, \&
  {Wilms}}]{dauser14}
{Dauser}, T., {Garcia}, J., {Parker}, M.~L., {Fabian}, A.~C., \& {Wilms}, J.
  2014, \mnras, 444, L100

\bibitem[{{Dauser} {et~al.}(2016){Dauser}, {Garc{\'{\i}}a}, {Walton},
  {Eikmann}, {Kallman}, {McClintock}, \& {Wilms}}]{dauser16}
{Dauser}, T., {Garc{\'{\i}}a}, J., {Walton}, D.~J., {et~al.} 2016, \aap, 590,
  A76

\bibitem[{{Done} {et~al.}(2004){Done}, {Wardzi{\'n}ski}, \&
  {Gierli{\'n}ski}}]{done04}
{Done}, C., {Wardzi{\'n}ski}, G., \& {Gierli{\'n}ski}, M. 2004, \mnras, 349,
  393

\bibitem[{{Fender} \& {Belloni}(2004)}]{fender04}
{Fender}, R. \& {Belloni}, T. 2004, \araa, 42, 317

\bibitem[{{Fender} {et~al.}(1999){Fender}, {Garrington}, {McKay}, {Muxlow},
  {Pooley}, {Spencer}, {Stirling}, \& {Waltman}}]{fender99}
{Fender}, R.~P., {Garrington}, S.~T., {McKay}, D.~J., {et~al.} 1999, \mnras,
  304, 865

\bibitem[{{Gallo} {et~al.}(2014){Gallo}, {Plotkin}, \& {Jonker}}]{gallo14}
{Gallo}, E., {Plotkin}, R.~M., \& {Jonker}, P.~G. 2014, \mnras, 438, L41

\bibitem[{{Gandhi} {et~al.}(2017){Gandhi}, {Bachetti}, {Dhillon}, {Fender},
  {Hardy}, {Harrison}, {Littlefair}, {Malzac}, {Markoff}, {Marsh}, {Mooley},
  {Stern}, {Tomsick}, {Walton}, {Casella}, {Vincentelli}, {Altamirano},
  {Casares}, {Ceccobello}, {Charles}, {Ferrigno}, {Hynes}, {Knigge},
  {Kuulkers}, {Pahari}, {Rahoui}, {Russell}, \& {Shaw}}]{gandhi17}
{Gandhi}, P., {Bachetti}, M., {Dhillon}, V.~S., {et~al.} 2017, Nature
  Astronomy, 1, 859

\bibitem[{{Garc{\'\i}a} {et~al.}(2014){Garc{\'\i}a}, {Dauser}, {Lohfink},
  {Kallman}, {Steiner}, {McClintock}, {Brenneman}, {Wilms}, {Eikmann},
  {Reynolds}, \& {Tombesi}}]{garcia14}
{Garc{\'\i}a}, J., {Dauser}, T., {Lohfink}, A., {et~al.} 2014, \apj, 782, 76

\bibitem[{{Greiner} {et~al.}(2001){Greiner}, {Cuby}, {McCaughrean},
  {Castro-Tirado}, \& {Mennickent}}]{greiner01a}
{Greiner}, J., {Cuby}, J.~G., {McCaughrean}, M.~J., {Castro-Tirado}, A.~J., \&
  {Mennickent}, R.~E. 2001, \aap, 373, L37

\bibitem[{{Hare} {et~al.}(2020){Hare}, {Tomsick}, {Buisson}, {Clavel},
  {Gandhi}, {Garcia}, {Grefenstette}, {Walton}, \& {Xu}}]{hare20}
{Hare}, J., {Tomsick}, J.~A., {Buisson}, D. J.~K., {et~al.} 2020, arXiv
  e-prints, arXiv:2001.03214

\bibitem[{{Heinz} {et~al.}(2016){Heinz}, {Corrales}, {Smith}, {Brandt},
  {Jonker}, {Plotkin}, \& {Neilsen}}]{heinz16}
{Heinz}, S., {Corrales}, L., {Smith}, R., {et~al.} 2016, \apj, 825, 15

\bibitem[{{Hjalmarsdotter} {et~al.}(2008){Hjalmarsdotter}, {Zdziarski},
  {Larsson}, {Beckmann}, {McCollough}, {Hannikainen}, \&
  {Vilhu}}]{hjalmarsdotter08}
{Hjalmarsdotter}, L., {Zdziarski}, A.~A., {Larsson}, S., {et~al.} 2008, \mnras,
  384, 278

\bibitem[{{Hjalmarsdotter} {et~al.}(2004){Hjalmarsdotter}, {Zdziarski},
  {Paizis}, {Beckmann}, \& {Vilhu}}]{hjalmarsdotter04}
{Hjalmarsdotter}, L., {Zdziarski}, A.~A., {Paizis}, A., {Beckmann}, V., \&
  {Vilhu}, O. 2004, in ESA Special Publication, Vol. 552, 5th INTEGRAL Workshop
  on the INTEGRAL Universe, ed. V.~{Schoenfelder}, G.~{Lichti}, \&
  C.~{Winkler}, 223

\bibitem[{{Hjellming} {et~al.}(2000){Hjellming}, {Rupen}, {Hunstead},
  {Campbell-Wilson}, {Mioduszewski}, {Gaensler}, {Smith}, {Sault}, {Fender},
  {Spencer}, {de la Force}, {Richards}, {Garrington}, {Trushkin}, {Ghigo},
  {Waltman}, \& {McCollough}}]{hjellming00}
{Hjellming}, R.~M., {Rupen}, M.~P., {Hunstead}, R.~W., {et~al.} 2000, \apj,
  544, 977

\bibitem[{{Houck}(2002)}]{houck02}
{Houck}, J.~C. 2002, in High Resolution X-ray Spectroscopy with XMM-Newton and
  Chandra, ed. G.~{Branduardi-Raymont}, 17

\bibitem[{{Hynes} {et~al.}(2019){Hynes}, {Robinson}, {Terndrup}, {Gand hi},
  {Froning}, {Wagner}, {Starrfield}, {Dhillon}, \& {Marsh}}]{hynes19}
{Hynes}, R.~I., {Robinson}, E.~L., {Terndrup}, D.~M., {et~al.} 2019, \mnras,
  487, 60

\bibitem[{{Iwakiri} {et~al.}(2019){Iwakiri}, {Negoro}, {Kawai}, {Sugizaki},
  {Yamaoka}, {Ueda}, {Nakajima}, {Maruyama}, {Aoki}, {Kobayashi}, {Nakahira},
  {Mihara}, {Tamagawa}, {Matsuoka}, {Sakamoto}, {Serino}, {Sugita}, {Nishida},
  {Yoshida}, {Tsuboi}, {Sasaki}, {Kawai}, {Sato}, {Shidatsu}, {Oeda},
  {Shiraishi}, {Ueno}, {Tomida}, {Ishikawa}, {Sugawara}, {Isobe}, {Shimomukai},
  {Tominaga}, {Tanimoto}, {Yamada}, {Ogawa}, {Setoguchi}, {Yoshitake},
  {Tsunemi}, {Yoneyama}, {Asakura}, {Ide}, {Yamauchi}, {Iwahori}, {Kurihara},
  {Kurogi}, {Miike}, {Kawamuro}, \& {Kawakubom}}]{iwakiri19}
{Iwakiri}, W., {Negoro}, H., {Kawai}, N., {et~al.} 2019, The Astronomer's
  Telegram, 12787, 1

\bibitem[{{Jenke} {et~al.}(2016){Jenke}, {Wilson-Hodge}, {Homan}, {Veres},
  {Briggs}, {Burns}, {Connaughton}, {Finger}, \& {Hui}}]{jenke16}
{Jenke}, P.~A., {Wilson-Hodge}, C.~A., {Homan}, J., {et~al.} 2016, \apj, 826,
  37

\bibitem[{{Jithesh} {et~al.}(2019){Jithesh}, {Maqbool}, {Dewangan}, \&
  {Misra}}]{jithesh19}
{Jithesh}, V., {Maqbool}, B., {Dewangan}, G.~C., \& {Misra}, R. 2019, The
  Astronomer's Telegram, 12805, 1

\bibitem[{{Kajava} {et~al.}(2018){Kajava}, {Motta},
  {S{\'a}nchez-Fern{\'a}ndez}, \& {Kuulkers}}]{kajava18}
{Kajava}, J.~J.~E., {Motta}, S.~E., {S{\'a}nchez-Fern{\'a}ndez}, C., \&
  {Kuulkers}, E. 2018, \aap, 616, A129

\bibitem[{{Kallman} {et~al.}(2019){Kallman}, {McCollough}, {Koljonen},
  {Liedahl}, {Miller}, {Paerels}, {Pooley}, {Sako}, {Schulz}, {Trushkin}, \&
  {Corrales}}]{kallman19}
{Kallman}, T., {McCollough}, M., {Koljonen}, K., {et~al.} 2019, \apj, 874, 51

\bibitem[{{Khargharia} {et~al.}(2010){Khargharia}, {Froning}, \&
  {Robinson}}]{khargharia10}
{Khargharia}, J., {Froning}, C.~S., \& {Robinson}, E.~L. 2010, \apj, 716, 1105

\bibitem[{{King} {et~al.}(2015){King}, {Miller}, {Raymond}, {Reynolds}, \&
  {Morningstar}}]{king15}
{King}, A.~L., {Miller}, J.~M., {Raymond}, J., {Reynolds}, M.~T., \&
  {Morningstar}, W. 2015, \apjl, 813, L37

\bibitem[{{King}(1993)}]{king93}
{King}, A.~R. 1993, \mnras, 260, L5

\bibitem[{{Kitamoto} {et~al.}(1989){Kitamoto}, {Tsunemi}, {Miyamoto},
  {Yamashita}, \& {Mizobuchi}}]{kitamoto89}
{Kitamoto}, S., {Tsunemi}, H., {Miyamoto}, S., {Yamashita}, K., \& {Mizobuchi},
  S. 1989, \nat, 342, 518

\bibitem[{{Koljonen} {et~al.}(2019){Koljonen}, {Vera}, {Lahteenmaki}, \&
  {Tornikoski}}]{koljonen19}
{Koljonen}, K., {Vera}, R., {Lahteenmaki}, A., \& {Tornikoski}, M. 2019, The
  Astronomer's Telegram, 12839, 1

\bibitem[{{Koljonen} {et~al.}(2011){Koljonen}, {Hannikainen}, \&
  {McCollough}}]{koljonen11}
{Koljonen}, K.~I.~I., {Hannikainen}, D.~C., \& {McCollough}, M.~L. 2011,
  \mnras, 416, L84

\bibitem[{{Koljonen} {et~al.}(2010){Koljonen}, {Hannikainen}, {McCollough},
  {Pooley}, \& {Trushkin}}]{koljonen10}
{Koljonen}, K.~I.~I., {Hannikainen}, D.~C., {McCollough}, M.~L., {Pooley},
  G.~G., \& {Trushkin}, S.~A. 2010, \mnras, 406, 307

\bibitem[{{Koljonen} {et~al.}(2018){Koljonen}, {Maccarone}, {McCollough},
  {Gurwell}, {Trushkin}, {Pooley}, {Piano}, \& {Tavani}}]{koljonen18}
{Koljonen}, K.~I.~I., {Maccarone}, T., {McCollough}, M.~L., {et~al.} 2018,
  \aap, 612, A27

\bibitem[{{Koljonen} \& {Maccarone}(2017)}]{koljonen17}
{Koljonen}, K.~I.~I. \& {Maccarone}, T.~J. 2017, \mnras, 472, 2181

\bibitem[{{Kuulkers} {et~al.}(1998){Kuulkers}, {Wijnands}, {Belloni},
  {M{\'e}ndez}, {van der Klis}, \& {van Paradijs}}]{kuulkers98}
{Kuulkers}, E., {Wijnands}, R., {Belloni}, T., {et~al.} 1998, \apj, 494, 753

\bibitem[{{Maccarone}(2002)}]{maccarone02}
{Maccarone}, T.~J. 2002, \mnras, 336, 1371

\bibitem[{{MacDonald} {et~al.}(2014){MacDonald}, {Bailyn}, {Buxton},
  {Cantrell}, {Chatterjee}, {Kennedy-Shaffer}, {Orosz}, {Markwardt}, \&
  {Swank}}]{macdonald14}
{MacDonald}, R.~K.~D., {Bailyn}, C.~D., {Buxton}, M., {et~al.} 2014, \apj, 784,
  2

\bibitem[{{Magdziarz} \& {Zdziarski}(1995)}]{magdziarz95}
{Magdziarz}, P. \& {Zdziarski}, A.~A. 1995, \mnras, 273, 837

\bibitem[{{Maitra} \& {Bailyn}(2006)}]{maitra06}
{Maitra}, D. \& {Bailyn}, C.~D. 2006, \apj, 637, 992

\bibitem[{{McCollough} {et~al.}(2016){McCollough}, {Corrales}, \&
  {Dunham}}]{mccollough16}
{McCollough}, M.~L., {Corrales}, L., \& {Dunham}, M.~M. 2016, \apjl, 830, L36

\bibitem[{{Middleton} {et~al.}(2018){Middleton}, {Walton}, {Alston}, {Dauser},
  {Eikenberry}, {Jiang}, {Fabian}, {Fuerst}, {Brightman}, {Marshall}, {Parker},
  {Pinto}, {Harrison}, {Bachetti}, {Altamirano}, {Bird}, {Perez},
  {Miller-Jones}, {Charles}, {Boggs}, {Christensen}, {Craig}, {Forster},
  {Grefenstette}, {Hailey}, {Madsen}, {Stern}, \& {Zhang}}]{middleton18}
{Middleton}, M.~J., {Walton}, D.~J., {Alston}, W., {et~al.} 2018, arXiv
  e-prints, arXiv:1810.10518

\bibitem[{{Miller} {et~al.}(2019){Miller}, {Balakrishnan}, {Reynolds},
  {Fabian}, {Kaastra}, \& {Kallman}}]{miller19}
{Miller}, J.~M., {Balakrishnan}, M., {Reynolds}, M., {et~al.} 2019, The
  Astronomer's Telegram, 12771, 1

\bibitem[{{Miller-Jones} {et~al.}(2009){Miller-Jones}, {Jonker}, {Dhawan},
  {Brisken}, {Rupen}, {Nelemans}, \& {Gallo}}]{millerjones09}
{Miller-Jones}, J.~C.~A., {Jonker}, P.~G., {Dhawan}, V., {et~al.} 2009, \apjl,
  706, L230

\bibitem[{{Miller-Jones} {et~al.}(2019){Miller-Jones}, {Tetarenko}, {Sivakoff},
  {Middleton}, {Altamirano}, {Anderson}, {Belloni}, {Fender}, {Jonker},
  {K{\"o}rding}, {Krimm}, {Maitra}, {Markoff}, {Migliari}, {Mooley}, {Rupen},
  {Russell}, {Russell}, {Sarazin}, {Soria}, \& {Tudose}}]{millerjones19}
{Miller-Jones}, J. C.~A., {Tetarenko}, A.~J., {Sivakoff}, G.~R., {et~al.} 2019,
  \nat, 569, 374

\bibitem[{{Morgan} {et~al.}(1997){Morgan}, {Remillard}, \&
  {Greiner}}]{morgan97}
{Morgan}, E.~H., {Remillard}, R.~A., \& {Greiner}, J. 1997, \apj, 482, 993

\bibitem[{{Morningstar} {et~al.}(2014){Morningstar}, {Miller}, {Reynolds}, \&
  {Maitra}}]{morningstar14}
{Morningstar}, W.~R., {Miller}, J.~M., {Reynolds}, M.~T., \& {Maitra}, D. 2014,
  \apjl, 786, L20

\bibitem[{{Motta} {et~al.}(2019){Motta}, {Williams}, {Fender}, {Titterington},
  {Green}, \& {Perrott}}]{motta19}
{Motta}, S., {Williams}, D., {Fender}, R., {et~al.} 2019, The Astronomer's
  Telegram, 12773, 1

\bibitem[{{Motta} {et~al.}(2017{\natexlab{a}}){Motta}, {Kajava},
  {S{\'a}nchez-Fern{\'a}ndez}, {Beardmore}, {Sanna}, {Page}, {Fender},
  {Altamirano}, {Charles}, {Giustini}, {Knigge}, {Kuulkers}, {Oates}, \&
  {Osborne}}]{motta17a}
{Motta}, S.~E., {Kajava}, J.~J.~E., {S{\'a}nchez-Fern{\'a}ndez}, C., {et~al.}
  2017{\natexlab{a}}, \mnras, 471, 1797

\bibitem[{{Motta} {et~al.}(2017{\natexlab{b}}){Motta}, {Kajava},
  {S{\'a}nchez-Fern{\'a}ndez}, {Giustini}, \& {Kuulkers}}]{motta17b}
{Motta}, S.~E., {Kajava}, J.~J.~E., {S{\'a}nchez-Fern{\'a}ndez}, C.,
  {Giustini}, M., \& {Kuulkers}, E. 2017{\natexlab{b}}, \mnras, 468, 981

\bibitem[{{Mu{\~n}oz-Darias} {et~al.}(2016){Mu{\~n}oz-Darias}, {Casares}, {Mata
  S{\'a}nchez}, {Fender}, {Armas Padilla}, {Linares}, {Ponti}, {Charles},
  {Mooley}, \& {Rodriguez}}]{munozdarias16}
{Mu{\~n}oz-Darias}, T., {Casares}, J., {Mata S{\'a}nchez}, D., {et~al.} 2016,
  \nat, 534, 75

\bibitem[{{Mu{\~n}oz-Darias} {et~al.}(2018){Mu{\~n}oz-Darias}, {Torres}, \&
  {Garcia}}]{munozdarias18}
{Mu{\~n}oz-Darias}, T., {Torres}, M.~A.~P., \& {Garcia}, M.~R. 2018, \mnras,
  479, 3987

\bibitem[{{Munoz-Darias} {et~al.}(2019){Munoz-Darias}, {Jimenez-Ibarra}, {Armas
  Padilla}, {Casares}, {Cuneo}, {Panizo-Espinar}, {Sanchez-Sierras}, \&
  {Torres}}]{munozdarias19}
{Munoz-Darias}, T., {Jimenez-Ibarra}, F., {Armas Padilla}, M., {et~al.} 2019,
  The Astronomer's Telegram, 12881, 1

\bibitem[{{Natalucci} {et~al.}(2015){Natalucci}, {Fiocchi}, {Bazzano},
  {Ubertini}, {Roques}, \& {Jourdain}}]{natalucci15}
{Natalucci}, L., {Fiocchi}, M., {Bazzano}, A., {et~al.} 2015, \apjl, 813, L21

\bibitem[{{Negoro} {et~al.}(2018){Negoro}, {Nakajima}, {Sakamaki}, {Maruyama},
  {Mihara}, {Nakahira}, {Yatabe}, {Takao}, {Matsuoka}, {Sakamoto}, {Serino},
  {Sugita}, {Kawakubo}, {Hashimoto}, {Yoshida}, {Kawai}, {Sugizaki},
  {Tachibana}, {Morita}, {Ueno}, {Tomida}, {Ishikawa}, {Sugawara}, {Isobe},
  {Shimomukai}, {Ueda}, {Tanimoto}, {Morita}, {Yamada}, {Tsuboi}, {Iwakiri},
  {Sasaki}, {Kawai}, {Sato}, {Tsunemi}, {Yoneyama}, {Yamauchi}, {Hidaka},
  {Iwahori}, {Kawamuro}, {Yamaoka}, \& {Shidatsu}}]{negoro18}
{Negoro}, H., {Nakajima}, M., {Sakamaki}, A., {et~al.} 2018, The Astronomer's
  Telegram, 11931, 1

\bibitem[{{Neilsen} {et~al.}(2019){Neilsen}, {Homan}, {Gendreau},
  {Arzoumanian}, {Steiner}, {Altamirano}, {Eikenberry}, {Remillard}, {Iwakiri},
  \& {Fabian}}]{neilsen19}
{Neilsen}, J., {Homan}, J., {Gendreau}, K., {et~al.} 2019, The Astronomer's
  Telegram, 12793, 1

\bibitem[{{Neilsen} \& {Lee}(2009)}]{neilsen09}
{Neilsen}, J. \& {Lee}, J.~C. 2009, \nat, 458, 481

\bibitem[{{Neustroev} {et~al.}(2014){Neustroev}, {Veledina}, {Poutanen},
  {Zharikov}, {Tsygankov}, {Sjoberg}, \& {Kajava}}]{neustroev14}
{Neustroev}, V.~V., {Veledina}, A., {Poutanen}, J., {et~al.} 2014, \mnras, 445,
  2424

\bibitem[{{Oosterbroek} {et~al.}(1997){Oosterbroek}, {van der Klis}, {van
  Paradijs}, {Vaughan}, {Rutledge}, {Lewin}, {Tanaka}, {Nagase}, {Dotani},
  {Mitsuda}, \& {Miyamoto}}]{oosterbroek97}
{Oosterbroek}, T., {van der Klis}, M., {van Paradijs}, J., {et~al.} 1997, \aap,
  321, 776

\bibitem[{{Orosz} {et~al.}(2001){Orosz}, {Kuulkers}, {van der Klis},
  {McClintock}, {Garcia}, {Callanan}, {Bailyn}, {Jain}, \&
  {Remillard}}]{orosz01}
{Orosz}, J.~A., {Kuulkers}, E., {van der Klis}, M., {et~al.} 2001, \apj, 555,
  489

\bibitem[{{Paerels} {et~al.}(2000){Paerels}, {Cottam}, {Sako}, {Liedahl},
  {Brinkman}, {van der Meer}, {Kaastra}, \& {Predehl}}]{paerels00}
{Paerels}, F., {Cottam}, J., {Sako}, M., {et~al.} 2000, \apjl, 533, L135

\bibitem[{{Pahari} {et~al.}(2015){Pahari}, {Misra}, {Dewangan}, \&
  {Pawar}}]{pahari15}
{Pahari}, M., {Misra}, R., {Dewangan}, G.~C., \& {Pawar}, P. 2015, \apj, 814,
  158

\bibitem[{{Parsignault} {et~al.}(1972){Parsignault}, {Gursky}, {Kellogg},
  {Matilsky}, {Murray}, {Schreier}, {Tananbaum}, {Giacconi}, \&
  {Brinkman}}]{parsignault72}
{Parsignault}, D.~R., {Gursky}, H., {Kellogg}, E.~M., {et~al.} 1972, Nature
  Physical Science, 239, 123

\bibitem[{{Pinto} {et~al.}(2019){Pinto}, {Mehdipour}, {Walton}, {Middleton},
  {Roberts}, {Fabian}, {Guainazzi}, {Soria}, {Kosec}, \& {Ness}}]{pinto19}
{Pinto}, C., {Mehdipour}, M., {Walton}, D.~J., {et~al.} 2019, arXiv e-prints,
  arXiv:1903.06174

\bibitem[{{Punsly} \& {Rodriguez}(2013)}]{punsly13}
{Punsly}, B. \& {Rodriguez}, J. 2013, \apj, 764, 173

\bibitem[{{Reid} {et~al.}(2014){Reid}, {McClintock}, {Steiner}, {Steeghs},
  {Remillard}, {Dhawan}, \& {Narayan}}]{reid14}
{Reid}, M.~J., {McClintock}, J.~E., {Steiner}, J.~F., {et~al.} 2014, \apj, 796,
  2

\bibitem[{{Revnivtsev} {et~al.}(2002){Revnivtsev}, {Gilfanov}, {Churazov}, \&
  {Sunyaev}}]{revnivtsev02}
{Revnivtsev}, M., {Gilfanov}, M., {Churazov}, E., \& {Sunyaev}, R. 2002, \aap,
  391, 1013

\bibitem[{{Richter}(1989)}]{richter89}
{Richter}, G.~A. 1989, Information Bulletin on Variable Stars, 3362, 1

\bibitem[{{Rodriguez} {et~al.}(2015){Rodriguez}, {Cadolle Bel},
  {Alfonso-Garz{\'o}n}, {Siegert}, {Zhang}, {Grinberg}, {Savchenko}, {Tomsick},
  {Chenevez}, {Clavel}, {Corbel}, {Diehl}, {Domingo}, {Gouiff{\`e}s},
  {Greiner}, {Krause}, {Laurent}, {Loh}, {Markoff}, {Mas-Hesse},
  {Miller-Jones}, {Russell}, \& {Wilms}}]{rodriguez15}
{Rodriguez}, J., {Cadolle Bel}, M., {Alfonso-Garz{\'o}n}, J., {et~al.} 2015,
  \aap, 581, L9

\bibitem[{{Roques} {et~al.}(2015){Roques}, {Jourdain}, {Bazzano}, {Fiocchi},
  {Natalucci}, \& {Ubertini}}]{roques15}
{Roques}, J.-P., {Jourdain}, E., {Bazzano}, A., {et~al.} 2015, \apjl, 813, L22

\bibitem[{{Rupen} {et~al.}(2002){Rupen}, {Dhawan}, \& {Mioduszewski}}]{rupen02}
{Rupen}, M.~P., {Dhawan}, V., \& {Mioduszewski}, A.~J. 2002, \iaucirc, 7928, 2

\bibitem[{{Rupen} {et~al.}(2003){Rupen}, {Mioduszewski}, \& {Dhawan}}]{rupen03}
{Rupen}, M.~P., {Mioduszewski}, A.~J., \& {Dhawan}, V. 2003, The Astronomer's
  Telegram, 172, 1

\bibitem[{{S{\'a}nchez-Fern{\'a}ndez}
  {et~al.}(2017){S{\'a}nchez-Fern{\'a}ndez}, {Kajava}, {Motta}, \&
  {Kuulkers}}]{sanchez17}
{S{\'a}nchez-Fern{\'a}ndez}, C., {Kajava}, J.~J.~E., {Motta}, S.~E., \&
  {Kuulkers}, E. 2017, \aap, 602, A40

\bibitem[{{Shaw} {et~al.}(2016){Shaw}, {Gandhi}, {Altamirano}, {Uttley},
  {Tomsick}, {Charles}, {F{\"u}rst}, {Rahoui}, \& {Walton}}]{shaw16}
{Shaw}, A.~W., {Gandhi}, P., {Altamirano}, D., {et~al.} 2016, \mnras, 458, 1636

\bibitem[{{Steeghs} {et~al.}(2013){Steeghs}, {McClintock}, {Parsons}, {Reid},
  {Littlefair}, \& {Dhillon}}]{steeghs13}
{Steeghs}, D., {McClintock}, J.~E., {Parsons}, S.~G., {et~al.} 2013, \apj, 768,
  185

\bibitem[{{Szostek} \& {Zdziarski}(2008)}]{szostek08}
{Szostek}, A. \& {Zdziarski}, A.~A. 2008, \mnras, 386, 593

\bibitem[{{Tachibana} {et~al.}(2014){Tachibana}, {Takagi}, {Serino}, {Morii},
  {Nakahira}, {Negoro}, {Ueno}, {Tomida}, {Kimura}, {Ishikawa}, {Nakagawa},
  {Mihara}, {Sugizaki}, {Sugimoto}, {Yoshikawa}, {Matsuoka}, {Kawai}, {Usui},
  {Yoshii}, {Yoshida}, {Sakamoto}, {Nakano}, {Kawakubo}, {Ohtsuki}, {Tsunemi},
  {Sasaki}, {Nakajima}, {Fukushima}, {Onodera}, {Suzuki}, {Ueda}, {Shidatsu},
  {Kawamuro}, {Hori}, {Tsuboi}, {Higa}, {Yamauchi}, {Yoshidome}, {Ogawa},
  {Yamada}, \& {Yamaoka}}]{tachibana14}
{Tachibana}, Y., {Takagi}, T., {Serino}, M., {et~al.} 2014, The Astronomer's
  Telegram, 5803, 1

\bibitem[{{Titarchuk}(1994)}]{titarchuk94}
{Titarchuk}, L. 1994, \apj, 434, 570

\bibitem[{{Tomsick} {et~al.}(1998){Tomsick}, {Lapshov}, \&
  {Kaaret}}]{tomsick98}
{Tomsick}, J.~A., {Lapshov}, I., \& {Kaaret}, P. 1998, \apj, 494, 747

\bibitem[{{Tomsick} {et~al.}(2014){Tomsick}, {Yamaoka}, {Corbel}, {Kalemci},
  {Migliari}, \& {Kaaret}}]{tomsick14}
{Tomsick}, J.~A., {Yamaoka}, K., {Corbel}, S., {et~al.} 2014, \apj, 791, 70

\bibitem[{{Trushkin} {et~al.}(2019){Trushkin}, {Nizhelskij}, {Tsybulev},
  {Bursov}, \& {Shevchenko}}]{trushkin19}
{Trushkin}, S.~A., {Nizhelskij}, N.~A., {Tsybulev}, P.~G., {Bursov}, N.~N., \&
  {Shevchenko}, A.~V. 2019, The Astronomer's Telegram, 12855, 1

\bibitem[{{Trushkin} {et~al.}(2020){Trushkin}, {Nizhelskij}, {Tsybulev},
  {Bursov}, \& {Shevchenko}}]{trushkin20}
{Trushkin}, S.~A., {Nizhelskij}, N.~A., {Tsybulev}, P.~G., {Bursov}, N.~N., \&
  {Shevchenko}, A.~V. 2020, The Astronomer's Telegram, 13442, 1

\bibitem[{{Uemura} {et~al.}(2002){Uemura}, {Kato}, {Ishioka}, {Tanabe},
  {Kiyota}, {Monard}, {Stubbings}, {Nelson}, {Richards}, {Bailyn}, \&
  {Santallo}}]{uemura02}
{Uemura}, M., {Kato}, T., {Ishioka}, R., {et~al.} 2002, \pasj, 54, L79

\bibitem[{{Uemura} {et~al.}(2014){Uemura}, {Moritani}, {Itoh}, {Akitaya},
  {Arai}, {Morihana}, {Honda}, \& {Matsumoto}}]{uemura14}
{Uemura}, M., {Moritani}, Y., {Itoh}, R., {et~al.} 2014, The Astronomer's
  Telegram, 5836, 1

\bibitem[{{Uttley} \& {Klein-Wolt}(2015)}]{uttley15}
{Uttley}, P. \& {Klein-Wolt}, M. 2015, \mnras, 451, 475

\bibitem[{{Vadawale} {et~al.}(2003){Vadawale}, {Rao}, {Naik}, {Yadav},
  {Ishwara-Chandra}, {Pramesh Rao}, \& {Pooley}}]{vadawale03}
{Vadawale}, S.~V., {Rao}, A.~R., {Naik}, S., {et~al.} 2003, \apj, 597, 1023

\bibitem[{{van Kerkwijk} {et~al.}(1996){van Kerkwijk}, {Geballe}, {King}, {van
  der Klis}, \& {van Paradijs}}]{vankerkwijk96}
{van Kerkwijk}, M.~H., {Geballe}, T.~R., {King}, D.~L., {van der Klis}, M., \&
  {van Paradijs}, J. 1996, \aap, 314, 521

\bibitem[{{Vasilopoulos} \& {Petropoulou}(2016)}]{vasilopoulos16}
{Vasilopoulos}, G. \& {Petropoulou}, M. 2016, \mnras, 455, 4426

\bibitem[{{Vilhu} {et~al.}(2009){Vilhu}, {Hakala}, {Hannikainen}, {McCollough},
  \& {Koljonen}}]{vilhu09}
{Vilhu}, O., {Hakala}, P., {Hannikainen}, D.~C., {McCollough}, M., \&
  {Koljonen}, K. 2009, \aap, 501, 679

\bibitem[{{Vrtilek} \& {Boroson}(2013)}]{vrtilek13}
{Vrtilek}, S.~D. \& {Boroson}, B.~S. 2013, \mnras, 428, 3693

\bibitem[{{Wagner} {et~al.}(1991){Wagner}, {Starrfield}, {Howell}, {Kreidl},
  {Bus}, {Cassatella}, {Bertram}, \& {Fried}}]{wagner91}
{Wagner}, R.~M., {Starrfield}, S.~G., {Howell}, S.~B., {et~al.} 1991, \apj,
  378, 293

\bibitem[{{Walton} {et~al.}(2017){Walton}, {Mooley}, {King}, {Tomsick},
  {Miller}, {Dauser}, {Garc{\'{\i}}a}, {Bachetti}, {Brightman}, {Fabian},
  {Forster}, {F{\"u}rst}, {Gandhi}, {Grefenstette}, {Harrison}, {Madsen},
  {Meier}, {Middleton}, {Natalucci}, {Rahoui}, {Rana}, \& {Stern}}]{walton17}
{Walton}, D.~J., {Mooley}, K., {King}, A.~L., {et~al.} 2017, \apj, 839, 110

\bibitem[{{Wijnands} \& {van der Klis}(2000)}]{wijnands00}
{Wijnands}, R. \& {van der Klis}, M. 2000, \apjl, 528, L93

\bibitem[{{Zdziarski} {et~al.}(2012){Zdziarski}, {Maitra}, {Frankowski},
  {Skinner}, \& {Misra}}]{zdziarski12}
{Zdziarski}, A.~A., {Maitra}, C., {Frankowski}, A., {Skinner}, G.~K., \&
  {Misra}, R. 2012, \mnras, 426, 1031

\bibitem[{{Zdziarski} {et~al.}(2013){Zdziarski}, {Mikolajewska}, \&
  {Belczynski}}]{zdziarski13}
{Zdziarski}, A.~A., {Mikolajewska}, J., \& {Belczynski}, K. 2013, \mnras, 429,
  L104

\bibitem[{{Zdziarski} {et~al.}(2010){Zdziarski}, {Misra}, \&
  {Gierli{\'n}ski}}]{zdziarski10}
{Zdziarski}, A.~A., {Misra}, R., \& {Gierli{\'n}ski}, M. 2010, \mnras, 402, 767

\bibitem[{{Zdziarski} {et~al.}(2016){Zdziarski}, {Segreto}, \&
  {Pooley}}]{zdziarski16}
{Zdziarski}, A.~A., {Segreto}, A., \& {Pooley}, G.~G. 2016, \mnras, 456, 775

\bibitem[{{{\.Z}ycki} {et~al.}(1999){{\.Z}ycki}, {Done}, \& {Smith}}]{zycki99}
{{\.Z}ycki}, P.~T., {Done}, C., \& {Smith}, D.~A. 1999, \mnras, 309, 561

\end{thebibliography}

\begin{table*}
\centering
\caption{Model A parameters for V404 Cyg epochs. Parameters without errors are kept frozen at the value shown. The dash marks components that are not part of the spectral model in a particular data set.}
\label{modela1}
\begin{tabular}{lccccccc}
\multicolumn{7}{c}{Model A: \textsc{c1} $\times$ \textsc{phabs} $\times$ \textsc{smedge} $\times$ \textsc{pcfabs} $\times$ (\textsc{relxillCp} + \textsc{xillverCp} + \textsc{gauss})} \\
\toprule
& & Epoch 1 & Epoch 2 & Epoch 3 & Epoch 4 & Epoch 5 \\
\midrule
\multicolumn{7}{l}{\textsc{phabs} $\times$ \textsc{smedge} $\times$ \textsc{pcfabs}} \\
\midrule
\hspace{0.1cm} N$_{\mathrm{H,1}}$ & 10$^{22}$ cm$^{-2}$ & 2.8$^{+0.6}_{-0.7}$ & 1.8$^{+0.5}_{-0.6}$ & 2.7$\pm$0.3 & 0.83--0.93 & 2.7$\pm$0.1 \\
\hspace{0.1cm} E & keV & 7.41$\pm$0.07 & 7.45$^{+0.07}_{-0.06}$ & -- & -- & -- \\
\hspace{0.1cm} $\tau$ & & 0.32$^{+0.05}_{-0.04}$ & 0.36$\pm$0.04 & -- & -- & -- \\
\hspace{0.1cm} $\sigma$ & keV & 1 & 1 & -- & -- & -- \\
\hspace{0.1cm} N$_{\mathrm{H,2}}$ & 10$^{22}$ cm$^{-2}$ & 39$\pm$2 & 35$\pm$2 & -- & -- & -- \\
\hspace{0.1cm} f$_{\mathrm{cov}}$ & & 0.82$\pm$0.01 & 0.75$\pm$0.01 & -- & -- & -- \\
\midrule
\multicolumn{7}{l}{\textsc{relxillCp}} \\
\midrule
\hspace{0.1cm} norm & $\times$10$^{-3}$ & 27$\pm$2 & 47$^{+5}_{-3}$ & 15$\pm$1 & 14.1$\pm$0.7 & 15.3$^{+0.9}_{-0.3}$ \\ 
\hspace{0.1cm} $\theta_{\mathrm{inc}}$ & deg & 22$^{+5}_{-6}$ & $>$78 & 32$\pm$1 & 35$\pm$1 & 37.1$\pm$0.4 \\
\hspace{0.1cm} R$_{\mathrm{in}}$ & R$_{\mathrm{ISCO}}$ & 5$^{+3}_{-2}$ & 20$^{+7}_{-4}$ & $<$1.1 & $<$1.2 & $<$1.1 \\ 
\hspace{0.1cm} z & & $<$0.013 & 0.031$\pm$0.004 & 0.011$\pm$0.002 & 0.013$^{+0.001}_{-0.002}$ & 0.019$\pm$0.002 \\
\hspace{0.1cm} $\Gamma$ & & 1.85$\pm$0.01 & 2.05$\pm$0.02 & 1.58$^{+0.01}_{-0.02}$ & 1.35$\pm$0.01 & 1.80$\pm$0.01 \\
\hspace{0.1cm} log $\xi$ & & 3.43$^{+0.04}_{-0.03}$ & 3.6$\pm$0.1 & 2.86$^{+0.05}_{-0.03}$ & 3.09$\pm$0.01 & 2.86$^{+0.01}_{-0.02}$ \\
\hspace{0.1cm} A$_{\mathrm{Fe}}$ & & 1.7$^{+0.2}_{-0.1}$ & 1.8$\pm$0.3 & 0.75$^{+0.02}_{-0.03}$ & 1.66$\pm$0.08 & 0.91$\pm$0.01 \\
\hspace{0.1cm} kT$_{\mathrm{e}}$ & keV & 3.9$\pm$0.2 & 2.5$\pm$0.1 & 50$^{+15}_{-3}$ & 17.5$\pm$0.2 & 90$^{+28}_{-13}$ \\
\midrule
\multicolumn{7}{l}{\textsc{xillverCp}} \\
\midrule
\hspace{0.1cm} log $\xi$ & & -- & -- & $<$1.7 & 2.35$\pm$0.02 & 1.7$^{+0.1}_{-0.2}$ \\
\hspace{0.1cm} norm & $\times$10$^{-3}$ & -- & --  & 26$\pm$3 & 44$\pm$2 & 14$^{+1}_{-3}$ \\ 
\midrule
\multicolumn{7}{l}{\textsc{gauss}} \\
\midrule
\hspace{0.1cm} E & keV & 6.52$^{+0.01}_{-0.04}$ &  6.48$^{+0.01}_{-0.04}$ & -- & -- & -- \\
\hspace{0.1cm} norm & $\times$10$^{-3}$ & -11$\pm$2 & -8$\pm$2 & -- & -- & -- \\
\midrule
\multicolumn{7}{l}{Constants$^{a}$} \\
\midrule
\hspace{0.1cm} \textsc{c1} & & 1.00 & 0.99 & 1.01 & 1.00 & 1.00 \\
\midrule
\multicolumn{7}{l}{Fluxes (10$^{-8}$ erg s$^{-1}$ cm$^{-2}$) } \\
\midrule
\hspace{0.1cm} Abs. & & 1.25$\pm$0.06 & 1.35$\pm$0.07 & 4.2$\pm$0.1 & 6.8$\pm$0.1 & 2.8$\pm$0.1 \\
\hspace{0.1cm} Unabs. & & 2.5$\pm$0.1 & 2.9$\pm$0.1 & 4.2$\pm$0.1 & 6.8$\pm$0.1 & 2.9$\pm$0.1 \\
\hspace{0.3cm} \textsc{relxill} & & 100\% & 100\% & 61\% & 45\% & 78\% \\
\hspace{0.3cm} \textsc{xillver} & & -- & -- & 39\% & 55\% & 22\% \\
\midrule
\multicolumn{7}{l}{Luminosities$^{b}$ and Eddington ratio$^{c}$ (3--79 keV)} \\
\midrule
\hspace{0.1cm} Abs. & 10$^{38}$ erg s$^{-1}$ & 0.09$\pm$0.02 & 0.09$\pm$0.03 & 0.29$\pm$0.05 & 0.46$\pm$0.07 & 0.19$\pm$0.03 \\
\hspace{0.1cm} Unabs. & 10$^{38}$ erg s$^{-1}$ & 0.17$\pm$0.03  & 0.20$\pm$0.03 & 0.29$\pm$0.05 & 0.46$\pm$0.07 & 0.20$\pm$0.03 \\
\hspace{0.3cm} L/L$_{\mathrm{Edd}}$ & & 0.015$^{+0.004}_{-0.002}$ & 0.017$^{+0.005}_{-0.002}$ & 0.025$^{+0.006}_{-0.004}$ & 0.041$^{+0.009}_{-0.006}$ & 0.017$^{+0.005}_{-0.002}$ \\
\midrule
\multicolumn{2}{l}{Both \nustar\/ detectors:} & (A/B) & (A/B) & (A/B) & (A/B) & (A/B) \\
\hspace{0.1cm} $\chi^{2}$/d.o.f & & 872/617 & 716/510 & 441/378 & 1088/827 & 952/747 \\
\hspace{0.1cm} $\chi_{\mathrm{red}}^{2}$ & & 1.41 & 1.40 & 1.17 & 1.32 & 1.27 \\
\addlinespace
\multicolumn{2}{l}{Single \nustar\/ detector:} & (A) & (B) & (A) & (B) & (B) \\
\hspace{0.1cm} $\chi^{2}$/d.o.f & & 335/306 & 292/244 & 207/189 & 460/400 & 454/362 \\
\hspace{0.1cm} $\chi_{\mathrm{red}}^{2}$ & & 1.10 & 1.20 & 1.10 & 1.15 & 1.25 \\
\bottomrule
\end{tabular}
\tablefoot{
\tablefoottext{a}{c1 is a constant between the FPMA and FPMB allowed to vary}
\tablefoottext{b}{The following distance estimates have been used: 2.39$\pm$0.14 kpc \citep{millerjones09}}
\tablefoottext{c}{The following mass estimates have been used: 9.0$^{+0.2}_{-0.6}$ M$_{\odot}$ \citep{khargharia10}}
}
\end{table*}

\begin{table*}
\centering
\caption{Model A parameters for GRS 1915$+$105, Cyg X-3 and V4641 Sgr. Parameters without errors are kept frozen at the value shown. The dash marks components that are not part of the spectral model in a particular data set.}
\label{modela2}
\begin{tabular}{lccccccccc}
\multicolumn{8}{c}{Model A: \textsc{c1} $\times$ \textsc{phabs} $\times$ \textsc{smedge} $\times$ \textsc{pcfabs} $\times$ (\textsc{xillverCp1/relxillCp} + \textsc{xillverCp2} + \textsc{gauss})} \\
\toprule
& & \multicolumn{3}{c}{GRS 1915$+$105} & Cyg X-3 & \multicolumn{2}{c}{V4641 Sgr} \\
\cmidrule(lr){3-5} \cmidrule(lr){6-6} \cmidrule(lr){7-8}
& & Epoch 1 & Epoch 2 & Epoch 3 & & Epoch 1 & Epoch 2 \\
\midrule
\multicolumn{8}{l}{\textsc{phabs} $\times$ \textsc{smedge} $\times$ \textsc{pcfabs}} \\
\midrule
\hspace{0.1cm} N$_{\mathrm{H,1}}$ & 10$^{22}$ cm$^{-2}$ & 5.5$\pm$0.1 & 5$\pm$1 & 7.2$\pm$0.4 & 8.2$^{+0.4}_{-0.5}$ & 6.2$\pm$0.6 & 3.3$\pm$0.5 \\
\hspace{0.1cm} E & keV & 7.34$^{+0.03}_{-0.04}$ & -- & -- & -- & -- & -- \\
\hspace{0.1cm} $\tau$ & & 0.15$^{+0.02}_{-0.01}$ & -- & -- & -- & -- & -- \\
\hspace{0.1cm} $\sigma$ & keV & 0.01 & -- & -- & -- & -- & -- \\
\hspace{0.1cm} N$_{\mathrm{H,2}}$ & 10$^{22}$ cm$^{-2}$ & -- & 41$\pm$5 & 78$\pm$7 & 44$\pm$3 & -- & -- \\
\hspace{0.1cm} f$_{\mathrm{cov}}$ & & -- & 0.58$\pm$0.04 & 0.57$\pm$0.04 & 0.54$\pm$0.02 & -- & -- \\
\midrule
\multicolumn{8}{l}{\textsc{xillverCp1/relxillCp}} \\
\midrule
\hspace{0.1cm} norm & $\times$10$^{-3}$ & 1.24$\pm$0.04 & 2.2$\pm$0.1 & 3.3$^{+0.3}_{-0.2}$ & 27$^{+6}_{-4}$ & 0.6$\pm$0.1 & 0.47$\pm$0.01 \\ 
\hspace{0.1cm} $\theta_{\mathrm{inc}}$ & deg & 70 & 70 & 70 & 59$\pm$5 & 70 & 70 \\
\hspace{0.1cm} R$_{\mathrm{in}}$ & & 3.9$^{+0.01}_{-0.02}$ & -- & -- & -- & 27$^{+43}_{-11}$ & 21$^{+7}_{-4}$\\
\hspace{0.1cm} z & & 0.124$^{+0.008}_{-0.02}$ & 0.010$\pm$0.002 & 0.021$\pm$0.005 & 0.003 & -- & -- \\
\hspace{0.1cm} $\Gamma$ & & 1.85$\pm$0.01 & 1.58$\pm$0.02 & 1.64$\pm$0.03 & 2.39$^{+0.03}_{-0.02}$ & 2.17$^{+0.08}_{-0.05}$ & 1.46$\pm$0.01 \\
\hspace{0.1cm} log $\xi$ & & 3.35$^{+0.02}_{-0.01}$ & 3.40$^{+0.03}_{-0.02}$ & 3.37$^{+0.03}_{-0.02}$ & 3.79$^{+0.07}_{-0.04}$ & 3.05$\pm$0.02 & 2.80$^{+0.02}_{-0.01}$ \\
\hspace{0.1cm} A$_{\mathrm{Fe}}$ & & 1.2$\pm$0.2 & 1.2$\pm$0.2 & 2.3$\pm$0.4 & $<$0.56 & 1.8$\pm$0.2 & 1.5$\pm$0.1 \\
\hspace{0.1cm} kT$_{\mathrm{e}}$ & keV & 400 & 15.6$\pm$0.6 & 12.9$\pm$0.4 & 33$^{+5}_{-6}$ & $>$75 & 9.3$\pm$0.1 \\
\midrule
\multicolumn{8}{l}{\textsc{xillverCp2}} \\
\midrule
\hspace{0.1cm} log $\xi$ & & -- & 2.00$^{+0.04}_{-0.11}$ & 2.1$^{+0.2}_{-0.1}$ & 2.88$^{+0.08}_{-0.07}$ & -- & -- \\
\hspace{0.1cm} norm & $\times$10$^{-3}$ & -- & 2.0$\pm$0.2 & 1.3$^{+0.5}_{-0.4}$ & 33$^{+5}_{-4}$ & -- & -- \\ 
\midrule
\multicolumn{8}{l}{\textsc{gauss}} \\
\midrule
\hspace{0.1cm} E & keV & 6.56$^{+0.01}_{-0.04}$ & -- & -- & -- & -- & --  \\
\hspace{0.1cm} norm & $\times$10$^{-3}$ & -0.78$\pm$0.05 & -- & -- & -- & -- & -- \\
\midrule
\multicolumn{8}{l}{Constants$^{a}$} \\
\midrule
\hspace{0.1cm} \textsc{c1} & & 1.01 & 1.02 & 1.01 & 1.01 & -- & -- \\
\midrule
\multicolumn{8}{l}{Fluxes (10$^{-8}$ erg s$^{-1}$ cm$^{-2}$) } \\
\midrule
\hspace{0.1cm} Abs. & & 0.216$\pm$0.005 & 0.155$\pm$0.005 & 0.151$\pm$0.008 & 0.59$\pm$0.05 & 0.06$\pm$0.01 & 0.10$\pm$0.01 \\
\hspace{0.1cm} Unabs. & & 0.237$\pm$0.005 & 0.180$\pm$0.005 & 0.20$\pm$0.01 & 0.93$\pm$0.08 & 0.07$\pm$0.01 & 0.12$\pm$0.01 \\
\hspace{0.3cm} \textsc{xillverCp1/relxillCp} & & 100\%  & 65\% & 88\% & 58\% & 100\% & 100\% \\
\hspace{0.3cm} \textsc{xillverCp2} & & -- & 35\% & 12\% & 42\% & -- & -- \\
\midrule
\multicolumn{8}{l}{Luminosities$^{b}$ and Eddington ratio$^{c}$ (3--79 keV)} \\
\midrule
\hspace{0.1cm} Abs. & 10$^{38}$ erg s$^{-1}$ & 0.19$^{+0.12}_{-0.07}$ & 0.14$^{+0.09}_{-0.05}$ & 0.13$^{+0.09}_{-0.05}$ & 0.4$^{+0.2}_{-0.1}$ & 0.03$\pm$0.01 & 0.05$\pm$0.01 \\
\hspace{0.1cm} Unabs. & 10$^{38}$ erg s$^{-1}$ & 0.21$^{+0.13}_{-0.08}$ & 0.16$^{+0.10}_{-0.06}$ & 0.17$^{+0.10}_{-0.06}$ & 0.6$^{+0.3}_{-0.2}$ & 0.03$\pm$0.01 & 0.05$^{+0.02}_{-0.01}$ \\
\hspace{0.3cm} L/L$_{\mathrm{Edd}}$ & & 0.013$^{+0.011}_{-0.006}$ & 0.010$^{+0.009}_{-0.005}$ & 0.010$^{+0.010}_{-0.004}$ & 0.10$^{+0.16}_{-0.07}$ & 0.004$^{+0.001}_{-0.002}$ & 0.007$\pm$0.002 \\
\midrule
\multicolumn{2}{l}{Both \nustar\/ detectors:} & (A/B) & (A/B) & (A/B) & (A/B) \\
\hspace{0.1cm} $\chi^{2}$/d.o.f & & 787/655 & 500/431 & 666/441 & 919/667 & 49/47 & 54/58 \\
\hspace{0.1cm} $\chi_{\mathrm{red}}^{2}$ & & 1.20 & 1.16 & 1.51 & 1.38 & 1.05 & 0.93 \\
\addlinespace
\multicolumn{2}{l}{Single \nustar\/ detector:} & (B) & (B) & (A) & (B) \\
\hspace{0.1cm} $\chi^{2}$/d.o.f & & 340/317 & 208/203 & 292/223 & 407/323 \\
\hspace{0.1cm} $\chi_{\mathrm{red}}^{2}$ & & 1.07 & 1.02 & 1.31 & 1.26\\
\bottomrule
\end{tabular}
\tablefoot{
\tablefoottext{a}{c1 is a constant between the \nustar\/ detectors FPMA and FPMB and is allowed to vary during the fitting.}
\tablefoottext{b}{The following distance estimates have been used: GRS 1915$+$105: 8.6$^{+2.0}_{-1.6}$ kpc \citep{reid14}, Cyg X-3: 7.4$\pm$1.1 kpc \citep{mccollough16}, V4641 Sgr: 6.2$\pm$0.7 kpc \citep{macdonald14}.}
\tablefoottext{c}{The following mass estimates have been used: GRS 1915$+$105: 12.4$^{+2.0}_{-1.8}$ M$_{\odot}$ \citep{reid14}, Cyg X-3: 2.4$^{+2.1}_{-1.1}$ M$_{\odot}$ \citep{zdziarski13}, V4641 Sgr: 6.4$\pm$0.6 M$_{\odot}$ \citep{macdonald14}.}
}
\end{table*}

\begin{table*}
\centering
\caption{Model B parameters for V404 Cyg epochs. Parameters without errors are kept frozen at the value shown. The dash marks components that are not part of the spectral model in a particular data set.}
\label{modelb1}
\begin{tabular}{lcccccc}
\multicolumn{7}{c}{Model B: \textsc{c1} $\times$ \textsc{phabs1} $\times$ \textsc{smedge} $\times$ (\textsc{c2} $\times$ \textsc{borus02(red)} + \textsc{c3} $\times$ \textsc{borus02(blue)}  + \textsc{phabs2} $\times$ \textsc{cabs} $\times$ \textsc{cutoffpl}} \\
\multicolumn{7}{c}{+ \textsc{c4} $\times$ \textsc{cutoffpl} + \textsc{pcfabs} $\times$ \textsc{bbody})} \\
\toprule
& & Epoch 1 & Epoch 2 & Epoch 3 & Epoch 4 & Epoch 5 \\
\midrule
\multicolumn{7}{l}{\textsc{phabs1} $\times$ \textsc{smedge}} \\
\midrule
N$_{\mathrm{H,ism}}$ & 10$^{22}$ cm$^{-2}$ & 0.83 & 0.83 & 0.83 & 0.83 & 0.83 \\
E & keV & 8.67$\pm$0.05 & 8.62$\pm$0.04 & -- & -- & 8.5$\pm$0.1 \\
$\tau$ & & 0.30$\pm$0.02 & 0.36$^{+0.02}_{-0.03}$ & -- & -- & 0.16$\pm$0.01 \\
$\sigma$ & keV & 0.5 & 0.5 & -- & -- & 0.5 \\
\midrule
\multicolumn{7}{l}{\textsc{borus02}} \\
\midrule
log N$_{\mathrm{H,tor}}$ & & 24.21$^{+0.05}_{-0.19}$ & 24.00$^{+0.06}_{-0.04}$ & 24.58$\pm$0.02 & 24.703$\pm$0.006 & 24.66$\pm$0.01 \\
cos($\theta_{\mathrm{tor}}$) & & $>$0.78 & $>$0.57 & 0.55$\pm$0.01 & 0.55$\pm$0.01 & 0.56$\pm$0.01 \\ 
cos($\theta_{\mathrm{inc}}$) & & 0.39 & 0.39 & 0.39 & 0.39 & 0.39 \\
$z$ & & 0.042$^{+0.004}_{-0.002}$ & 0.056$^{+0.004}_{-0.003}$ & 0.014$^{+0.004}_{-0.002}$ & 0.014$\pm$0.001 & 0.022$\pm$0.002 \\
\midrule
\multicolumn{7}{l}{\textsc{phabs2} $\times$ \textsc{cabs} $\times$ \textsc{cutoffpl}} \\
\midrule
log N$_{\mathrm{H,los}}$ & & 24.36$^{+0.04}_{-0.05}$ & 24.24$^{+0.02}_{-0.03}$ & 23.53$^{+0.02}_{-0.03}$ & 23.58$^{+0.02}_{-0.03}$ & 23.57$\pm$0.01 \\
norm & & 262$^{+201}_{-112}$ & 92$^{+19}_{-15}$ & 3.2$\pm$0.5 & 6.1$^{+0.8}_{-0.6}$ & 14.2$^{+0.9}_{-0.8}$ \\
$\Gamma$ & & 2.6 & 2.6 & 1.92$\pm$0.05 & 2.05$^{+0.04}_{-0.03}$ & 2.50$\pm$0.02 \\
E$_{\mathrm{cut}}$ & keV & 21.8$^{+0.7}_{-0.3}$ & 26.4$^{+0.9}_{-1.3}$ & 94$^{+23}_{-14}$ & 62$^{+5}_{-3}$ & $>$474 \\
\midrule
\multicolumn{7}{l}{\textsc{pcfabs} $\times$ \textsc{bbody}} \\
\midrule
log N$_{\mathrm{H,bb}}$ & & 23.782$^{+0.008}_{-0.009}$ & 23.73$\pm$0.01 & -- & -- & -- \\
cov & & 0.821$^{+0.006}_{-0.005}$ & 0.76$^{+0.02}_{-0.01}$ & -- & -- & -- \\ 
norm & & 0.32$\pm$0.01 & 0.44$^{+0.02}_{-0.01}$ & -- & -- & -- \\
kT & keV & 1.120$\pm$0.005 & 0.95$\pm$0.01 & -- & -- & -- \\
\midrule
\multicolumn{7}{l}{Constants$^{a}$} \\
\midrule
\textsc{c1} & & 1.00 & 0.99 & 1.01 & 1.00 & 1.00 \\ 
\textsc{c2} & & 0.11$^{+0.13}_{-0.08}$ & 0.07$^{+0.03}_{-0.02}$ & 7.7$\pm$0.7 & 15.7$\pm$0.8 & 7.3$\pm$0.4 \\
\textsc{c3} & & 0.05$^{+0.05}_{-0.03}$ & 0.10$^{+0.04}_{-0.03}$ & 2.2$^{+0.7}_{-0.5}$ & 8.1$\pm$0.6 & 3.7$\pm$0.4 \\
\textsc{c4} & & 0 & 0.034$^{+0.008}_{-0.007}$ & 0.24$^{+0.03}_{-0.02}$ & 0.22$\pm$0.02 & 0.185$^{+0.008}_{-0.007}$ \\
\midrule
\multicolumn{7}{l}{Fluxes (10$^{-8}$ erg s$^{-1}$ cm$^{-2}$) } \\
\midrule
Abs. & & 1.4$\pm$0.3 & 1.36$\pm$0.07 & 4.2$\pm$0.5 & 7.0$\pm$0.6 & 2.8$\pm$0.1 \\
Unabs. & & 23$\pm$6 & 9.8$\pm$0.7 & 5.0$\pm$0.5 & 8.1$\pm$0.5 & 4.0$\pm$0.1 \\
\hspace{0.2cm} LOS & & 90\% & 74\% & 33\% & 25\% & 50\% \\
\hspace{0.2cm} Scattered & & 2\% & 2\% & 59\% & 70\% & 40\% \\
\hspace{0.2cm} Leaked & & -- & 2\% & 8\% & 5\% & 10\% \\
\hspace{0.2cm} BB & & 8\% & 22\% & -- & -- & -- \\
\midrule
\multicolumn{7}{l}{Luminosities$^{b}$ and Eddington ratio$^{c}$ (3--79 keV)} \\
\midrule
Abs. & 10$^{38}$ erg s$^{-1}$ & 0.09$\pm$0.03 & 0.09$\pm$0.02 & 0.29$\pm$0.08 & 0.48$\pm$0.10 & 0.19$\pm$0.03 \\
Unabs. & 10$^{38}$ erg s$^{-1}$ & 1.6$\pm$0.2 & 0.67$\pm$0.13 & 0.34$\pm$0.08 & 0.55$\pm$0.10 & 0.28$\pm$0.04 \\
\hspace{0.2cm} L/L$_{\mathrm{Edd}}$ & & 0.14$^{+0.07}_{-0.05}$ & 0.06$^{+0.02}_{-0.01}$ & 0.03$\pm$0.01 & 0.05$\pm$0.01 & 0.02$\pm$0.01 \\
\midrule
\multicolumn{2}{l}{Both \nustar\/ detectors:} & (A/B) & (A/B) & (A/B) & (A/B) & (A/B) \\
\hspace{0.1cm} $\chi^{2}$/d.o.f & & 877/619 & 758/510 & 451/379 & 1143/828 & 945/746 \\
\hspace{0.1cm} $\chi_{\mathrm{red}}^{2}$ & & 1.42 & 1.49 & 1.19 & 1.38 & 1.27 \\
\addlinespace
\multicolumn{2}{l}{Single \nustar\/ detector:} & (A) & (B) & (A) & (B) & (B) \\
\hspace{0.1cm} $\chi^{2}$/d.o.f & & 392/308 & 297/244 & 207/189 & 504/401 & 448/362 \\
\hspace{0.1cm} $\chi_{\mathrm{red}}^{2}$ & & 1.27 & 1.22 & 1.10 & 1.26 & 1.24 \\
\bottomrule
\end{tabular}
\tablefoot{
\tablefoottext{a}{c1 is a constant between the FPMA and FPMB allowed to vary}
\tablefoottext{b}{The following distance estimates have been used: 2.39$\pm$0.14 kpc \citep{millerjones09}}
\tablefoottext{c}{The following mass estimates have been used: 9.0$^{+0.2}_{-0.6}$ M$_{\odot}$ \citep{khargharia10}}
}
\end{table*}

\begin{table*}
\centering
\caption{Model B parameters for GRS 1915$+$105, Cyg X-3 and V4641 Sgr. Parameters without errors are kept frozen at the value shown. The dash marks components that are not part of the spectral model in a particular data set.}
\label{modelb2}
\begin{tabular}{lccccccc}
\multicolumn{8}{c}{Model B: \textsc{c1} $\times$ \textsc{phabs1} $\times$ \textsc{smedge} $\times$ (\textsc{c2} $\times$ \textsc{borus02(red)} + \textsc{c3} $\times$ \textsc{borus02(blue)}  + \textsc{phabs2} $\times$ \textsc{cabs} $\times$ \textsc{cutoffpl}} \\
\multicolumn{8}{c}{+ \textsc{const4} $\times$ \textsc{cutoffpl} +  \textsc{gauss1} +  \textsc{gauss2})} \\
\toprule
& & \multicolumn{3}{c}{GRS 1915$+$105} & Cyg X-3 & \multicolumn{2}{c}{V4641 Sgr} \\
\cmidrule(lr){3-5} \cmidrule(lr){6-6} \cmidrule(lr){7-8}
& & Epoch 1 & Epoch 2 & Epoch 3 & & Epoch 1 & Epoch 2 \\
\midrule
\multicolumn{8}{l}{\textsc{phabs1} $\times$ \textsc{smedge}} \\
\midrule
N$_{\mathrm{H,ism}}$ & 10$^{22}$ cm$^{-2}$ & 3.5 & 3.5 & 3.5 & 3.5 & 0.2 & 0.2 \\
E & keV & 7.13$^{+0.03}_{-0.02}$ & 8.8$\pm$0.1 & 8.47$^{+0.06}_{-0.05}$ & 8.86$^{+0.04}_{-0.06}$ & 9.3$\pm$0.1 & 9.2$\pm$0.1 \\
$\tau$ & & 0.85$^{+0.02}_{-0.01}$ & 0.48$\pm$0.05 & 1.07$^{+0.08}_{-0.07}$ & 0.37$^{+0.03}_{-0.02}$ & 0.78$^{+0.05}_{-0.06}$ & 0.46$\pm$0.04 \\
$\sigma$ & keV & 2.0 & 2.0 & 2.7 & 0.7$\pm$0.1 & 0.5 & 0.5 \\
\midrule
\multicolumn{8}{l}{\textsc{borus02}} \\
\midrule
log N$_{\mathrm{H,tor}}$ & & 24.37$\pm$0.02 & 24.40$\pm$0.04 & 24.45$^{+0.05}_{-0.06}$ & 24.03$\pm$0.02 & 25.2$^{+0.2}_{-0.9}$ & $>$25.4 \\
cos($\theta_{\mathrm{tor}}$) & & $>$0.94 & 0.80$^{+0.05}_{-0.13}$ & 0.57$\pm$0.02 & $>$0.74 & $<$0.36 & 0.36$\pm$0.01 \\ 
cos($\theta_{\mathrm{inc}}$) & & 0.6 & 0.68$^{+0.06}_{-0.16}$ & 0.45$^{+0.01}_{-0.02}$ & 0.75$^{+0.02}_{-0.04}$ & 0.35 & 0.35 \\
$z$ & & 0.057$\pm$0.008 & 0.007$^{+0.006}_{-0.003}$ & $<$0.004 & 0.003 & 0.029$^{+0.004}_{-0.003}$ & 0.016$\pm$0.002 \\
\midrule
\multicolumn{8}{l}{\textsc{phabs2} $\times$ \textsc{cabs} $\times$ \textsc{cutoffpl}} \\
\midrule
log N$_{\mathrm{H,los}}$ & & 23.35$\pm$0.01 & 23.67$\pm$0.02 & 23.71$\pm$0.02 & 23.52$\pm$0.01 & 22.8$\pm$0.1 & 22.2$^{+0.2}_{-0.3}$ \\
norm & & 0.65$^{+0.03}_{-0.04}$ & 0.20$\pm$0.02 & 0.44$^{+0.04}_{-0.03}$ & 5.4$^{+0.6}_{-0.4}$ & 0.18$\pm$0.02 & 0.05$\pm$0.01 \\
$\Gamma$ & & 2.27$\pm$0.02 & 1.60$^{+0.02}_{-0.03}$ & 1.80$^{+0.02}_{-0.01}$ & 2.32$\pm$0.03 & $>$2.54 & 1.64$^{+0.06}_{-0.05}$ \\
E$_{\mathrm{cut}}$ & keV & $>$388 & 23.6$^{+0.5}_{-0.12}$ & 20.8$\pm$0.4 & 26$^{+2}_{-1}$ & $>$74 & 20 \\
\midrule
\multicolumn{8}{l}{\textsc{gauss1} +  \textsc{gauss2}} \\
\midrule
E$_{1}$ & keV & 6.56$\pm$0.04 & 6.76$\pm$0.08 & 6.70$\pm$0.02 & 6.67 & -- & -- \\
$\sigma_{1}$ & keV & 0.002 & 0.002 & 0.002 & 0.002 & -- & -- \\
norm & $\times$10$^{-3}$ & -0.79$\pm$0.04 & 0.20$^{+0.04}_{-0.03}$ & 0.55$^{+0.04}_{-0.03}$ & 6$\pm$0.2 & -- & -- \\
E$_{2}$ & keV & 8.36$\pm$0.08 & -- & -- & 7.8 & -- & -- \\
$\sigma_{2}$ & keV & 0.002 & -- & -- & 0.002 & -- & -- \\
norm & $\times$10$^{-3}$ & 0.2$\pm$0.03 & -- & -- & 0.6$\pm$0.1 & -- & -- \\
\midrule
\multicolumn{8}{l}{Constants} \\
\midrule
\textsc{c1} & & 1.01 & 1.02 & 1.00 & 1.01 & -- & -- \\ 
\textsc{c2} & & 1.4$\pm$0.1 & 0.92$^{+0.04}_{-0.15}$ & 2.4$^{+0.2}_{-0.3}$ & 0.73$^{+0.09}_{-0.08}$ & -- & -- \\
\textsc{c3} & & $<$0.18 & $<$0.42 & $<$0.23 & 0.5$\pm$0.1 & 43$^{+7}_{-16}$ & 30$^{+7}_{-6}$  \\
\textsc{c4} & & 0.59$\pm$0.02 & 0.18$\pm$0.01 & 0.14$\pm$0.01 & 0.26$\pm$0.01 & -- & -- \\
\midrule
\multicolumn{8}{l}{Fluxes (10$^{-8}$ erg s$^{-1}$ cm$^{-2}$) } \\
\midrule
Abs. & & 0.22$\pm$0.01 & 0.15$\pm$0.02 & 0.15$\pm$0.02 & 0.60$\pm$0.04 & 0.07$\pm$0.02 & 0.12$\pm$0.02 \\
Unabs. & & 0.31$\pm$0.01 & 0.24$\pm$0.02 & 0.28$\pm$0.02 & 1.11$\pm$0.05 & 0.08$\pm$0.02 & 0.13$\pm$0.02 \\
\hspace{0.2cm} LOS & & 52\% & 55\% & 58\% & 67\% & 26\% & 20\% \\
\hspace{0.2cm} Scattered & & 17\% & 35\% & 32\% & 16\% & 74\% & 80\% \\
\hspace{0.2cm} Leaked & & 31\% & 10\% & 10\% & 17\% & 0\% & 0\% \\
\midrule
\multicolumn{8}{l}{Luminosities$^{a}$ and Eddington ratio$^{b}$ (3--79 keV)} \\
\midrule
Abs. & 10$^{38}$ erg s$^{-1}$ & 0.19$^{+0.12}_{-0.07}$ & 0.13$^{+0.09}_{-0.05}$ & 0.13$^{+0.09}_{-0.05}$ & 0.4$^{+0.2}_{-0.1}$ & 0.03$^{+0.02}_{-0.01}$ & 0.06$^{+0.01}_{-0.02}$ \\
Unabs. & 10$^{38}$ erg s$^{-1}$ & 0.27$^{+0.16}_{-0.09}$ & 0.21$^{+0.14}_{-0.08}$ & 0.25$^{+0.16}_{-0.10}$ & 0.7$^{+0.3}_{-0.2}$ & 0.03$^{+0.02}_{-0.01}$ & 0.06$\pm$0.01 \\
\hspace{0.2cm} L/L$_{\mathrm{Edd}}$ & & 0.02$\pm$0.01 & 0.013$^{+0.012}_{-0.006}$ & 0.016$^{+0.015}_{-0.008}$ & 0.12$^{+0.18}_{-0.07}$ & 0.004$^{+0.003}_{-0.002}$ & 0.007$^{+0.003}_{-0.002}$ \\
\midrule
\multicolumn{2}{l}{Both \nustar\/ detectors:} & (A/B) & (A/B) & (A/B) & (A/B) \\
\hspace{0.1cm} $\chi^{2}$/d.o.f & & 818/650 & 440/427 & 716/439 & 907/663 & 50/44 & 64/56 \\
\hspace{0.1cm} $\chi_{\mathrm{red}}^{2}$ & & 1.26 & 1.03 & 1.63 & 1.37 & 1.13 & 1.15 \\
\addlinespace
\multicolumn{2}{l}{Single \nustar\/ detector:} & (B) & (B) & (A) & (B) \\
\hspace{0.1cm} $\chi^{2}$/d.o.f & & 348/312 & 180/198 & 244/220 & 392/318 \\
\hspace{0.1cm} $\chi_{\mathrm{red}}^{2}$ & & 1.12 & 0.91 & 1.11 & 1.23 \\
\bottomrule
\end{tabular}
\tablefoot{
\tablefoottext{a}{The following distance estimates have been used: GRS 1915$+$105: 8.6$^{+2.0}_{-1.6}$ kpc \citep{reid14}, Cyg X-3: 7.4$\pm$1.1 kpc \citep{mccollough16}, V4641 Sgr: 6.2$\pm$0.7 kpc \citep{macdonald14}.}
\tablefoottext{b}{The following mass estimates have been used: GRS 1915$+$105: 12.4$^{+2.0}_{-1.8}$ M$_{\odot}$ \citep{reid14}, Cyg X-3: 2.4$^{+2.1}_{-1.1}$ M$_{\odot}$ \citep{zdziarski13}, V4641 Sgr: 6.4$\pm$0.6 M$_{\odot}$ \citep{macdonald14}. For Cyg X-3 we use Eddington ratio that is twice the normal due to the accreting matter being helium-dominated.}
}
\end{table*}

\end{document}